\documentclass[mathpazo]{cicp}

\usepackage{epsfig,rotating}
\usepackage[top=3truecm, bottom=2.5truecm, left=2.5truecm, right=2.5truecm]{geometry}
\usepackage[dvips,usenames]{color}
\usepackage{graphicx}
\usepackage{epsf}
\usepackage{epsfig}
\usepackage{subfigure}
\usepackage{multirow}
\usepackage{lscape}
\usepackage{amsmath}
\usepackage{tabularx}
\usepackage{textcomp}
\usepackage{algorithm,algorithmic}

\newcounter{question}
\usepackage[pdfpagemode=UseNone,pdfstartview=FitH,colorlinks=true,citecolor=OliveGreen,linkcolor=red]{hyperref}

\newcommand{\DistMesh}{{\sf DistMesh}}
\newcommand{\Matlab}{{\sf MATLAB}}
\newcommand{\MatlabReg}{\Matlab\textsuperscript{\textregistered}}
\newcommand{\Ansys}{{\sf ANSYS\textsuperscript{\textregistered}}}
\newcommand{\myunit}[1]{\mathrm{#1}}
\newcommand{\mymum}{\myunit{\mbox{$\mu$}m}}
\newcommand{\leavethisout}[1]{}
\newcommand{\Wb}{W_b}
\newcommand{\Hb}{H_b}
\newcommand{\Db}{D_b}
\newcommand{\stiff}{\kappa}
\newcommand{\stress}{\sigma}
\newcommand{\stressAdh}{\Sigma_{adh}}
\newcommand{\stressCoh}{\Sigma_{coh}}
\newcommand{\stressCohExt}{\stressCoh^{ext}}
\newcommand{\stressCohInt}{\stressCoh^{int}}
\newcommand{\Ascale}{{\cal A}}
\newcommand{\deltah}{\delta_{h}}
\newcommand{\deltamin}{\epsilon_{min}}
\newcommand{\deltasub}{\epsilon^{sub}}
\newcommand{\deltaext}{\epsilon^{ext}}
\newcommand{\testcase}[1]{{\sc #1}}
\newcommand{\dzero}{d^0}
\newcommand{\kodzero}{\stiff_{bio}/\dzero}
  \renewcommand{\kodzero}{\widetilde{\stiff}_{bio}}
\newcommand{\strain}{\varepsilon}
\newcommand{\straincrit}{\strain_{max}}
\newcommand{\area}{A}
\newcommand{\surf}{S}
\newcommand{\REA}{{\cal P}}
\newcommand{\dREA}{\partial\REA}
\newcommand{\half}{\frac{1}{2}}
\newcommand{\thalf}{\frac{3}{2}}

\newcommand{\abcd}{\mbox{a-b-c-d}}
\newcommand{\apbpcpdp}{\mbox{a$\,'$b$\,'$c$\,'$d$\,'$}}
\renewcommand{\abcd}{\mbox{\itshape abcd}}
\renewcommand{\apbpcpdp}{\mbox{\itshape a$\,'$b$\,'$c$\,'$d$\,'$}}

\newcommand{\algcode}[1]{{\tt #1}}
\newcommand{\myparagraph}[1]{\subsubsection{#1}}

\usepackage{xspace}
\usepackage[norefeq,refpage]{nomencl}

\newcommand{\nomheading}[1]{\mbox{\hspace{-33pt}}{\bfseries\itshape #1:}}

\renewcommand{\nomgroup}[1]{
  \ifthenelse{\equal{#1}{A}}{\item \nomheading{Roman letters}}%
  {\ifthenelse{\equal{#1}{G}}{\vspace*{0.1cm}\item \nomheading{Greek letters}}%
    {\ifthenelse{\equal{#1}{T}}{\vspace*{0.1cm}\item \nomheading{Superscripts}}%
      {\ifthenelse{\equal{#1}{S}}{\vspace*{0.1cm}\item \nomheading{Subscripts}}{}}}}%
  {}}

\newcommand{\nomunits}[1]{\ ($\myunit{#1}$)}

\begin{document}

\title{Simulating biofilm deformation and detachment with the immersed boundary method}

\author[R.~Sudarsan et.~al.]{%
  Rangarajan Sudarsan\affil{1}\comma\corrauth,
  Sudeshna Ghosh\affil{2},
  John M.~Stockie\affil{2}\ and
  Hermann J.~Eberl\affil{1}}
\address{%
  \affilnum{1}\
  Department of Mathematics and Statistics and Biophysics
  Interdepartmental Graduate Program, University of Guelph, ON,
  N1G~2W1, Canada.\\
  \affilnum{2}\
  Department of Mathematics, Simon Fraser University, Burnaby, BC,
  V5A~1S6, Canada.}
\emails{{\tt rsudarsa@uoguelph.ca} (R.~Sudarsan), {\tt sud1800@yahoo.co.in}
  (S.~Ghosh), {\tt stockie@math.sfu.ca} (J.~M.~Stockie), {\tt
    heberl@uoguelph.ca} (H.~J.~Eberl)}

\begin{abstract}
  We apply the immersed boundary (or IB) method to simulate deformation and
  detachment of a periodic array of wall-bounded biofilm colonies in response
  to a linear shear flow.  The biofilm material is represented as a network
  of Hookean springs that are placed along the edges of a triangulation of
  the biofilm region. The interfacial shear stress, lift and drag forces
  acting on the biofilm colony are computed by using fluid stress jump method
  developed by Williams, Fauci and Gaver [\emph{Disc.\ Contin.\ Dyn.\ Sys.\
  B} {11}(2):519--540, 2009], with a modified version of their exclusion
  filter.\
  Our detachment criterion is based on the novel concept of an averaged
  equivalent continuum stress tensor defined at each IB point in the biofilm
  which is then used to determine a corresponding von~Mises yield
  stress; wherever this yield stress exceeds a given critical threshold
  the connections to that node are severed, thereby signalling the onset
  of a detachment event. In order to capture the deformation and
  detachment behaviour of a biofilm colony at different stages of
  growth, we consider a family of four biofilm shapes with varying
  aspect ratio.  Our numerical simulations focus on the behaviour of
  weak biofilms (with relatively low yield stress threshold) and
  investigate features of the fluid-structure interaction such as
  locations of maximum shear and increased drag. The most important
  conclusion of this work is that the commonly employed detachment
  strategy in biofilm models based only on interfacial shear stress can
  lead to incorrect or inaccurate results when applied to the study of
  shear induced detachment of weak biofilms.  Our detachment strategy
  based on equivalent continuum stresses provides a unified and
  consistent IB framework that handles both sloughing and erosion modes
  of biofilm detachment, and is consistent with strategies employed in
  many other continuum based biofilm models.

  \leavethisout{
    Our simulations indicate that (a) in contrast to stiff biofilm
    colonies with large $\kodzero$ values, weaker ones
    experienced increased drag; (b) reduced spacing between the colonies
    reduced the drag by as much as 50 - 100\%\ and changed the profile
    of interfacial shear stress; (c) von~Mises stress inside different
    biofilm colonies exhibited high values along the biofilm fluid
    interface and around the front and back corner regions where the
    biofilm colony is attached to the wall; (d) For weak biofilm
    colonies with higher aspect ratio, in the deformed configuration
    poor correlation is seen between the computed interfacial shear
    stress and interfacial von~Mises stress; this indicating that
    detachment done only using interfacial shear stress as done in
    models in literature will lead to erroneous results; (e) The overall
    significant contribution in our work is the development of the
    equivalent continuum stress based detachment strategy which provides
    a unified framework to treat both biofilm sloughing and erosion mode
    of detachment in the immersed boundary setup and is consistent with
    what is used in other continuum mechanics based biofilm models.}
\end{abstract}


\maketitle




\section{Introduction}
\label{intro}

The subject of this work is the flow-induced deformation of a biofilm
colony, which is a meso\-scale collection of bacterial cells held together
by an extracellular polymeric network (EPS) that is secreted by the cells.  The
dimensions of a biofilm colony can be anywhere from tens to hundreds of
microns, whereas the size of an individual bacterial cell making up the
colony is on the order of 1--5 microns; our focus is on continuum models
that treat the biofilm as a viscoelastic solid continuum rather than
incorporating the dynamics of individual bacteria.  The flow-induced
deformations of the biofilm colony affect the fluid dynamic forces acting
on it, and thereby also alter both the extent and the mode of detachment
(i.e., sloughing or erosion) that may be experienced by the biofilm.  We
are particularly interested in understanding whether biofilm colonies
gain any protection against detachment when they are in close proximity
to other colonies.  To this end, our aim is to develop a robust
numerical method for simulating the interaction between a biofilm colony
and the surrounding fluid that is capable of capturing the different modes
of biofilm detachment.

Our approach is based on the immersed
boundary (or IB) method, in which the biofilm continuum is replaced by a
network of Hookean springs.  Although the IB method has already been
used by several authors for studying biofilm deformation and detachment
\cite{alpkvist2007, hammond}, our approach of computing an equivalent
continuum stress at each IB node is markedly different and using it to
initiate detachment provides us with a way of handling detachment in a
manner that is consistent with other continuum mechanics based models
such as~\cite{duddu2009}.


\subsection{Bacterial biofilms}
\label{bacbiofilm}

Bacterial biofilms are aggregations of microbes that grow on surfaces in
an aqueous environment. They form when bacterial cells suspended in the
fluid attach themselves to a surface and begin producing an
extracellular polymeric substance in which the growing bacteria
cells embed themselves.  Biofilms play contrasting roles depending on
the scenarios in which they are encountered: in waste-water treatment
\cite{nicolella20001} and environmental engineering they play a helpful
role; whereas biofilms encountered on medical devices or food-processing
equipment can be detrimental in the sense that they cause infections
\cite{costerton21051999} or compromise food safety \cite{shi2009407}.

In these and other applications, biofilms experience vastly different
physical conditions (temperature and pH), hydrodynamics (ranging from
creeping to turbulent flow) and chemical environments (rich or
sparse in nutrients, or saturated with antibiotics).  Their adaptability
to diverse conditions is believed to derive at least in part
from the mechanical and chemical protection provided by the gel-like EPS
layer.  It is important from an engineering standpoint to understand the
mechanical properties of biofilms so that we can
devise effective methods for not only enhancing biofilm growth and survival
but also removing them from surfaces.  Consequently, over the past
10 years significant effort has been expended to develop novel
rheological methods for measuring mechanical properties of biofilms
\cite{guelon2011advances}.  Such experiments have typically arrived at
different conclusions on how to characterize biofilms, with some
concluding that they behave as elastic \cite{ohashi1996novel} or
viscoelastic solids \cite{hohne2009flexible}, while others liken
biofilms to viscoelastic fluids \cite{towler2003}.  Furthermore, even
within the same material class, measured material parameter values can
vary over a fairly wide range.  The general consensus is that biofilms
behave as (visco-)elastic solids when the applied fluid shear stress is
at or near the stresses at which the biofilm was grown, while at higher
values of shear stress the biofilm can yield and behave as a
viscoelastic fluid.  In this paper, we restrict ourselves to the low
shear stress case and model the biofilm mechanically as a viscoelastic
solid embedded within a viscous fluid.

\subsection{Mathematical models of biofilm growth, deformation and detachment}
\label{biofilm_mech_detach}

Mathematical models for simulating biofilm growth must take into account a
wide range of mechanistic and other dynamical processes, including growth and
death of bacteria, attachment of cells to the substratum, transport of
solutes (nutrients, metabolic products, antibiotics) within the surrounding
fluid and the biofilm itself, reaction kinetics, and removal of bacterial
cells as clusters (sloughing) or as individual cells (erosion) when the
biofilm EPS matrix weakens in response to fluid shear or chemical treatments.
The earliest models developed in the 1980's \cite{kissel1984numerical,
wanner1986multispecies} assumed that the biofilm is one-dimensional, while
many subsequent modelling efforts have attempted to capture the 2D or 3D
morphology that develops during biofilm growth processes. The treatment of
the hydrodynamics and its interaction with the biofilm has received varying
degrees of treatment in these models, ranging from some models that consider
biofilm transport processes in isolation and use a specified nutrient
concentration boundary layer thickness at the biofilm-fluid interface,
whereas other models perform full fluid flow and nutrient transport
calculations with or without incorporating fluid-structure interaction
effects.  A variety of approaches have been developed to model the growth and
spreading of the biomass, including individual-based models
\cite{kreft2001individual}, cellular automata \cite{wimpenny1997unifying},
continuum models \cite{alpkvista2007multidimensional, duddu2009,
klapper2002finger} and phase field models \cite{zhang2008phase}.  An
exhaustive review of the different modeling approaches can be found in the
article \cite{wang2010} and performance benchmark comparisons are available
in \cite{eberl2006mathematical}.

Among the more complete models are those that capture multi-dimensional
growth and fluid flow
\cite{duddu2009, 
  eberl2001new, eberl2008exposure, picioreanu2001}, although the role played by
hydrodynamics in inducing biofilm deformation and detachment has most
often received only ad~hoc or approximate treatment.  These
approximations are aimed at capturing the solid mechanics governing the
dynamics of the deforming biofilm while simplifying as much as possible
the complex fluid flow that surrounds it.  The first attempt at solving
the solid mechanics problem inside the biofilm and using it to initiate
detachment was in \cite{picioreanu2001} where the biofilm was treated as
a linearly elastic material and a von~Mises yield stress
criterion was used to initiate detachment; however, this work neglected the
effects of biofilm deformation. In contrast, the particle-based biofilm
model in \cite{xavier2005} ignored the fluid and proposed a method for
initiating detachment using a \emph{detachment speed function} that is
based on the normal velocity of the biofilm-fluid interface.  A similar
approach has been used in \cite{smith2007extended}, 
which includes the effects of both biofilm growth and flow by
employing a detachment speed that depends on the interfacial shear
stress.  This approach to initiating detachment rests on the assumption
that regions where the interfacial shear stress is highest correspond to
locations where the biofilm strain is also high.

One of the aims of the current study is to verify the validity of this
last assumption by considering a full fluid-structure interaction
simulation that is capable of determining stresses in both fluid and
biofilm.  We do not explicitly model biofilm growth but rather mimic the
effects of growth by considering a family of biofilm colony shapes of
different aspect ratio, where each colony size corresponds to a
different instant of time during the growth of the same colony.  In each
case, we investigate how the deformation affects both the mode and the
extent of biofilm detachment. This is in contrast with other approaches
\cite{duddu2009, picioreanu2001} where the solid mechanics are simulated
but any deformations of the biofilm are neglected.

\subsection{Fluid-structure interaction in biofilms}
\label{imb_biofilm}


During the last several years, significant progress has been made in the
study of fluid-structure interaction (FSI) in biofilms, driven by the
increased availability of experimental data on biofilm mechanical properties
and the motivation to understand the role they play in biofilm survival.
Most FSI studies neglect biofilm growth by taking advantage of a natural
separation of time scales, in that growth processes are very slow in relation
to fluid motion and biofilm deformation. More recently, a phase field method
has been applied successfully in 2D \cite{lindley2012} and 3D
\cite{seeluangsawat2011} to simulate biofilm growth coupled with deformation
arising from interaction with the surrounding flowing fluid treating the
biofilm continuum as a multiphase polymeric gel. With the exception of these
two works, the most common approach in the biofilm FSI literature combines a
Lagrangian discretization of the biofilm with an Arbitrary Lagrangian
Eulerian (or ALE) formulation for the fluid. The first study of this kind
appeared in \cite{towler2007} where they studied the deformation of a 2D
hemispherical biofilm colony placed in a turbulent flow using the \Ansys\
commercial software package.  More recently \cite{taherzadeh2012}, a similar
ALE approach was used to study flow-induced oscillations of 2D biofilm
streamers and their effect on mass transfer.  A more realistic 3D biofilm
model was studied in \cite{bol2009} that used sliced 3D confocal laser
scanning microscopy data to construct the colony shapes, and employed a
nonlinear hyper-elastic constitutive model that accommodates detachment
based on a von~Mises yield stress criterion.

The aforementioned approaches have the advantage of being
well-established in the literature and capable of easily incorporating
experimental parameters that measure biofilm rheology. The
primary disadvantage is their high computational cost and algorithmic
complexity that result from needing to constantly re-mesh the fluid
domain as the biofilm colony deforms. Moreover, this re-meshing cost
increases enormously if detachment is incorporated in the model.

Motivated by the desire to develop a simpler and more efficient computational
approach, Alpkvist and Klapper \cite{alpkvist2007} proposed an alternate FSI
strategy based on the immersed boundary (or IB) method in which the biofilm
is discretized at a set of moving Lagrangian points.  These IB points move
relative to an underlying fixed Cartesian grid on which the fluid equations
are solved.  The elastic properties of the biofilm are captured by
distributing forces onto the fluid that derive from a network of Hookean
springs joining the IB points.  An incompressible fluid pervades both fluid
and biofilm regions and the biofilm inherits the density and viscosity of the
surrounding fluid, so that the biofilm is actually a visco-elastic composite
material consisting of elastic spring forces and fluid viscous forces.  This
approach has the clear advantage that no re-meshing of the fluid grid is
required.  To illustrate the versatility of their approach, Alpkvist and
Klapper simulated the deformation of both 2D and 3D biofilm structures, while
also incorporating a simple detachment criterion based on cutting individual
springs when they are stretched beyond a critical length.  We remark that
other IB models for biofilms were developed prior to \cite{alpkvist2007},
namely the work of Dillon and collaborators \cite{dillon1996,dillon2000};
however, these authors were concerned with slow flow and the dynamics of
individual bacterial cells aggregating and settling on the substratum, and
hence the results are relevant to different phenomena occurring on much
smaller spatial scales and at much earlier stages of biofilm formation.

A number of other IB approaches have since appeared, such as \cite{vo2010}
who performed 3D simulations of biofilm deformation, comparing Hookean
(elastic, spring-only) and Kelvin-Voigt (visco-elastic, spring plus dashpot)
models for the biofilm material. Using a parametric study of spring stiffness
and damping coefficient and comparisons with experiments, they established
that realistic biofilm deformation behaviour can be obtained using the IB
method.  More recently, Hammond et al.~\cite{hammond} developed more detailed
2D and 3D IB models for fragmentation of a biofilm colony in which the
location of actual bacterial cells from 3D images was used to determine
coordinates of IB points.  They employed a similar spring network and
detachment strategy as in \cite{alpkvist2007}, but they allowed the density
of the biofilm to differ from that of the fluid, and in subsequent work
\cite{hammond2012} also extended their approach to handle variable
viscosity.

Despite the increasing popularity of the IB method in biofilm FSI
studies, two main challenges remain to be addressed before the method is
capable of simulating realistic biofilm deformation and detachment.  The
first relates to connecting values of the IB spring parameters (elastic
stiffness and damping coefficients) to actual biofilm material
properties.  Despite the parametric study in~\cite{vo2010} that showed
it was possible to determine suitable parameters for given biofilm
colony shape and IB spring network topology, there remains as yet no
\emph{a~priori} method for determining IB parameters for a given
biofilm.

The second challenge relates to initiating detachment in the IB
framework in a way that is consistent with other more established
continuum-mechanics-based biofilm studies mentioned in
Section~\ref{biofilm_mech_detach}.  Whereas IB methods have so far used
strain in any given spring as a measure for initiating detachment,
this is not a true measure of strain in the continuum mechanics context.
Indeed, Hammond et al.\ \cite{hammond} demonstrated that when using
spring strain as a detachment criterion the resulting biofilm shape
is sensitive to the critical strain parameter, and so it is unclear how
to choose this parameter to match a given set of biofilm mechanical
properties.

\subsection{Objectives and outline}
\label{imb_objectives}

In this paper, we aim to address the challenges identified at the end of the
preceding section by developing an immersed boundary approach for simulating
biofilms that is capable of capturing realistic deformation and detachment
behaviours.  We begin with a 2D IB model inspired by that of Alpkvist and
Klapper~\cite{alpkvist2007}, and extend this work guided by two main
objectives.  Our first objective is to develop a novel approach for
initiating biofilm detachment that is consistent with methods employed in
biofilm models using continuum mechanics based calculations to enact
detachment.  In this way, we can retain the simpler spring network
representation of the biofilm continuum and the advantages it offers, while
also implementing a more physically realistic criterion for detachment.  Our
second major objective is to investigate the effect of flow-induced
deformation on both the mechanical stability and mode of detachment
experienced by weak biofilm structures having realistic shapes that resemble
those grown under mass transfer limited conditions.  This will allow us to
better understand such fundamental questions as how flow-induced deformation
affects the forces acting on biofilms, and how spatial clustering of biofilm
colonies can shield them from detachment by reducing the hydrodynamic shear
forces.

Our modelling approach incorporates the work done in several previous
computational studies of two-dimensional biofilms in~\cite{sudarsan2005,
  xu2008computational} wherein we investigated the fluid shear-induced
detachment forces (drag and lift) acting on rigid biofilm colonies that
are both uniformly and non-uniformly spaced.  In contrast to these
earlier studies that were restricted to values of shear rate exceeding
10\;$\myunit{cm/s}$, we focus in this paper on more flexible weak
biofilm colonies immersed in a slower shear flow having shear rate less
than 1\;$\myunit{cm/s}$.

The organization of the remainder of this paper is as follows.  In
Section~\ref{ibmethod}, we define the problem geometry and parameters, and
describe the governing equations and corresponding numerical scheme for our
basic IB model framework.  Section~\ref{model-stress-force} contains the
novel biofilm-related aspects of our IB model where we derive our approach
for calculating fluid shear stress along the biofilm-colony fluid interface and
from that the drag/lift forces acting on the biofilm colony.  In
Section~\ref{comp-force} we describe how we implement a modified
version of the \emph{exclusion filter} devised by Williams et al.\
in~\cite{williams2009} for accurately approximating interfacial shear
stresses on curved immersed boundaries, and then in
Section~\ref{eqcont-node-stress} we describe how we adopt the concept of a
continuum stress around each IB node in the biofilm material.
Section~\ref{algo-detachment} discusses how these quantities are used to
obtain a more realistic detachment criterion for biofilm colonies in the IB
context.  Finally, in Section~\ref{Results} we perform a number of numerical
tests that validate our numerical approach and also demonstrate the advantage
of our IB model for simulating realistic deformation and detachment events in
fluid shear-induced biofilm dynamics.

\section{Immersed boundary (IB) method}
\label{ibmethod}

The IB method is both a mathematical formulation and a numerical method for
simulating the complex interaction between a deformable solid material and a
surrounding fluid.  The approach dates back to work of
Peskin~\cite{peskin1977numerical} who originally developed the approach to
simulate blood flow interacting with heart muscle, and more recent
theoretical and computational developments are summarized nicely in the
review paper \cite{peskin2003}.  The IB approach has been applied extensively
to the study FSI in bio-fluid mechanics, including such problems as amoeboid
locomotion~\cite{bottino1998a}, platelet
aggregation~\cite{fogelson2004platelet}, and sperm
motility~\cite{fauci1995sperm}.

The IB method is a mixed Eulerian-Lagrangian approach wherein the fluid
equations are discretized on a fixed, rectangular (Eulerian) mesh
whereas the elastic structure is defined by a set of (Lagrangian) IB
points that move relative to the underlying fluid mesh as the structure
deforms.  The effect of the immersed boundary on the fluid is
represented using a singular source term in the fluid momentum
equations, which is distributed onto the underlying fluid grid by
writing it as the convolution of an elastic force density with
a regularized delta function.

\subsection{Problem geometry}
\label{problem-geom}

\begin{figure}[bthp]
  \footnotesize\sffamily
  \centering
  \includegraphics[width=0.55\textwidth,clip,bb=0 0 496 504]{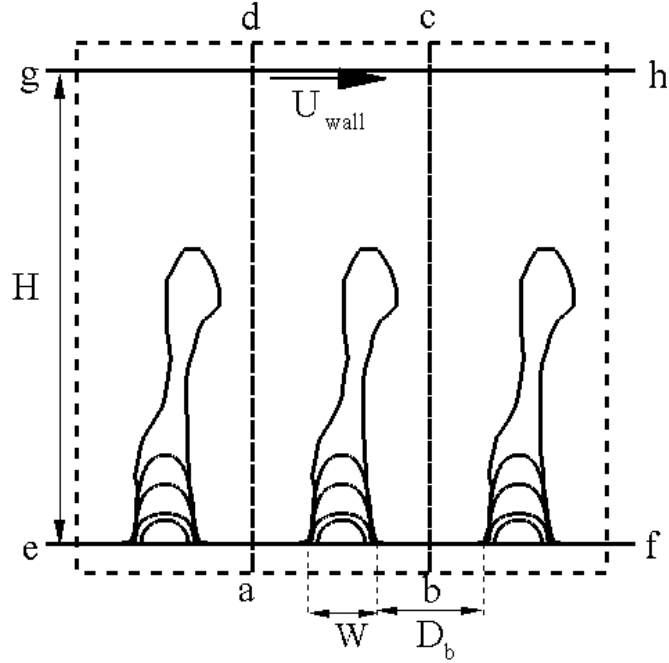}
  \caption{The problem domain consisting of a stationary bottom wall
    (e-f) separated by a distance $H$ from a top wall (g-h) that moves
    at velocity $(U_{wall},0)$. Biofilm colonies of width $\Wb$ and
    increasing height represent shapes at different stages of growth.
    The walls are treated with IB points and the whole flow region is
    embedded in a slightly larger computational domain (\abcd) that is
    periodic in both $x$ and $y$.}
  \label{Fig1-geometry}
\end{figure}

A diagram of the problem domain is given in Fig.~\ref{Fig1-geometry}, which
depicts shape of a biofilm colony encountered during different stages of
growth. The domain is delimited by two horizontal walls separated by a
distance $H$, where the bottom wall (serving as the biofilm substrate) is
held stationary while the top wall moves horizontally with constant velocity
$U_{wall}$.  The domain is filled with a viscous, incompressible fluid so
that in the absence of biofilm, the top wall will generate a flow that over
time approaches a steady-state corresponding to linear (planar) shear with
shear rate $G={U_{wall}}/{H}$.
\nomenclature[ah]{$H$}{vertical wall separation distance \nomunits{cm}}%
\nomenclature[auwall]{$U_{wall}$}{top wall velocity \nomunits{cm/s}}%
\nomenclature[ag]{$G$}{shear rate $=U_{wall}/H$ \nomunits{1/s}}%

Biofilm colonies having identical shape and size are placed along the
bottom wall with a uniform spacing of $\Db$ between adjacent
colonies. In order to simulate biofilms at different growth stages
occuring under mass transfer limited conditions, we select four
representative biofilm shapes with increasing aspect ratio that have
fixed width but increasing height.  Motivated by shapes at early stages
of growth observed in experiments~\cite{donlan2002biofilms,
  klausen2003involvement} and 3D simulations~\cite{duddu2009,
  eberl2001new, klapper2002finger, zhang2008phase}, we choose an
idealized biofilm shape described by the equation
\begin{gather}
  \label{superellipse}
  \left( \frac{x}{{\Wb}/{2}}\right)^{n_1} + \left(
    \frac{y}{\Hb}\right)^{n_2} = 1
  \quad \text{with $n_1,n_2 > 0$ and $y \geqslant 0$},
\end{gather}
which is half of a \emph{super-ellipse} with width $\Wb$ and height
$\Hb$.  In the case when $n_1=n_2=2$, Eq.~\eqref{superellipse} reduces
to half of a regular ellipse, while further constraining ${\Wb} = 2\Hb$
yields a semi-circle with radius $\Hb$.
\nomenclature[ax]{$\mathbf{x}$}{Cartesian coordinates $=(x,y)$ \nomunits{cm}}%
\nomenclature[adb]{$\Db$}{biofilm horizontal separation distance \nomunits{cm}}%
\nomenclature[ahb]{$\Hb$}{biofilm colony width \nomunits{cm}}%
\nomenclature[awb]{$\Wb$}{biofilm colony height \nomunits{cm}}%

We believe that this choice is a reasonable parameterization of the
typical finger-like shapes that dominate earlier stages of
biofilm growth, in contrast with later stages that exhibit a classical
\emph{mushroom-shaped} colony having an ellipsoidal head on top of a thin
cylindrical stem.  For the purpose of this study, we take values of
$n_1=n_2=2.5$ and $\Wb=50\;\mymum$, and select a sequence of three initial
shapes with $\Hb=25$, 50 and 75\;$\mymum$ respectively. In
addition, we consider two other shapes: a semi-circle with
diameter 40\;$\mymum$ (that is, $n_1=n_2=2$, $\Wb=40\;\mymum$ and
$\Hb = 20\;\mymum$); and an irregular mushroom-like structure with
$\Wb \approx 60\;\mymum$ and $\Hb \approx 280\;\mymum$, whose shape is
extracted from figures in~\cite{alpkvist2007}. The latter mushroom
shape is the tallest biofilm colony depicted in Fig.~\ref{Fig1-geometry}
and is chosen to illustrate the effectiveness of the detachment criteria
developed in this study by direct comparisons with the results from
\cite{alpkvist2007,hammond}.

Rather than imposing wall boundary conditions directly on the fluid, we
simplify the fluid solver by treating the top and bottom walls using
immersed boundaries and embed the problem domain
inside a slightly larger fluid domain denoted by dashed lines in
Fig.~\ref{Fig1-geometry}.  Because we are interested in studying
repeating arrays of biofilms colonies, periodic boundary
conditions are imposed in both the $x$ and $y$ directions so that the
computational domain is $\Omega = [0,W]\times[0,H]$ with $W=\Wb+\Db$
(labeled \abcd\ in the figure), and it contains a single biofilm colony.
\nomenclature[gw]{$\Omega$}{fluid domain}%

\subsection{Governing equations}
\label{imb-form}

The domain $\Omega$ is filled with a Newtonian incompressible fluid
having constant density $\rho$ ($\myunit{g/cm^3}$) and dynamic viscosity
$\mu$ ($\myunit{g/cm\;s}$).  Denote the infinitesimally thin immersed
boundaries representing the fixed bottom wall and moving top wall by
$\Gamma_{bot}$ and $\Gamma_{top}$ respectively, and let $\Gamma_{bio}$
represent the solid elastic structure corresponding to the biofilm
colony.  Note that $\Gamma_{bio}$ is actually a composite material that
consists of the elastic force-generating material and fluid that
co-exist within the same region.  We also assume for the purposes of
this study that the biofilm is neutrally buoyant and has the same
density as the fluid (although it is straightforward to extend the IB
approach to deal with variable density
problems~\cite{ghosh2013,hammond}).  The fluid is therefore governed at
all points $\mathbf{x}=(x,y)\in\Omega$ by the incompressible
Navier-Stokes equations 
\begin{gather}
  \label{mom-eq}
  \rho \frac{\partial \mathbf{u}}{\partial t} +
  \rho \mathbf{u}\cdot\nabla\mathbf{u} = - \nabla p + \mu
  \nabla^2\mathbf{u} + \mathbf{f},
\end{gather}
\begin{gather}
  \label{cont-eq}
  \nabla\cdot\mathbf{u} = 0,
\end{gather}
where $\mathbf{u}(\mathbf{x},t)$ ($\myunit{cm/s}$) is the fluid
velocity and $p(\mathbf{x},t)$ is pressure ($\myunit{g/cm\;s^2}$).
\nomenclature[grho]{$\rho$}{fluid density \nomunits{g/cm^3}}%
\nomenclature[gmu]{$\mu$}{fluid dynamic viscosity \nomunits{g/cm\;s}}%
\nomenclature[ggamma]{$\Gamma$}{immersed boundary}%
\nomenclature[sbot]{$bot$}{bottom wall}%
\nomenclature[stop]{$top$}{top wall}%
\nomenclature[sbio]{$bio$}{biofilm}%
\nomenclature[au]{$\mathbf{u}$}{fluid velocity \nomunits{cm/s}}%
\nomenclature[ap]{$p$}{fluid pressure \nomunits{g/cm\;s^2}}%
\nomenclature[af]{$\mathbf{f}$}{fluid force \nomunits{g/cm\;s^2}}%

The effect of the solid boundaries on the fluid is encompassed in the
fluid forcing term $\mathbf{f}$, which we consider next.  Assume that
the configuration of the solid material making up both channel walls and
biofilm is described by a function $\mathbf{X}(\mathbf{q},t)$, where
$\mathbf{q}$ is a generalized (dimensionless) parameterization that is
either a scalar ($\mathbf{q} = s$) in the case of the walls
$\Gamma_{bot}$ and $\Gamma_{top}$, or else a vector ($\mathbf{q}=(r,s)$)
for a solid region like $\Gamma_{bio}$.  The IB force in both cases is
specified in terms of a discrete network of IB points connected by
springs, and more detail on the precise form of these spring-force
connections for walls and biofilm is provided later in
Sections~\ref{meshgen} and~\ref{imb-forcedensity}.  Assume that the
force generated by any deformed configuration can be described by an IB
force density function $\mathbf{F}(\mathbf{X}(\mathbf{q},t)$ depending
on the current stretched configuration of the spring network.  Then the
fluid force $\mathbf{f}$ may be determined by spreading the force
density at IB points onto the fluid using a delta function convolution
\begin{gather}
  \label{imb-spread}
  \mathbf{f}(\mathbf{x},t) = \int_{\Gamma} \mathbf{F}(\mathbf{X},t) \
  \delta(\mathbf{x}-\mathbf{X}(\mathbf{q},t))\; d\mathbf{q} ,
\end{gather}
\nomenclature[gdelta]{$\delta,\,\delta_h$}{continuous, discrete delta function \nomunits{1/cm}}%
where $\delta(\mathbf{x})=\delta(x) \delta(y)$ is the Cartesian
product of two 1D Dirac delta functions and $\Gamma=\Gamma_{bio}\cup
\Gamma_{top} \cup \Gamma_{bot}$ represents the set of all immersed
boundaries.
\nomenclature[afi]{$\mathbf{F}$}{immersed boundary force density \nomunits{g/cm^2\, s^2}}%
\nomenclature[axi]{$\mathbf{X}$}{immersed boundary configuration $=(X,Y)$ \nomunits{cm}}%
\nomenclature[aq]{$\mathbf{q}$}{immersed boundary parameterization}%

The final equation required to close the system is an evolution
equation for the immersed boundaries, which we assume move with the same
velocity as the  surrounding fluid
\begin{gather}
  \label{imb-noslip}
  \frac{\partial \mathbf{X}}{\partial t} = \int_{\Omega}
  \mathbf{u}(\mathbf{x},t) \delta(\mathbf{x}-\mathbf{X}(\mathbf{q},t) \;
  d\mathbf{x}.
\end{gather}
This is simply another way of stating the no-slip condition for a
deformable boundary at location $\mathbf{X}(\mathbf{q},t)$.

\subsection{Numerical algorithm}
\label{imb-nummethod}

We now describe the basic numerical algorithm for solving
Eqs.~\eqref{mom-eq}--\eqref{imb-noslip}, which is a semi-implicit scheme
very similar to the one employed in~\cite{stockie1997analysis}.  The
fluid domain $\Omega$ is discretized on a regular Cartesian mesh with
coordinates $\mathbf{x}_{ij}=(x_i,y_j) = (ih_x, jh_y)$ for $i=0, 1,
\dots, N_x$ and $j=0, 1, \dots, N_y$, where $h_x=W/N_x$ and $h_y=H/N_y$
are constant grid spacings in the $x$ and $y$ directions (and we assume
for simplicity that $h_x=h_y$).  The time interval of interest $[0,T]$
is likewise divided into a sequence of $N_t$ equally-spaced points
denoted $t_{n} = n \Delta t$ for $n = 0,1,\ldots, N_{t}$, where $\Delta
t=T/N_t$ is the time step.  Let the discrete values of the velocity and
pressure be denoted by $\mathbf{u}_{ij}^{n}$ and $p^n_{ij}$
respectively.\
\nomenclature[ahx]{$h_x, h_y$}{fluid grid spacings in $x$ and $y$}%
\nomenclature[anx]{$N_x, N_y$}{number of fluid grid points in $x$ and $y$}%
\nomenclature[anx]{$N_b$}{number of immersed boundary points}%
Suppose that the immersed boundary is described by a set of $N_b$
IB points whose locations at any time $t_n$ are given by
$\mathbf{X}^n_{\ell} = (X^n_\ell,Y^n_\ell)$ for $\ell = 1, 2 \ldots,
N_b$.  The corresponding force densities are denoted by
$\mathbf{F}^n_{\ell}$, with the precise specification of the immersed
boundary discretization and force density calculation
$\mathbf{F}^n_\ell$ being given in the the following two sections.

We now describe our algorithm for updating fluid grid quantities
$\mathbf{u}^{n}_{ij}$ and $p^{n}_{ij}$ and the IB configuration
$\mathbf{X}_{\ell}^{n}$ from time $t_n$ to time $t_{n+1}$.  The
algorithm proceeds in four main steps:
\begin{description}
\item[Step 1:] Compute the force density $\mathbf{F}_{\ell}^{n}$
  based on the current IB configuration $\mathbf{X}^{n}_{\ell}$ as
  described in Section~\ref{imb-forcedensity}.

\item[Step 2:] Spread the force density onto fluid grid points using
  a discrete representation of the delta-function convolution in
  Eq.~\eqref{imb-spread}
  \begin{gather}
    \label{alg-spreading}
    \mathbf{f}^{n}_{ij} = \sum_{\ell=1}^{N_b} \mathbf{F}_{\ell}^{n}
    \deltah(\mathbf{x}_{ij} - \mathbf{X}^{n}_{\ell}) \Ascale,
  \end{gather}
\nomenclature[aascale]{$\Ascale$}{area scaling factor \nomunits{cm\;\mbox{or}\;cm^2}}%
  where $\deltah (\mathbf{x})$ is a regularized delta function given by
  \begin{gather}
    \label{alg-regdelta}
    \deltah (\mathbf{x}) = \frac{1}{h_x h_y} \phi\left(
      \frac{x}{h_x}\right) \phi\left( \frac{y}{h_y}\right) ,
  \end{gather}
  with
  \begin{gather}
    \phi(r) =
    \begin{cases}
      \frac{1}{4}\left(1 + \cos(\frac{\pi r}{2}) \right), & \text{if }
      \lvert r \rvert \leqslant 2, \\
      0, & \text{otherwise}.
    \end{cases}
  \end{gather}
  The scaling factor $\Ascale$ has units of length for forces generated
  by the 1D wall interfaces (for which \eqref{alg-spreading}
  approximates a line integral) whereas $\Ascale$ has units of area for
  the 2D biofilm region. To ensure that Eq.~\eqref{alg-spreading} is a
  consistent representation of the corresponding integrals under grid
  refinement, $\Ascale$ is inversely proportional to the number of
  IB points.  Details on the precise expression used for
  $\Ascale$ in each case are provided in Section~\ref{imb-forcedensity}.

\item[Step 3:] Integrate the Navier-Stokes equations using Chorin's
  split-step projection scheme:
  \begin{enumerate}
    \renewcommand{\theenumi}{\alph{enumi}}
  \item Compute an intermediate velocity $\mathbf{u}^{(1)}_{ij}$ by
    updating the velocity only for the contribution from the IB
    elastic force:
    \begin{gather}
      \label{alg-step3a}
      \rho \left( \frac{\mathbf{u}^{(1)}_{ij} -
          \mathbf{u}_{ij}^{n}}{\Delta t}\right) =
      \mathbf{f}^{n}_{ij}.
    \end{gather}

  \item Compute intermediate velocities $\mathbf{u}^{(2)}_{ij}$ and
    $\mathbf{u}^{(3)}_{ij}$ by applying convection and diffusion
    terms using an alternating direction implicit (ADI) approach:
    \begin{gather}
      \label{alg-ADI-step1}
      \rho \left(\frac{\mathbf{u}^{(2)}_{ij} -
          \mathbf{u}^{(1)}_{ij}}{\Delta t}
        + u^{n}_{ij} D^{0}_x \mathbf{u}_{ij}^{(2)} \right) = \mu
      D^{+}_x D^{-}_x \mathbf{u}^{(2)}_{ij} ,
      \\
      \label{alg-ADI-step2}
      \rho \left(\frac{\mathbf{u}^{(3)}_{ij} -
          \mathbf{u}^{(2)}_{ij}}{\Delta t}
        + u^{n}_{ij} D^{0}_y \mathbf{u}_{ij}^{(3)} \right) = \mu
      D^{+}_x D^{-}_x \mathbf{u}^{(3)}_{ij}.
    \end{gather}
    The operators $D_x^+$ and $D_x^-$ refer to the standard first-order
    forward and backward difference approximations of the
    $x$-derivative, and $D_x^0$ is the standard second-order centered
    difference approximation.  Analogous definitions apply for the
    $y$-derivative approximations $D^+_y$, $D^-_y$ and $D^0_y$.
    Equations~\eqref{alg-ADI-step1} and~\eqref{alg-ADI-step2} thus
    represent periodic tridiagonal linear systems for the intermediate
    velocities.

  \item Project the intermediate velocity $\mathbf{u}^{(3)}_{ij}$ onto
    the space of divergence-free vector fields by first solving a
    Poisson equation    for the pressure $p^{n+1}_{ij}$
    \begin{gather}
      \label{alg-PPE}
      \nabla_h \cdot \nabla_h p^{n+1}_{ij} = \frac{\rho}{\Delta t}
      \nabla_h \mathbf{u}^{(3)}_{ij},
    \end{gather}
    where $\nabla_h=(D_x^0,D_y^0)$ is a centered approximation of the
    gradient operator and the discrete Laplacian $\nabla_h\cdot\nabla_h$
    yields a wide finite difference stencil that spans four grid points
    in each direction as described
    in~\cite{stockie1997analysis}. Because of the periodic boundary
    conditions, the pressure Poisson equation is solved most efficiently
    using a fast Fourier transform (FFT).  Finally, the velocity
    projection is completed via the correction
    \begin{gather}
      \label{alg-finalvel}
      \mathbf{u}_{ij}^{n+1} = \mathbf{u}_{ij}^{(3)} - \frac{\Delta
        t}{\rho} \nabla_h p_{ij} .
    \end{gather}
  \end{enumerate}

\item[Step 4:] Evolve the immersed boundary to time $t_{n+1}$ using
  \begin{gather}
    \label{alg-ibcoordupdate}
    \mathbf{X}_{\ell}^{n+1} = \mathbf{X}_{\ell}^{n} + \Delta t \sum_{i,j}
    \mathbf{u}_{ij}^{n+1} \deltah (\mathbf{x}_{ij} -
    \mathbf{X}_{\ell}^{n}) h_x h_y .
  \end{gather}
\end{description}
The IB algorithm explained above is first-order accurate in both time
and space.  Despite the use of second-order differences for spatial
derivatives, the spatial accuracy reduces to first order owing to the
particular choice of interpolation used for spreading the fluid velocity
onto the immersed boundary~\cite{mori-2008}.

This algorithm has the advantage that it is simple and easy to code,
although it suffers from a fairly strict stability restriction on the
time step owing to the explicit treatment of the IB forcing term in Step
1 of the algorithm.  We implement the above algorithm using \MatlabReg, and
when that is combined with the extra cost of implementing a realistic
detachment criterion (see details Section~\ref{algo-detachment}) our
approach is restricted to fairly short-time simulations.  The scope of
this paper is therefore limited to the study of biofilm deformation and
detachment in the early stages up until a quasi-steady biofilm
configuration in reached.  Any full-scale implementation of the detachment criteria
described in this study would therefore benefit from a more efficient IB
implementation such as the fully implicit approach
in~\cite{mori2008implicit} or one of the parallel approaches developed
in either~\cite{griffith2007adaptive} or~\cite{wiensStockie2015}.

\subsection{Discrete representation of walls and biofilm}
\label{meshgen}

In this section, we describe the discretization of the immersed
boundaries (walls and biofilm) in terms of a network of IB points
connected by springs.  Section~\ref{imb-forcedensity} will then provide
details of the force density calculations based on this discrete IB
configuration.

We begin with the horizontal walls at locations $y=0$ and $H$ (labeled
e-f and g-h in Fig.~\ref{Fig1-geometry}) that are each replaced
by a 1D periodic array of IB points stretching across the
domain.  Adjacent wall points are not explicitly connected
to each other, but instead each IB point is connected by a very stiff
spring to a corresponding \emph{tether} or \emph{target point} that
initially occupies the same position as the IB point.  The tether forces
generated by these springs constrain the IB points to remain close to
their target locations and hence mimic a solid wall.  The initial IB
point spacing $h_{wall}$ is chosen so that $h_{wall} < \half
\max(h_x,h_y)$, which helps to control numerical errors that would
otherwise lead to significant leakage of fluid between IB
points~\cite{peskin2003}.

Next we consider the discrete representation of the biofilm colony,
which is obtained by triangulating the biofilm region, placing IB points
at the nodal locations, and constructing a network of Hookean springs
corresponding to the edges in the triangulation.  The spring stiffness
is chosen to approximate the elastic properties of an actual biofilm
material.  The biofilm is attached to the lower wall by connecting
points along the bottom of the colony to a corresponding wall IB point
with a stiff spring, where each pair of points initially occupies the
same location.  To ensure that such a spring network mimics a
mechanically isotropic biofilm, we use an initial triangulation that has
approximately uniform shape and size of the elements.  For this purpose,
we use the open-source \Matlab\ package \DistMesh\ of Persson and
Strang~\cite{persson2004simple}.
The average spacing between biofilm IB points
is initially chosen to be roughly equal to one-third of the fluid
grid spacing.  This choice is motivated by Vo et
al.~\cite{vo2010} and ensures that the convective flow inside the
biofilm is negligible, so that the biofilm acts as an impermeable
material.  Nevertheless, we note that during the course of our biofilm
simulations as the biofilm deforms and IB points reach their maximum
separation, it is possible for portions of the biofilm colony to
experience a small apparent permeability.  We will return to
this point when devising a method for approximating the interfacial
shear stress in Section~\ref{comp-force}.

Fig.~\ref{Fig2-mesh} depicts triangulations of two initial biofilm
shapes used in this study, one a super-ellipse with height $\Hb=75\;
\mymum$, and the other a mushroom-shaped biofilm colony with height
$\Hb\approx 280\; \mymum$.  A coarser mesh is depicted than what is
actually used so that the shape and distribution of triangles are
evident.
\setlength{\unitlength}{\textheight}
\begin{figure}[bthp]
  \footnotesize\sffamily
  \centering
  \begin{tabular}{cccc}
    \multicolumn{2}{c}{(a) Super-ellipse shape, with filleted corners
      at the wall.}
    & \multicolumn{2}{c}{(b) Mushroom shape.}\\[-1.6cm]
    \begin{picture}(0.06,0.4)
      \put(0.05,0.02){\vector(0,1){0.21}}
      \put(0.05,0.225){\vector(0,-1){0.21}}
      \put(0,0.12){\text{75~$\mymum$}}
    \end{picture}
    &
    \includegraphics[height=0.3\textheight,clip,bb=0 0 665 715]{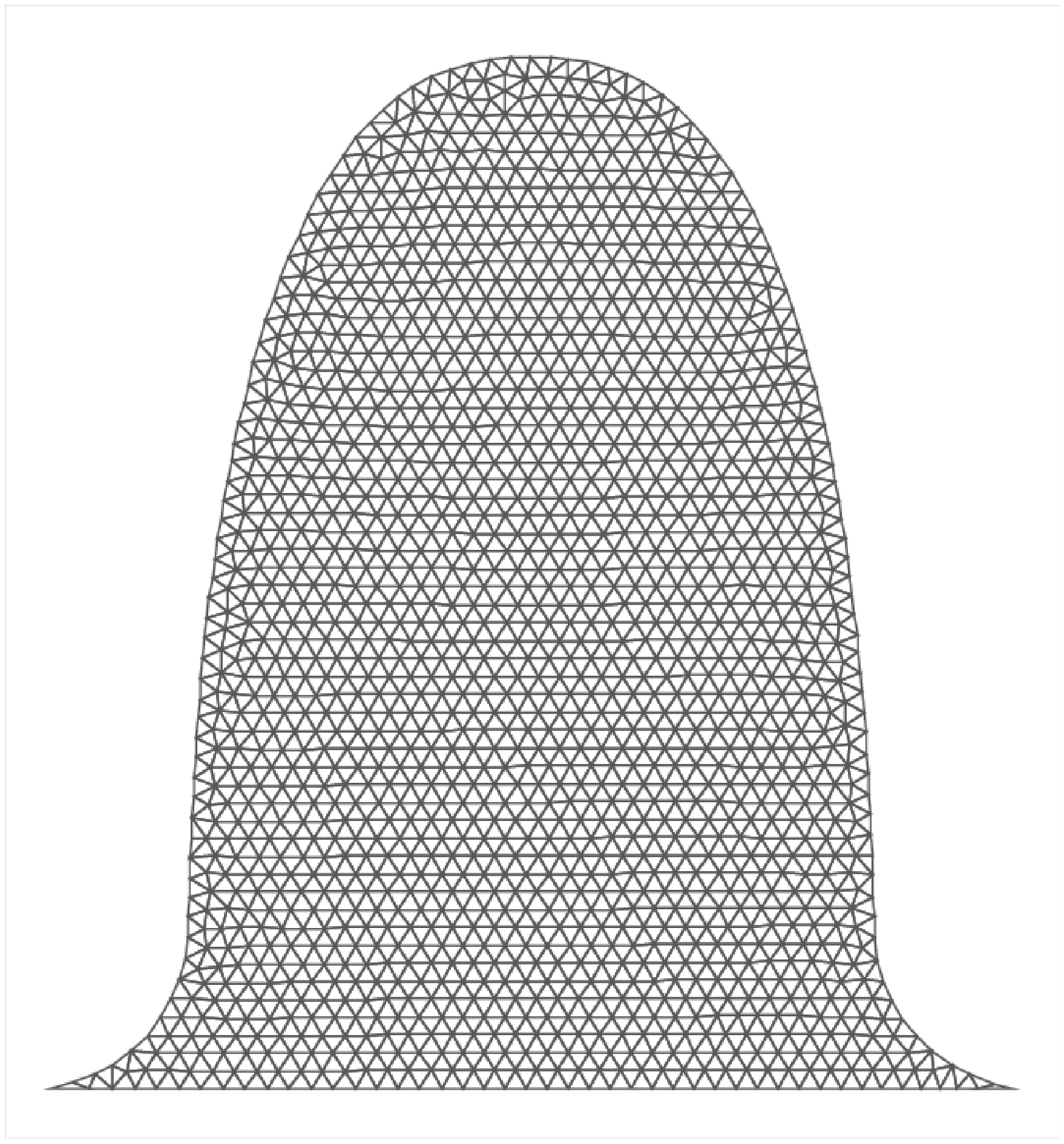} &
    \begin{picture}(0.06,0.4)
      \put(0.06,0.02){\vector(0,1){0.28}}
      \put(0.06,0.3){\vector(0,-1){0.28}}
      \put(0,0.16){\text{280~$\mymum$}}
    \end{picture}
    &
    \includegraphics[height=0.4\textheight,clip,bb=0 0 215 715]{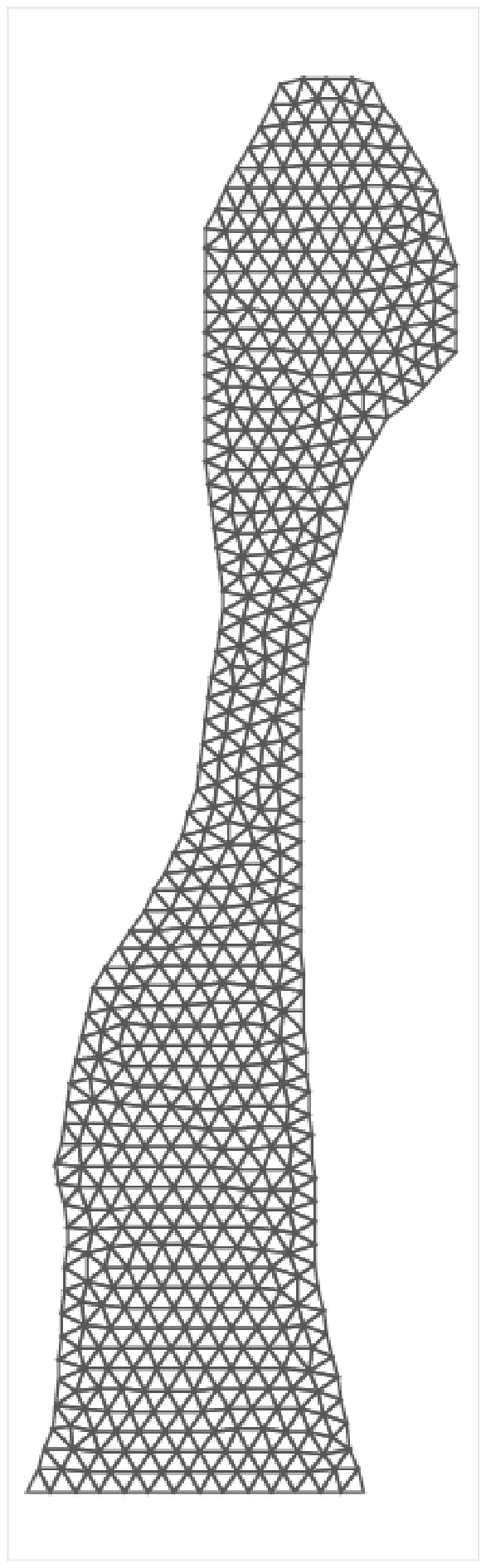}
  \end{tabular}
  \caption{Quasi-uniform triangulations generated by \DistMesh\ for:
    (a)~a super-elliptical biofilm colony with height $\Hb=75\;\mymum$
    and mean edge length 1.5~$\mymum$; and (b)~a taller
    mushroom-shaped colony with mean edge length 4~$\mymum$.}
  \label{Fig2-mesh}
\end{figure}
IB simulations with a super-ellipse show that the sharp 90 degree
corners at the intersection between the biofilm colony periphery and the
bottom wall cause localized discontinuities in the slope of the biofilm
colony-fluid interface to develop at these locations as the biofilm
colony begins to deform. We believe this is due to the
corner-type singularities appearing in the fluid that are
communicated to the immersed structure via delta function
interpolation. In actual biofilms, such sharp corners seldom occur and
so we instead smooth out the corners using a quarter-circular curve or
\emph{fillet}.  As depicted in Fig.~\ref{Fig2-mesh}(a), the size of the
fillet is kept small so as to minimize its influence on either the
biofilm or the corresponding elastic forces.

\subsection{Discrete force density calculation}
\label{imb-forcedensity}

As indicated earlier, the force density function consists of
contributions from two classes of immersed boundaries: horizontal rigid
walls ($\mathbf{F}_{wall}$) and elastic deformable biofilm regions
($\mathbf{F}_{bio}$).  The force contribution in both cases is
calculated using a discrete specification that is defined in terms of
the current configuration of IB points in either walls or biofilm.
\nomenclature[swall]{$wall$}{wall points}%

\myparagraph{Top and bottom wall forces}

The discrete wall force arising at any given wall IB point is determined
from the stretched state of the spring of zero resting length that connects
it to the corresponding tether point, yielding a force density
\begin{gather}
  \label{wall-tether}
  \mathbf{F}^{n}_{wall,\ell} = \stiff_{wall} \
  (\mathbf{X}^n_{wall,\ell} - \mathbf{X}^n_{teth,\ell}) ,
\end{gather}
\nomenclature[steth]{$teth$}{tether points}%
\nomenclature[gteth]{$\stiff$}{spring stiffness \nomunits{g/cm\;s^2}}%
where $\stiff_{wall}$ ($\myunit{g/cm^2\;s^2}$) is the spring stiffness,
and $\mathbf{X}^n_{wall,\ell}$ and $\mathbf{X}^n_{teth,\ell}$ are the
coordinates of the wall at tether points at time level $n$.  For the
stationary bottom wall all tether points are fixed in time, whereas the
top wall tether points move with a given constant velocity
$({U}_{wall}, 0)$ according to
\begin{gather}
  \label{eqn-topwall}
  \mathbf{X}^{n+1}_{teth,\ell} = \mathbf{X}^{n}_{teth,\ell} +
  ({U}_{wall},0) \Delta t \mod (W,H),
\end{gather}
where the ``modulo'' operator ensures that IB points remain inside the
domain by imposing the periodic boundary condition in the horizontal
direction.  By choosing $\stiff_{wall}$ sufficiently large, we ensure
that the wall IB points do not deviate significantly from their tether
point locations throughout a simulation.  For both sets of wall IB
points, we set the integral scaling factor in Eq.~\eqref{alg-spreading}
equal to the tether point spacing, $\Ascale=h_{wall}$.

\myparagraph{Biofilm forces}

The force density at an IB point within the biofilm region is determined
by summing up the elastic force contributions coming from all springs
connected to that node in the triangulation. Any given spring link is
identified by an index pair $\ell,m$ corresponding to two IB points with
coordinates $\mathbf{X}_{\ell}^{n}$ and $\mathbf{X}_{m}^{n}$. The
elastic force density contribution at node $\ell$ due to spring $\ell,m$
acts in the direction of the vector $\mathbf{d}_{\ell m}^n =
\mathbf{X}_{\ell}^n - \mathbf{X}_{m}^n$ joining nodes $\ell$ and $m$.
By denoting the resting length of this spring $\dzero_{\ell m}$, the
force density $\mathbf{F}_{bio,\ell}^{n}$ at node $\ell$ owing to all
springs attached to that node can be written as~\cite{alpkvist2007}
\begin{gather}
  \label{eqn-biofilmforce}
  \mathbf{F}_{bio,\ell}^{n} = \frac{\stiff_{bio}}{\dzero}
  \sum_{m=1}^{N_b} \mathbb{I}_{\ell m}
  \frac{\mathbf{d}_{\ell m}^n}{d_{\ell m}^n} \frac{(d_{\ell m}^n -
  \dzero_{\ell m})}{\dzero_{\ell m} } ,
\end{gather}
\nomenclature[ad]{$\dzero$}{spring resting length \nomunits{cm}}%
where $\stiff_{bio}$ is the spring stiffness coefficient ($\myunit{g
  /cm\;s^2}$), $\dzero$ is the average spring resting length
and $d_{\ell m}=\|\mathbf{d}_{\ell m}\|$.  The symbol
$\mathbb{I}_{\ell m}$ is a square connectivity matrix of
dimension $N_b \times N_b$ whose entries are either 1 or 0 depending on
whether or not nodes $\ell$ and $m$ are connected.  Implicit in this
notation is the fact that the summation is only done over pairs of nodes
that are connected by an edge in the network.  The scaling factor
$\Ascale$ in Eq.~\eqref{alg-spreading} for the biofilm force spreading
term is taken equal to the average area of a triangle in the biofilm at
its rest state, which is equal to the total initial area divided by
$N_b$ (and hence has units of $\myunit{cm^2}$).  Finally, following the
arguments of Alpkvist and Klapper~\cite{alpkvist2007}, we can ensure
that the biofilm deformation is independent of grid refinement by
scaling the spring stiffness value with the nodal mean distance $\dzero$
and setting $\kodzero = \stiff_{bio} / \dzero$.

\section{Interfacial shear stress, drag and lift forces, and detachment}
\label{model-stress-force}

\subsection{Computing forces on the biofilm-fluid interface}
\label{comp-force}

Determining the forces acting on a biofilm colony as it deforms in
response to fluid shear is of fundamental importance in this study.  The
most common global measures of hydrodynamic force in such FSI
simulations are the drag and lift force. In addition, we are interested
in calculating the local interfacial shear stress along the
biofilm-fluid interface, owing to its essential role in determining both
the mode and extent of biofilm colony detachment. Drag and lift forces
are typically calculated in the IB framework by summing the
corresponding components of the IB force along the
interface~\cite{ghosh2013, lai2000immersed}.  Instead, we follow the
approach of Williams, Fauci and Gaver~\cite{williams2009} (which we
refer from this point on as WFG) wherein the traction force is first
calculated from the interfacial shear stress and then integrated along
the biofilm-fluid interface.  Our aim in this section is therefore to
first obtain an expression for interfacial shear stress, and then to
derive expressions for the drag and lift forces.

Evaluating interfacial stress in the IB framework is complicated
not only by the diffuse nature of immersed boundaries owing to
regularized IB forces, but also because of the spatial averaging inherent
in the velocity no-slip boundary condition.  If not handled
appropriately, both of these steps can introduce large errors in
interfacial shear stress.  This issue was studied by
WFG~\cite{williams2009} who proposed two methods for calculating the
tangential interfacial shear stress based on the equation
\begin{gather}
  \label{eqn-shearstress}
  \underbrace{{\mathbf{t}} \cdot [\boldsymbol{\stress}]
    \cdot {\mathbf{n}}}_\textrm{Jump in FS} =
  \underbrace{-\frac{{\mathbf{t}} \cdot \mathbf{F}}{\lvert
      \frac{\partial \mathbf{X}}{\partial s}\rvert}}_\textrm{WS} .
\end{gather}
\nomenclature[at]{$\mathbf{t}$}{unit tangent vector to the biofilm-fluid interface}%
\nomenclature[an]{$\mathbf{n}$}{unit normal vector to the biofilm-fluid interface}%
\nomenclature[gsig]{$\sigma$}{fluid stress tensor \nomunits{g/cm\;s^2}}%
Here, $\mathbf{X}(s,t)$ is a parametric representation of the
biofilm-fluid interface, $\mathbf{n} =
\frac{\partial\mathbf{X}}{\partial s}/|\frac{\partial
  \mathbf{X}}{\partial s}|$ is the unit normal vector
(directed outward from the biofilm into the surrounding fluid),
$\mathbf{t}$ is the counter-clockwise tangent vector,
$\boldsymbol{\stress} = -p \mathbb{1} + \mu (\nabla\mathbf{u} +
\nabla\mathbf{u}^{T})$ is the fluid stress tensor, and square
brackets $[ \cdot ]$ denote the jump in a quantity across the interface.
WFG proposed evaluating the tangential stress using
either side of Eq.~\eqref{eqn-shearstress}: the left hand side requires
calculating the jump in fluid stress across the biofilm-fluid interface,
and so is referred to as the \emph{FS method}; whereas the right hand
side involves local IB force densities, and is called the wall stress or
\emph{WS method}.
WFG performed IB simulations of 2D Poiseuille flow in a channel, with
and without obstructions, and drew the following conclusions about the
relative merits of these two methods:
\begin{itemize}
\item On a curved boundary, the WS method over-estimates shear stress in
  comparison with the FS method.
\item In order to maximize accuracy with the FS method, the fluid shear
  stress associated with an IB point on the interface should be
  evaluated at a point located inside the domain, directed along the
  normal vector and separated from the boundary by a distance equal to
  the fluid grid spacing.  The fluid grid cell within which this point
  falls is called the \emph{interpolation box} because it defines a set
  of 4 fluid grid points that will be used for interpolating the stress.
\item For a curved immersed boundary, the estimate for shear stress at
  certain IB points can be adversely affected when the boundary
  intersects the interpolation box and a portion of the interpolation
  box lies outside the fluid region.  Therefore, only those stresses
  determined using interpolation boxes that lie entirely on one side of
  the interface should be included, and for this purpose WFG designed an
  \emph{exclusion filter} that omits any such unwanted stress
  contributions.
\end{itemize}
We implemented the WFG exclusion filter approach and found that while it
works well for the simple immersed boundaries considered
in~\cite{williams2009}, the accuracy of the shear stress calculation
degrades for the highly curved interfaces that are so common in biofilm
applications.  We therefore developed a modified version of the
exclusion filter mentioned above that handles highly curved boundaries
in a robust fashion.

\begin{figure}[bthp]
  \footnotesize\sffamily
  \centering
  \includegraphics[width=0.8\textwidth,clip,bb=0 0 511 312,clip]{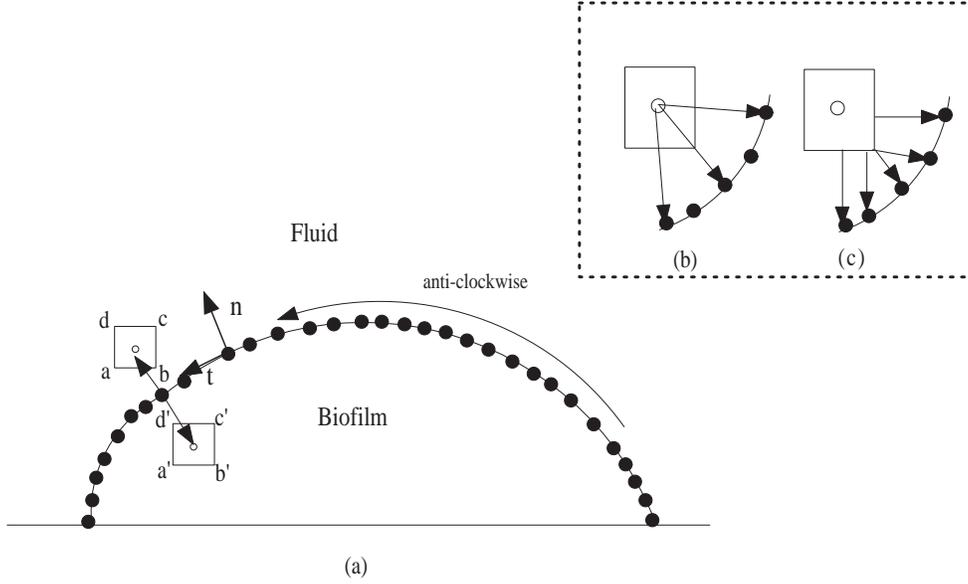}
  \caption{(a) The exclusion filter uses fluid grid cells
    \abcd\ and \apbpcpdp\ as the two interpolation boxes,
    determined by extending an IB point along the outward and inward
    normals.  For the sake of clarity, IB points lying inside the
    biofilm are not shown here. Distances from the interpolation box
    to the IB can be calculated in two ways as shown in the inserts: (b)
    measured from the centroid of the interpolation box, or (c) measured
    as the shortest distance from the edges.}
  \label{Fig3-filter_design}
\end{figure}

We begin by explaining the WFG exclusion filter in reference to
Fig.~\ref{Fig3-filter_design}, which depicts the IB points
$\mathbf{X}^{int}_{\ell}$ for $\ell=1, 2, \dots, N^{int}$ lying along a
biofilm-fluid interface, ordered in the counter-clockwise direction.
For each IB point, the outward unit normal $\mathbf{n}_\ell$ is computed
using a local cubic Lagrange interpolation procedure as explained
in~\cite{zhao2001front}.  We then identify two points that will be used
to evaluate the stress, $\mathbf{XE}^{out}_{\ell} =
\mathbf{X}^{int}_{\ell} + h_x \mathbf{n}_\ell$ and
$\mathbf{XE}^{in}_{\ell} = \mathbf{X}^{int}_{\ell} - h_x
\mathbf{n}_\ell$, which lie on either side of the interface located a
distance $h_x$ along the outward/inward normals from $\mathbf{X}^{int}$,
respectively (assuming here that $h_x=h_y$).  The interpolation boxes or
fluid grid cells in which these two evaluation points
$\mathbf{XE}^{out}_{\ell}$ and $\mathbf{XE}^{in}_{\ell}$ lie are denoted
\abcd\ and \apbpcpdp, and their respective centroids by
$\mathbf{XC}^{out}_{\ell}$ and $\mathbf{XC}^{in}_{\ell}$.
\nomenclature[axe]{$\mathbf{XE}$}{stress evaluation point}%
\nomenclature[axc]{$\mathbf{XC}$}{centroid of the interpolation box}%
\nomenclature[amd]{${MD}$}{minimum distance to the interpolation box}%
\nomenclature[tint]{$int$}{interface}%
Then, we calculate the minimum distance from the biofilm-fluid interface
to the centroids of the two interpolation boxes (referring to
Fig.~\ref{Fig3-filter_design}(b)):
\begin{align}
  MD^{out}_{\ell} = \min_m \left( D^{out}_{\ell m} \right)
  \quad \text{and} \quad
  MD^{in}_{\ell}  = \min_m \left( D^{in}_{\ell m} \right),
\end{align}
where $m$ ranges over $1,2,\ldots, N^{int}$, and $D^{out}_{\ell m}$ and
$D^{in}_{\ell m}$ represent the distance between IB point $m$ and the
centroid of the two interpolation boxes associated with the
$\ell^{\mathrm{th}}$ IB point.

The main principle behind the WFG exclusion filter~\cite{williams2009}
is to base stress calculations only on those points that are most
representative of the fluid flow around the biofilm colony, first by
excluding interpolation boxes that are cut by the interface, and second
by choosing stress evaluation points $\mathbf{XE}$ that are as far from
the interface as possible (so that they are least affected by the
regularized IB force).  To this end, we identify IB points $\ell$ for
which $MD_\ell$ has a local maximum, while also requiring that no
interpolation box be used for more than one stress calculation.

As mentioned above, when this WFG filter is applied to biofilms with
highly curved interfaces, the accuracy of the stress approximation
degrades.  One cause of this degradation is that certain portions of the
interface may end up with few IB points included, leading to poor
resolution.  To address this problem, we propose a modification of the
WFG exclusion filter that is implemented in
Algorithm~\ref{algo:exclusion-filter} for stresses on the outer (fluid)
side of the interface; only minor changes are required for stresses on
the inner (biofilm) side.  This new filter makes the following
three modifications to the basic WFG exclusion filter:
\begin{itemize}
\item In step~\ref{algo:step1}, check whether each interpolation box
  is cut by the biofilm-fluid interface.

\item In step~\ref{algo:step2}, compute the minimum distance between the
  biofilm-fluid interface and the \emph{edges} of the interpolation box
  \abcd\ as shown in Fig.~\ref{Fig3-filter_design}(c), rather than the
  centroid in the original filter.  This provides additional resolution
  by differentiating between nearby IB points that are equidistant from
  the centroid, since points are not equally-spaced from the edges of
  the box (compare Figs.~\ref{Fig3-filter_design}(b,c)). The minimum
  distance calculation from the edge of an interpolation box is
  performed using a fast and elegant algorithm
  from~\cite{arvo1991graphics} that introduces no significant additional
  cost.

\item In step~\ref{algo:step4}, the strong local maximum criterion is
  relaxed and we instead introduce a new user-specified filtering
  parameter $\deltamin$. This change ensures (for an appropriate choice
  of $\deltamin$) that we retain some IB points that would otherwise
  have been rejected by the original WFG filter.
\end{itemize}
\begin{algorithm}[htbp]
  \caption{Modified WFG exclusion filter, which determines a list of IB
    points (\algcode{pt\_list}) and interpolation boxes
    (\algcode{box\_list}) for stress calculations.}
  \label{algo:exclusion-filter}
  \begin{algorithmic}[1]
    \FORALL{interface IB points $\mathbf{X}^{int}_\ell$}
       \STATE Compute the outward unit normal $\mathbf{n}_\ell$.
       \STATE Locate the evaluation point $\mathbf{XE}^{out}_\ell =
          \mathbf{X}^{int}_{\ell} + h_x \mathbf{n}_\ell$.
       \STATE Identify the interpolation box \abcd\ corresponding to
          the fluid cell containing $\mathbf{XE}^{out}_\ell$.
       \STATE Add \abcd\ to the sorted list \algcode{box\_list} and remove
          duplicates.
    \ENDFOR
    \FORALL{interpolation boxes $\abcd \in$~\algcode{box\_list}}
       \IF{\abcd\ \label{algo:step1} is intersected by any of the
          $N^{int}-1$ line segments comprising the biofilm-fluid interface}
          \STATE Remove \abcd\ from \algcode{box\_list}.
       \ELSE
          \STATE Determine \label{algo:step2} the minimum distance
             $MD^{out}_\ell$ from the edges of box \abcd\ to the
             biofilm-fluid interface, and the corresponding IB point
             $\ell$.
          \STATE Store $MD^{out}_\ell$ and \abcd\ in the data structure
             for IB point $\ell$.
       \ENDIF
    \ENDFOR
    \STATE Initialize \algcode{pt\_list} to contain a list of all
       interface IB points.
    \FORALL{$\mathbf{X}_\ell \in$~\algcode{pt\_list}}
       \STATE Let \abcd\ be the interpolation box corresponding to
           $\mathbf{X}_\ell$.
       \IF{either of the neighbouring IB points $\ell\pm 1$ has the same
           interpolation box \abcd \break
           \OR the \label{algo:step4} minimum distance $MD_\ell< \deltamin$}
              \STATE Remove $\mathbf{X}_\ell$ from \algcode{pt\_list}.
       \ENDIF
    \ENDFOR
  \end{algorithmic}
\end{algorithm}

After executing Algorithm~\ref{algo:exclusion-filter}, we obtain a list
of IB points at which fluid stress tensor components can be estimated
inside corresponding interpolation boxes.  We first determine
approximations of the velocity derivatives at the corners of each
interpolation box using one-sided second-order differences, where the
points included in the difference stencils are chosen to lie entirely
inside (or outside) the biofilm colony as determined by the unit normal
to the interface.  We then take the velocity derivatives and pressures,
and apply bicubic interpolation~\cite{press1992numerical} to determine
corresponding values at the stress evaluation point, which are then
combined to obtain the stress tensor components, $\stress_{11}$,
$\stress_{22}$ and $\stress_{12}=\stress_{21}$. In most cases, the
set of IB points selected for computing the stress tensor
components inside and outside the biofilm colony are different;
therefore, computing stress jumps requires that stresses be
interpolated onto points in each set.  We may then obtain all necessary
values of the the stress jump $[\boldsymbol{\sigma}]$ and the tangential
component of the interfacial shear stress, $\mathbf{t} \cdot
[\boldsymbol{\sigma}] \cdot \mathbf{n}$.  Finally, the drag and lift
forces are computed by integrating the jump in shear stress along the
biofilm-fluid interface using
\begin{gather}
  \label{eq-liftdrag}
  \begin{pmatrix}
    f_D \\ f_L
  \end{pmatrix}
  = \int \left( n_1, n_2 \right) \cdot
  \begin{pmatrix}
    \lbrack\stress_{11}\rbrack & \lbrack\stress_{12}\rbrack   \\
    \lbrack\stress_{21}\rbrack & \lbrack\stress_{22}\rbrack
  \end{pmatrix}
  \; d\surf .
\end{gather}
\nomenclature[fdfl]{$f_D,\, f_L$}{drag and lift force}%

\subsection{Computing the averaged equivalent continuum stress inside
  the biofilm}
\label{eqcont-node-stress}

In our IB model, the biofilm continuum is replaced by a network of
discrete springs wherein the elastic restoring forces arising from
stretched/compressed springs take the place of stress and strain in
an real elastic continuum.  The most natural way to simulate biofilm
detachment within such a spring network representation is to
cut any spring links for which the local strain exceeds a critical value, as
explained in~\cite{alpkvist2007,hammond2012}.  However, this approach
suffers from several drawbacks.  First of all, the force resulting from
stretching or contraction of a 1D spring element cannot accurately
capture the actual strain in an elastic continuum and consequently
there is no direct way to determine a critical spring strain threshold
based on measured biofilm mechanical properties.  This
contrasts markedly with other approaches such as~\cite{bol2009, duddu2009,
  picioreanu2001} that discretize the solid mechanics equations directly
and employ a more physically
realistic von~Mises yield stress criterion to initiate detachment. In
addition, there is no reliable way in the spring network approach to
determine spring parameters so as to ensure that different
triangulations exhibit similar detachment dynamics under the same flow
conditions.

In order to bridge this gap between continuum mechanics-based biofilm
models and the discrete IB spring-based model, we introduce the
notion of a stress tensor defined at each node in the
network. This is accomplished by assuming that there is an equivalent
continuum representative elementary area or REA surrounding each node,
within which we compute an average value of the stress
tensor components in terms of the spring forces acting on that node.
The primary motivation for this definition comes from the Discrete
Element Method (DEM) for computing microstructural stress in a granular
medium~\cite{bagi1996stress, fortin2003construction}. In the DEM, the
dynamics of a granular medium are determined by treating each
grain separately and solving the governing force balance equations
under the combined action of grain contact forces, body forces and
external forces.  In place of grains we have IB points, and our Hookean
spring connections replace the contact forces between the grains.

\myparagraph{Constructing the REA around each IB point}

\begin{figure}[bthp]
  \footnotesize\sffamily
  \centering
  \includegraphics[width=0.6\textwidth,clip,bb=163 94 621 545]{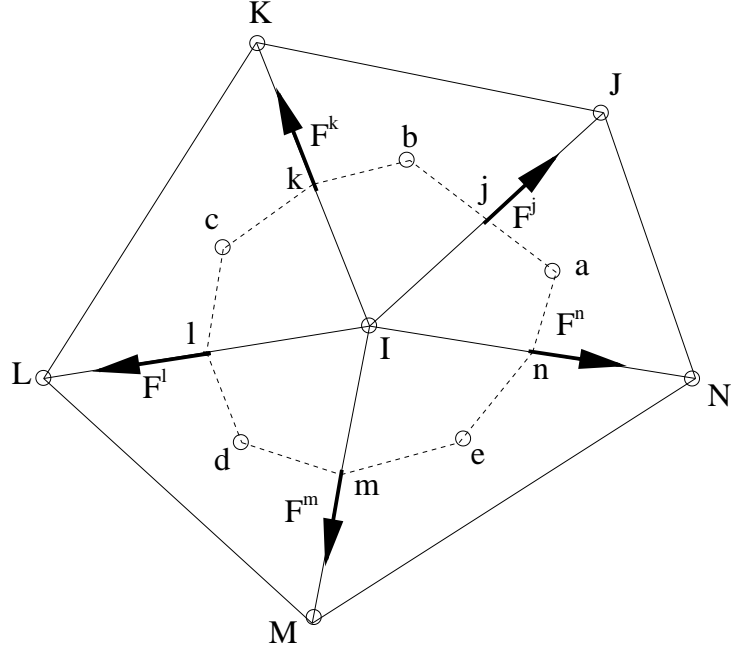}
  \caption{The representative elementary area (REA) surrounding IB node
    $I$ corresponds to the polygonal region $\REA$ (shown using dashed
    lines) connecting the centroids ($a,b,c,d,e$) of the surrounding
    triangles to mid-points ($j,k,l,m,n$) of the edges emanating from
    node $I$.}
  \label{Fig4-eqcontstress}
\end{figure}

In contrast with DEM simulations of granular media that identify an REA
with a Voronoi cell constructed from a Delaunay
triangulation~\cite{bagi1996stress}, we employ instead a control volume
finite element method construction \cite{voller2009basic} that is based on a
(non-Delaunay) triangulation generated by \DistMesh~\cite{persson2004simple}.
For each biofilm point labelled $I$ in the triangulation, the corresponding
REA is constructed by joining with straight lines the centroids of all
surrounding triangles to the mid-points of the corresponding edges
(springs) emanating from node $I$ as depicted in
Fig.~\ref{Fig4-eqcontstress}.  If there are $m$ springs
connected to node $I$, then the corresponding REA is a polygon with $2m$
sides.  By this construction, we note that the area $\area$ of
the REA around a node (which is required for calculating the stress tensor
later in Eq.~\eqref{eq-avgeqstress_final}) is one-third of the total area of
all triangles surrounding the node.  This REA
construction extends in a straightforward manner to volumes in three
dimensions.

\myparagraph{Computing the equivalent continuum stress tensor}

Consider the REA surrounding a point $\mathbf{X}^{I}$ in the
triangulation, pictured as a dashed polygonal region in
Fig.~\ref{Fig4-eqcontstress}.  Denote the REA and its boundary by
$\REA$ and $\dREA$ respectively, and let $\area$ refer to the area of
$\REA$.  Suppose that an equivalent continuum material occupies the REA,
with stress field having Cartesian components $\stress^{eq}_{ij}$ for
$i,j=1,2$.  Then the average stress over $\REA$ is
\begin{gather}
  \label{eq-avgeqstress}
  \overline{\stress}_{ij}^{eq} = \frac{1}{\area} \iint_\REA
  \stress^{eq}_{ij}\; d\area,
\end{gather}
\nomenclature[teq]{$eq$}{equivalent}%
\nomenclature[aprea]{$\REA$}{representative elementary area or REA}%
which can be manipulated to obtain
\begin{gather*}
  \overline{\stress}_{ij}^{eq}
  = \frac{1}{\area} \iint_\REA \stress^{eq}_{kj} x_{i,k} \; d\area
  = \frac{1}{\area} \iint_\REA \left[ (\stress^{eq}_{kj} \
    x_i)_{,k}  - x_i \ \stress^{eq}_{kj,k} \right]\; d\area,
\end{gather*}
where the subscript ``$,\,k$'' denotes a $k$-component derivative and
the Einstein summation convention is assumed for repeated indices.  The
divergence theorem may then be applied to the first term to get
\begin{gather}
  \overline{\stress}_{ij}^{eq}
  = \frac{1}{\area} \left(\oint_{\dREA} \stress^{eq}_{kj}
    x_i n_k\; d\surf -
    \iint_\REA x_i \ \stress^{eq}_{kj,k}\; d\area \right).
  \label{eq:stress-ij-eq}
\end{gather}

Now consider the two types of force that can act on the REA: a surface
traction force $T_j(\mathbf{x})$ that acts at points on the REA
boundary, and a body force $g_j(\mathbf{x})$ acting at interior points.
Imposing a force balance on the boundary yields
\begin{gather}
  \label{eq-bceqcont}
  \stress^{eq}_{ij} n_i(\mathbf{x}) = T_j(\mathbf{x})
\end{gather}
at points $\mathbf{x}\in\dREA$, where $n_i$ denotes the
\emph{outward-pointing} unit normal vector to $\dREA$.  We note here
that the convention in solid mechanics is to use the inward normal
(which ensures compressive stresses are positive), however we break this
convention for the sake of consistency with rest of the text.  Balancing
forces in the interior of the REA gives
\begin{gather}
  \label{eq-linmom}
  \stress^{eq}_{ij,i} \;\; + \;\; \rho g_j = \rho a_j,
\end{gather}
where $a_j$ is the $j$--component of acceleration.  Substituting
Eqs.~\eqref{eq-bceqcont}--\eqref{eq-linmom} into \eqref{eq:stress-ij-eq}
then yields
\begin{gather}
  \label{eq-avgstressderiv}
  \overline{\stress}_{ij}^{eq}
  = \frac{1}{\area} \left(\oint_{\dREA}  x_i T_j\; d\surf\;\; +\;\;
    \iint_\REA \rho x_i (g_j - a_j)\; d\area \right).
\end{gather}

We now introduce notation for IB nodes $\mathbf{X}^\alpha$,
$\alpha=J,K,L,M,N$, that are immediate neighbors of node $I$ in the
triangulation shown in Fig.~\ref{Fig4-eqcontstress}, along with
corresponding edge vectors $\mathbf{E}^\beta$ for $\beta=j,k,l,m,n$
directed outward along springs (with
$\mathbf{E}^j=\mathbf{X}^J-\mathbf{X}^I$ when $\beta=j$, for example).
The edge mid-points are then denoted by $\mathbf{Z}^\beta=\mathbf{X}^I +
\mathbf{E}^\beta/2$, with corresponding spring forces
$\mathbf{F}^\beta$.  If we associate the boundary traction force
$\mathbf{T}(\mathbf{x})$ for $\mathbf{x}\in\dREA$ with the spring forces
$\mathbf{F}^{\beta}$, then the boundary integral term in
\eqref{eq-avgstressderiv} may be rewritten in component form as
\begin{gather}
  \label{contstress_bc}
  \oint_{\dREA} x_i T_j\; d\surf = \sum_{\beta \in \dREA} Z^{\beta}_i
  F^{\beta}_{j}.
\end{gather}
Substituting this expression into~\eqref{eq-avgstressderiv} yields
\begin{align}
  \label{eq-avgstressderiv2}
  \overline{\stress}_{ij}^{eq}  &=
  \frac{1}{\area} \left(\sum_{\beta \in \dREA} Z^{\beta}_i F^{\beta}_{j} +
    \iint_\REA \rho x_i (g_j - a_j)\; d\area \right).
\end{align}

We assume in this paper that the biofilm is neutrally buoyant and that
the equivalent continuum stress is computed in a steady-state
configuration for which inertial forces are negligible; consequently,
the integral term in Eq.~\eqref{eq-avgstressderiv2} is zero. The
remaining summation term can be further simplified by substituting
$\mathbf{Z}^\beta=\mathbf{X}^I + \mathbf{E}^\beta/2$ and using the
equilibrium condition
\begin{gather}
  \label{xi-fbeta}
  X^I_i \sum_{\beta \in \dREA} F^{\beta}_{j} = 0,
\end{gather}
to obtain
\begin{gather}
  \label{eq-avgeqstress_final}
  \left( \overline{\stress}_{ij}^{eq} \right)_I = \frac{1}{2\area}
  \sum_{\beta \in \dREA} E^{\beta}_i F^{\beta}_{j}.
\end{gather}
We note that this averaged equivalent continuum stress tensor is
guaranteed to be
symmetric ($\overline{\stress}_{ij}^{eq} =
\overline{\stress}_{ji}^{eq}$), which should be contrasted with the
analogous derivation for granular media where the DEM approach leads to
a non-symmetric stress tensor owing to contact forces with
a nonzero moment about grain centers \cite{bagi1996stress}.
This does not happen in our IB spring network because elastic spring
forces always act along lines connecting IB points and hence do not
generate any such moments.

For our choice of polygonal REA around node $I$, the area $A$
in~\eqref{eq-avgeqstress_final} is equal to one-third of the total area
of all triangles surrounding the node.  If we apply our stress
calculation method to a biofilm that only deforms and experiences no
detachment, the number of springs connected to each node is constant and
a simple data structure can be used to store the information needed to
compute stress.  Furthermore, the area $A$ and edge vector components
${E}_i^\beta$ are easily updated using the current IB node
coordinates. However, in the more complicated case with detachment,
extra geometric information must be stored along with the IB point data;
for example, we must keep track of the changing number springs
connecting each IB node as well as the number of active triangles around
each node. The necessary changes to the code and data structures are
straightforward, and the added computational cost is negligible in
comparison to that for the IB algorithm.

\subsection{Biofilm detachment criterion}
\label{algo-detachment}

As mentioned earlier, we handle biofilm detachment using a yield stress
criterion from the von~Mises stress theory, which is one of many
approaches used to model failure of a ductile material
subjected to external loading~\cite{boresi1993advanced}.  The von~Mises
yield stress criterion has been employed successfully in the context of
biofilm modeling \cite{bol2009, duddu2009, picioreanu2001}, even though
the biofilm composite (composed of cells, EPS and fluid) is not
strictly a ductile material.  In two dimensions, the von~Mises yield
stress at any IB point inside the biofilm region can be expressed in
terms of the averaged equivalent continuum stress tensor components in
Eq.~\eqref{eq-avgeqstress_final} as
\begin{gather}
  \label{eq-vonmises}
  (\stress_{von})^2 = (\overline{\stress}^{eq}_{11})^2 -
  \overline{\stress}^{eq}_{11} \overline{\stress}^{eq}_{22} +
  (\overline{\stress}^{eq}_{22})^2 + 3 (\overline{\stress}^{eq}_{12})^2.
\end{gather}
\nomenclature[gsigavon]{$\sigma_{von}$}{von~Mises yield stress \nomunits{g/cm\;s^2}}%
The value of $\stress_{von}$ is then compared to some threshold yield
stress, which is a measure of the biofilm cohesive strength. If
$\stress_{von}$ exceeds the threshold, then detachment is initiated by
severing all springs connecting it to neighbouring IB points.  This
should be contrasted with other approaches
based on edge strain \cite{alpkvist2007, hammond2012}, wherein
springs are cut individually based on some threshold strain value
and an IB point is only detached from the colony when all springs
connected to it are severed.

The detachment process in the case of biofilms is complicated somewhat
by the fact that yield strength varies
throughout the biofilm colony.  For example, the strength with which the
base of the biofilm colony adheres to the substratum, which we call the
biofilm adhesive strength $\stressAdh$, is several orders of magnitude
larger than the cohesive strength of the bulk biofilm material.
Furthermore, the bulk cohesive strength also varies since the
portion of the colony nearest the biofilm-fluid interface has a stress
threshold $\stressCohExt$ that is significantly smaller than the value
$\stressCohInt$ in the interior.  Consequently, we have three
threshold values satisfying $\stressCohExt < \stressCohInt < \stressAdh$,
which leads to a natural separation of the biofilm into three zones --
substratum, interface and interior.
\nomenclature[gsigzadh]{$\stressAdh$}{biofilm adhesive stress threshold
  near the substratum}%
\nomenclature[gsigzcohe]{$\stressCohExt$}{biofilm cohesive stress
  threshold near the biofilm-fluid interface}%
\nomenclature[gsigzcohi]{$\stressCohInt$}{biofilm cohesive stress
  threshold in the interior}%

\begin{figure}[bthp]
  \footnotesize\sffamily
  \centering
  \subfigure[]{%
    \includegraphics[height=0.25\textheight,clip,bb=80 80 520 475]{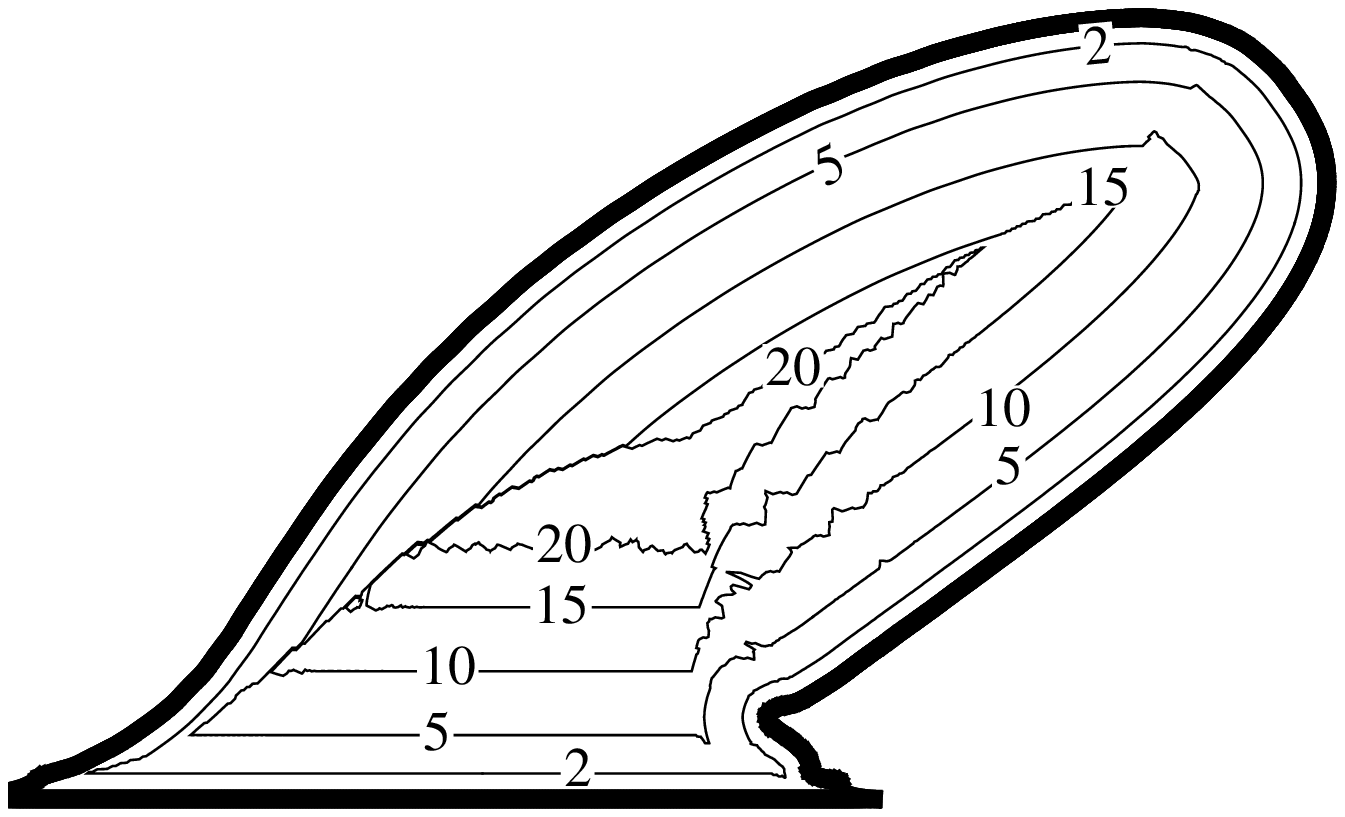}
    \label{fig4b:subfigure1}}
  \qquad \qquad
  \subfigure[]{%
    \includegraphics[height=0.36\textheight,clip,bb=120 65 500 720]{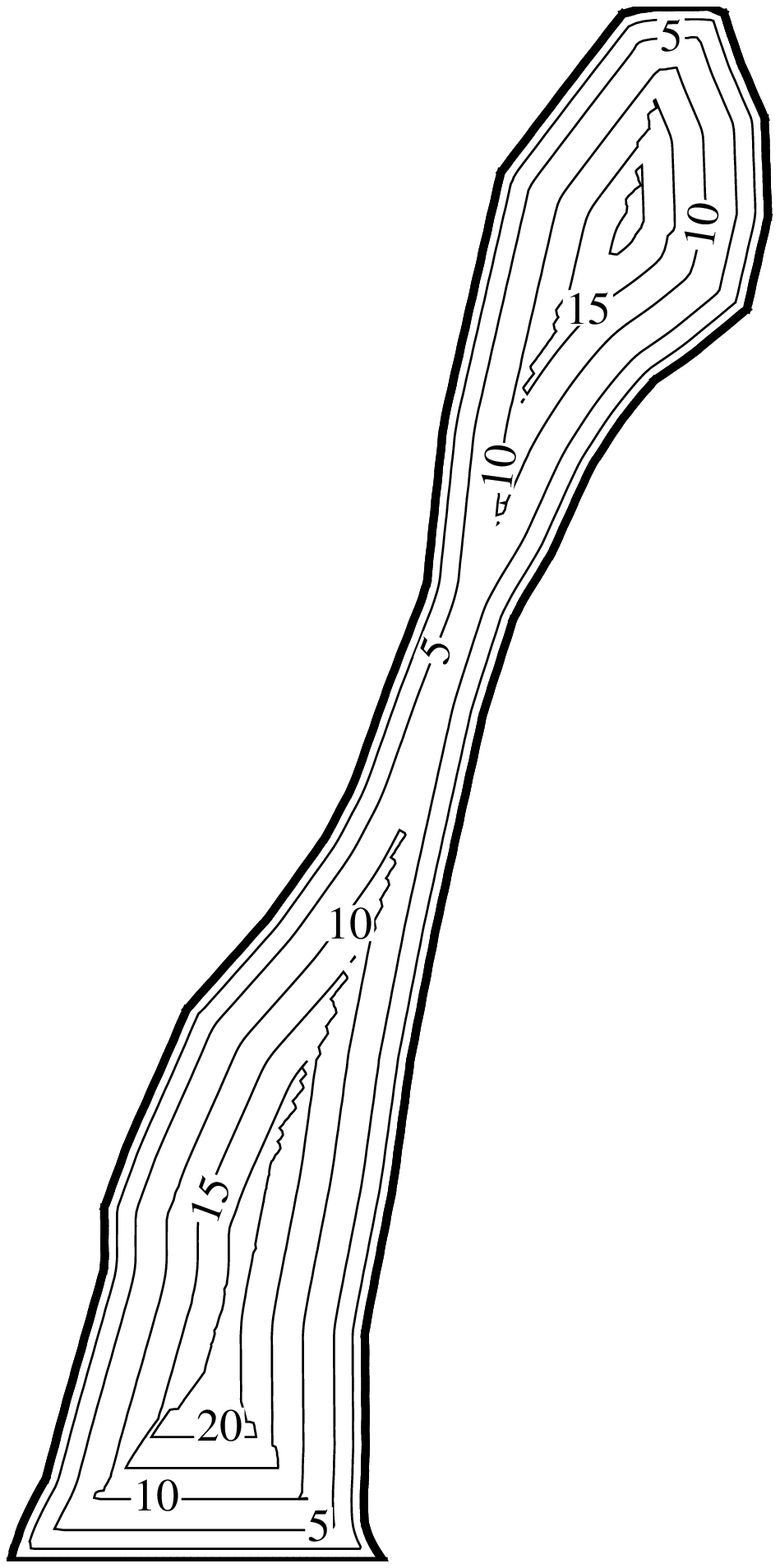}
    \label{fig4b:subfigure2}}
  \caption{Contours of the distance function $D_\ell^{ext}$ (labels in
    $\myunit{cm}$) from any given IB point to the biofilm-fluid
    interface, computed using the algorithm in~\cite{elias2007simple}.
    Two biofilm colonies are shown near steady-state: (a) the elliptical
    configuration \testcase{Sup75}, and (b) a mushroom-shaped colony.}
  \label{Fig4b-distfunction}
\end{figure}

With this in mind, we propose the following detachment strategy.  First,
for any given IB point $\mathbf{X}_\ell$ we calculate the shortest
distance from point $\ell$ to the substratum and to the
biofilm-fluid interface, denoted by $D_\ell^{sub}$ and $D_\ell^{ext}$
respectively.  We then select a yield stress criterion to be
imposed by determining which zone the IB point belongs to:
\begin{description}
\item[Zone~1:] consists of all IB points near the substratum that
  satisfy $D_\ell^{sub} \leqslant \deltasub$, where $\deltasub$ is a
  user-specified parameter.  In this case, the point detaches whenever
  $\stress_{von} \geqslant \stressAdh$.
\item[Zone~2:] consists of any of remaining IB points near the
  biofilm-fluid interface that satisfy $D_\ell^{ext} \leqslant
  \deltaext$.  Here, the point detaches when $\stress_{von} \geqslant
  \stressCohExt$.
\item[Zone~3:] consists of all remaining interior biofilm points, which
  detach if $\stress_{von} \geqslant \stressCohInt$.
\end{description}
Calculating the distance $D_\ell^{sub}$ is trivial because the bottom
wall in our numerical simulations is parallel to the $x$--axis.
However, the calculation of $D_\ell^{ext}$ is more involved owing to the
irregular biofilm shape and also because the colony deforms in time,
hence requiring that $D_\ell^{ext}$ be recalculated in each time step.
Therefore, care must be taken in order to design an efficient algorithm
for estimating $D_\ell^{ext}$; for this purpose we employ a finite
element-based signed distance function for triangles developed
in~\cite{elias2007simple}, which is an extension of the fast marching
method.  The cost of this algorithm can be optimized by only calculating
the distance function at IB points lying within a narrow band near the
interface.  As an illustration, Fig.~\ref{Fig4b-distfunction} shows
sample contour plots of $D_\ell^{ext}$ for two different biofilm colony
shapes.

The treatment of biofilm detachment is performed after completing Step~1
(the force calculation step) in the immersed boundary algorithm.  The
steps in the yield stress based detachment process are summarized in
Algorithm~\ref{algo:detach}. We note that in the implementation outlined
here, we make use of three integer arrays of ``status flags'' -- named
\algcode{*\_STATUS} with \algcode{* = INODE, ISPRING, ITRI} -- one each
for IB nodes, springs and triangles.  These flags are set to either 0 or
1 depending on whether the status is detached or active, respectively.

\begin{algorithm}[htbp]
  \caption{Biofilm detachment algorithm based on equivalent continuum
    stress and the von~Mises yield stress criterion.}
  \label{algo:detach}
  \begin{algorithmic}[1]
    \STATE For each IB node $\ell$, compute the distance from the
    substratum, $D_\ell^{sub}$.

    \STATE Identify IB nodes $\ell$ that lie on
    the interface, and assign $D_\ell^{ext}=0$ there.

    \STATE At all remaining IB nodes, compute the distance function
    $D_\ell^{ext}$ using the algorithm in~\cite{elias2007simple}.

    \FORALL{IB nodes $\mathbf{X}_\ell$ that are active (with
      \algcode{INODE\_STATUS = 1})}

      \STATE Compute the REA area $A$ as the sum of areas of all active
        triangles (with \algcode{ITRI\_STATUS = 1}) neighbouring node
        $\ell$.

      \STATE Use all active springs (with \algcode{ISPRING\_STATUS = 1})
        to calculate the equivalent continuum stress components
        $\stress^{eq}_{ij}$ from \label{alg:step2}
        Eq.~\eqref{eq-avgeqstress_final}.

      \STATE Compute the von~Mises yield stress $\stress_{von}$
        using Eq.~\eqref{eq-vonmises}.

      \IF{$D_\ell^{sub} \leqslant \deltasub$}
         \STATE Let \algcode{THRESHOLD = }$\stressAdh$.
      \ELSIF{$D_\ell^{ext} \leqslant \deltaext$}
         \STATE Let \algcode{THRESHOLD = }$\stressCohExt$.
      \ELSE
         \STATE Let \algcode{THRESHOLD = }$\stressCohInt$.
      \ENDIF

      \IF{$\stress_{von}\geqslant$ \algcode{THRESHOLD}}
         \STATE Let \algcode{INODE\_STATUS = 0} (detached).
         \STATE For each spring adjacent to this node, set
           \algcode{ISPRING\_STATUS = 0}.
         \STATE Use \algcode{INODE\_STATUS} and
           \algcode{ISPRING\_STATUS} to update the status of the triangle associated with the detached node to inactive (\algcode{ITRI\_STATUS = 0}).
       \ENDIF
    \ENDFOR
  \end{algorithmic}
\end{algorithm}

\section{Numerical simulations}
\label{Results}

\subsection{Model parameters}
\label{modelparam}

Table~\ref{Table1} summarizes all parameter values used in our
numerical simulations of biofilm--fluid interaction.  The parameters are
separated naturally into the following categories:
\begin{table}[htbp]
  \centering
  \caption{Parameter values for the various numerical test cases.}
  \label{Table1}
  \begin{tabularx}{0.9\linewidth}{|llX|}
    \hline
    \multicolumn{2}{|l}{Description}  & Values \\\hline\hline
    \multicolumn{3}{|l|}{\emph{Fluid domain:}} \\
    \qquad
    & Domain height        & $H = 3 \times \Hb$ \\
    & Colony spacing       & $\Db =  50,\ 150,\ 250,\ 400\;\mymum\;(\Wb,3\Wb,5\Wb,8\Wb)$ \\\hline
    \multicolumn{3}{|l|}{\emph{Biofilm shape (width and height):}} \\
    & Semi-circle:         & $\Wb = 40\;\mymum$, $\Hb = 20\;\mymum$ (\testcase{Semi20}) \\
    & Super-ellipse:       & $\Wb = 50\;\mymum$, $\Hb = 25\;\mymum$ (\testcase{Sup25}) \\
    &                      & $\Wb = 50\;\mymum$, $\Hb = 50\;\mymum$ (\testcase{Sup50}) \\
    &                      & $\Wb = 50\;\mymum$, $\Hb = 75\;\mymum$ (\testcase{Sup75}) \\\hline
    \multicolumn{3}{|l|}{\emph{Fluid/biofilm grid:}} \\
    & Fluid grid spacings  & $h_x=h_y=0.5\;\mymum$ (\testcase{Semi20}, \testcase{Sup25}) \\
    &                      & $h_x=h_y = 0.75\;\mymum$ (\testcase{Sup50}, \testcase{Sup75})  \\
    & IB wall point spacing& $h_{wall} = \frac{1}{4} \min(h_x,\;h_y)$ \\
    & Biofilm spring rest-length & $\dzero = 0.15\;\mymum $ (\testcase{Semi20}, \testcase{Sup25}) \\
    &                      & $\dzero = 0.225\;\mymum$ (\testcase{Sup50}, \testcase{Sup75})  \\\hline
    \multicolumn{3}{|l|}{\emph{Fluid/biofilm material properties:}} \\
    & Fluid density        & $\rho =  1.0\;\myunit{g/cm^3}$ \\
    & Fluid viscosity      & $\mu = 0.01\;\myunit{g/cm\;s}$ \\
    & Shear rate           & $G = 0.625\;\myunit{s^{-1}}$ \\
    & Biofilm spring stiffness & $\stiff_{bio}= 0.75 \dzero,\; 7.5 \dzero,\; 75 \dzero \;\; \myunit{g/cm^2\;s^2}$ \\
    & Wall spring stiffness& $\stiff_{wall} = 10^5\;\myunit{g/cm^2\;s^2}$  \\ \hline
\end{tabularx}
\end{table}
\begin{itemize}
\item \emph{Biofilm geometry:} We took four different biofilm colony
  shapes, one a semi-circle of radius $20\;\mymum$ and width $\Wb=40$
  (labeled \testcase{Semi20}), and three (semi-)super-ellipses having
  width $\Wb=50\;\mymum$ and height $\Hb=25$, $50$ and $75\;\mymum$
  (labeled \testcase{Sup25}, \testcase{Sup50}, \testcase{Sup75}
  respectively). These dimensions are representative of typical biofilm
  colonies and also capture a range of aspect ratios observed
  in early stages of biofilm colony growth.

\item \emph{Fluid domain:} The vertical spacing $H$ between top and
  bottom channel walls is set to triple the height of the biofilm colony
  ($H=3\Hb$) in order to minimize boundary effects. Simulations reveal
  that at our results are relatively insensitive to changes in $H$. The
  width of the fluid domain is $\Wb + \Db$, and values of $\Db=\Wb$,
  $3\Wb$, $5\Wb$ and $8\Wb$ were used to study the effect of colony
  spacing.

\item \emph{Fluid grid:} We chose relatively small values of fluid grid
  spacing $h_x$ and $h_y$ that permit accurate resolution of the biofilm
  colony.  In particular, we aimed to ensure that recirculating eddies
  arising from flow separation are well captured.  A constant time step
  of $\Delta t=10^{-5}\;\myunit{s}$ was used in all simulations.

\item \emph{Biofilm grid:} The mean spacing $\dzero$ between biofilm IB
  points was chosen to satisfy $\dzero \leqslant \frac{1}{3} \min(h_x,\;
  h_y)$ as in~\cite{vo2010} so as to avoid numerical errors due to
  leakage of fluid between IB points.  This value of $\dzero$ is
  provided to the \DistMesh\ code as a measure of average edge length
  for the biofilm triangulation; for example, the \testcase{Sup75}
  biofilm colony with $\dzero=0.225\;\mymum$ yields a triangulation with
  81,449 IB nodes and $222,659$ edges.

\item \emph{Fluid material properties:} We take fluid parameters
  consistent with water: $\rho=1\;\myunit{g/cm^3}$ and
  $\mu=0.01\;\myunit{g/cm\;s}$.  The shear rate was set to
  $G=0.625\;\myunit{s^{-1}}$ for all simulations, which is
  high enough to induce large deformations in weak biofilm colonies
  while also attaining a steady state over a relatively short time period
  (roughly 2--8~$\myunit{s}$). This shear rate corresponds to a Reynolds
  number $Re={\rho G \Wb^2}/{\mu}=1.5625\times 10^{-3}$ for the case
  $\Wb=50\;\mymum$, where we have used the width of the biofilm colony
  as a length scale.

\item \emph{Biofilm material properties:} To mimic weak biofilms with
  varying mechanical strength, we choose several values of the IB spring
  stiffness corresponding to $\kodzero=0.75$, 7.5 and 75, where we
  recall that $\kodzero = {\stiff_{bio}}/{\dzero}$.  The wall spring
  stiffness $\stiff_{wall}=10^5$ is chosen much larger so that wall
  points do not move appreciably.
\end{itemize}
In summary, we consider four different biofilm shapes, four colony
spacings and three values of the spring constant, corresponding to a
total of 48 simulations with a single value of shear rate
$G=0.625\;\myunit{s^{-1}}$.

\subsection{Model validation: Channel with a rigid bump}

Before applying our immersed boundary algorithm to the biofilm test
cases outlined in the previous section, we first validate our numerical
approach using a simpler set-up consisting of a rectangular channel
containing a rigid semi-circular bump on the bottom wall. The same
problem was considered in Williams et al.~\cite{williams2009} as an
illustration of their exclusion filter approach for interfacial stress
calculation.

\begin{figure}[bthp]
  \footnotesize\sffamily
  \centering
  \includegraphics[width=0.8\textwidth,clip,bb=60 175 581 626]{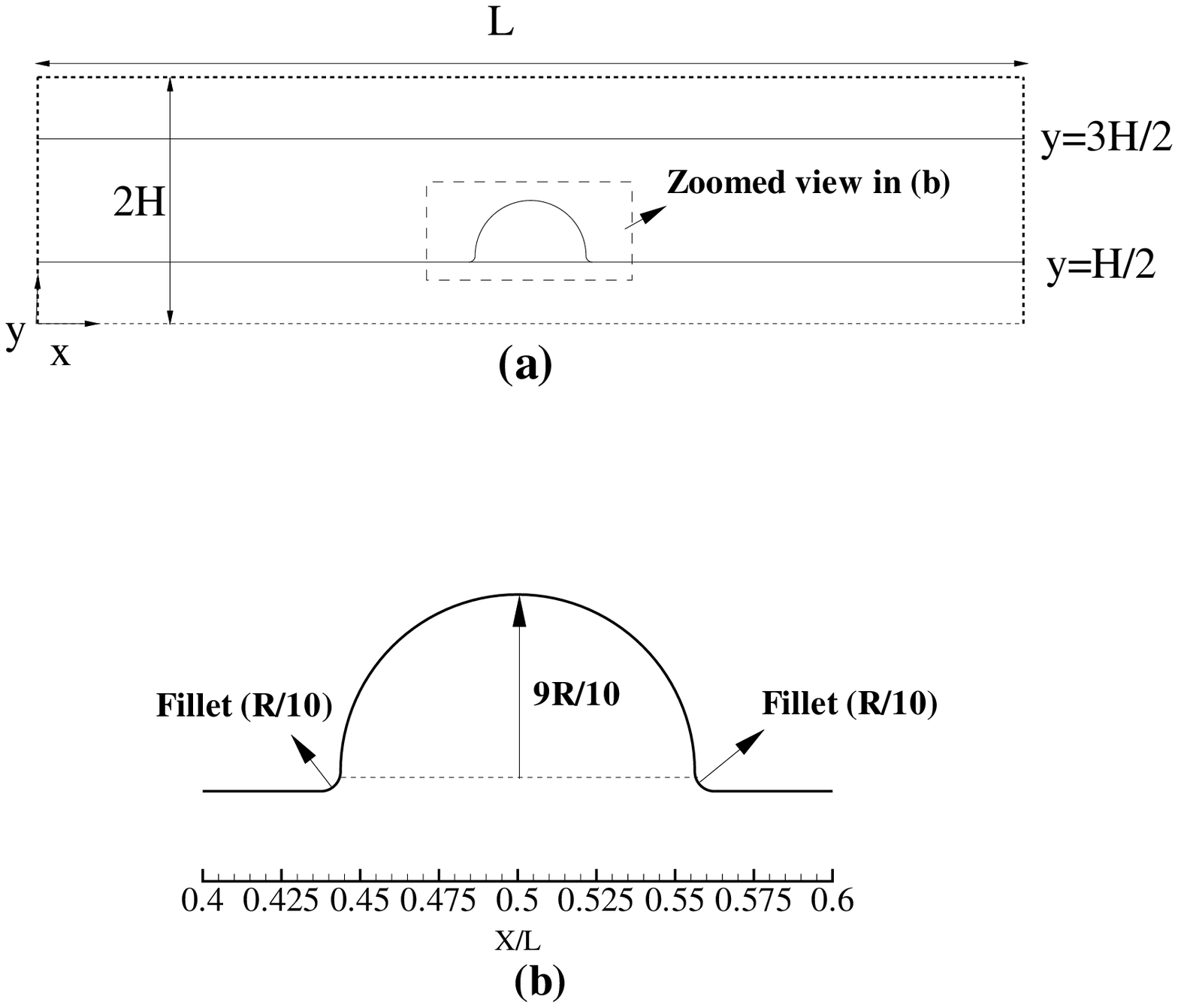}
  \caption{(a) Computational domain for the channel flow with a rigid
    bump.  The fluid domain is a box of size $L \times 2H$ containing
    two parallel, horizontal walls at height $y=\frac{H}{2}$ and
    $\frac{3H}{2}$.  Attached to the bottom wall is a filleted
    (smoothed) semi-circular bump.  The flow is driven from left to
    right through the central channel by imposing a constant (positive)
    fluid body force. (b) Zoomed view of the semi-circular obstruction.}
  \label{Fig5-faucibump}
\end{figure}

The problem geometry is shown in Fig.~\ref{Fig5-faucibump}, consisting
of a channel of width $H$ constructed from two parallel horizontal walls
along $y=\frac{H}{2}$ and $y=\frac{3H}{2}$.  The channel is embedded
within a larger rectangular fluid domain of size $L\times 2H$ and
periodic boundary conditions are imposed the outer boundaries.  A
constant fluid body force $\mathbf{f}_{B}(\mathbf{x}) = (\Delta P/L,\;
0)$ is applied at each fluid grid point inside the channel, which in the
absence of any obstruction would generate a parabolic Poiseuille flow
with pressure difference $\Delta P$ across the channel from left to right.
However, an obstruction is introduced within the channel consisting of an
immersed boundary in the shape of a semi-circle of radius
$\frac{9R}{10}$ and centered at $(\frac{L}{2}, \; \frac{H}{2} +
\frac{R}{10})$.  The corners connecting the bump to the bottom wall are
smoothed using quarter-circular fillets with radius $\frac{R}{10}$ as
shown in Fig.~\ref{Fig5-faucibump}(b) that serve to regularize
the IB shape and avoid flow irregularities near sharp corners.

The channel walls and semi-circular obstruction are represented using IB
points, each of which is connected by a single spring to a tether point
that is fixed in space, and the spring stiffness is chosen large enough
that the IB points do not move appreciably.  The spacing $h_{wall}$
between adjacent IB points is chosen so that $h_{wall} \leqslant \half
\min(h_x,\; h_y)$, which aims to minimize any numerical errors arising
from leakage of fluid across the immersed boundaries.

We choose parameter values the same as in~\cite{williams2009}, namely
$H=0.1\;\myunit{m}$, $L = 0.8\;\myunit{m}$, $R = 0.05\; \myunit{m}$,
$\Delta P/L = -10^{5}\;\myunit{kg/m^2}$, $\mu = 25\;\myunit{kg/m\;s}$,
$\rho = 1\;\myunit{kg/m^3}$ and $\stiff=10^{9}\;\myunit{kg/m^2\;s^2}$.
In contrast with the CGS units used in the rest of this paper, we employ
MKS units in this section only for ease of comparison with the results
in~\cite{williams2009}.  The fluid domain is discretized on a uniform
grid of $512 \times 128$ points and we use a constant time step
$\Delta t = {0.16 \rho h_{wall}^2}/{\mu}$.  The Reynolds number based on
channel height is $Re = {\rho \Delta P H^3}/({8 \mu^2 L}) \approx
0.02$, which indicates that inertial effects are negligible and permits
us to make comparisons with the Stokes flow solution from Gaver and
Kute~\cite{gaver1998theoretical}.

\begin{figure}[bthp]
  \footnotesize\sffamily
  \centering
  \subfigure[]{%
    \includegraphics[width=0.40\textwidth,clip,bb=0 0 420 420]{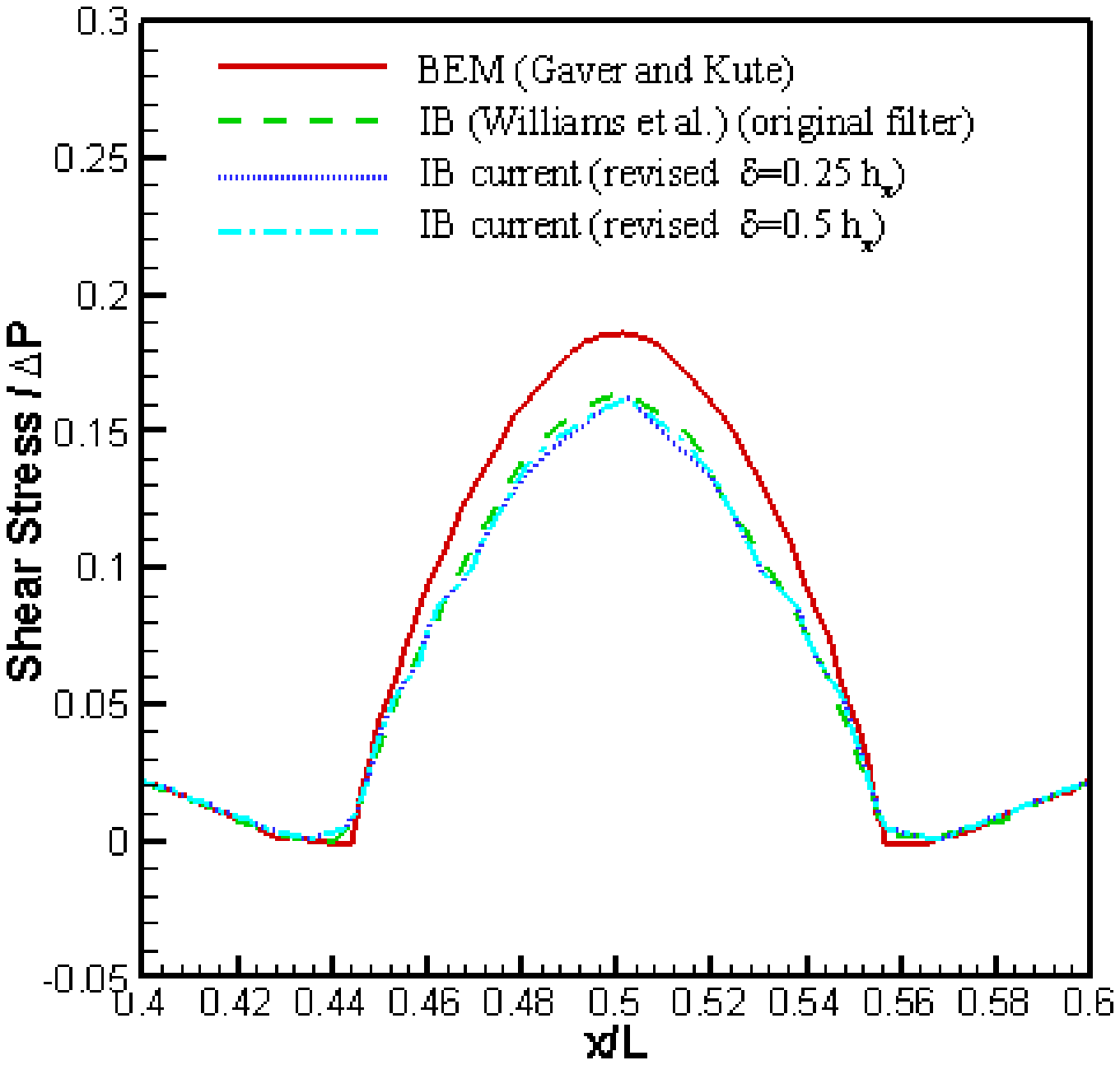}
    \label{fig6:subfigure1}}
  \subfigure[]{%
    \includegraphics[width=0.44\textwidth,clip,bb=0 0 442 434]{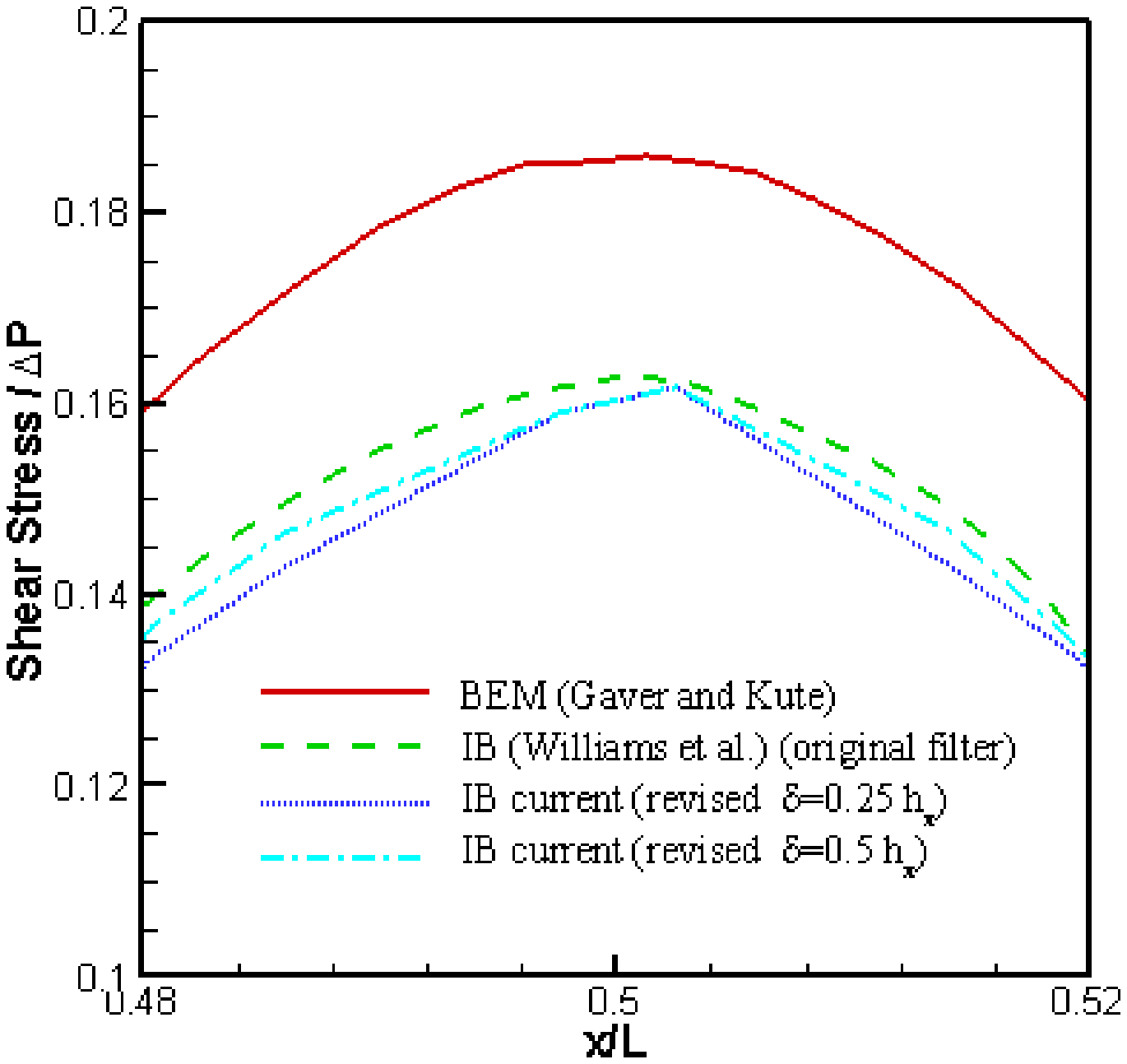}
    \label{fig6:subfigure2}}
  \caption{(a) Dimensionless interfacial shear stress computed along a
    semi-circular obstruction in a channel.  For comparison,
    results are included from another IB approach~\cite{williams2009}
    and the boundary element method~\cite{gaver1998theoretical}.  (b)
    Close-up view near the apex of the bump where shear stress attains
    a maximum.}
  \label{Fig6-validation_shearstress}
\end{figure}

The computed interfacial shear stress, non-dimensionalized by $\Delta
P$, is shown in Fig.~\ref{Fig6-validation_shearstress}(a) along the
central portion of the bottom wall including the semi-circular bump.
Results are presented for two values of the exclusion filter
parameter, $\deltamin = 0.15 h_x$ and $0.4 h_x$. Also included in this
figure are numerical results computed with two other methods, namely the
WFG method with an FS-based exclusion filter~\cite{williams2009}, and
the boundary element computations of Gaver and
Kute~\cite{gaver1998theoretical}.  The results from our revised
exclusion filter are within approximately 5\%\ of WFG's results, with
the differences most pronounced on either side of the crest of the bump
(see Fig.~\ref{Fig6-validation_shearstress}(b)).  The correspondence with
WFG's results improves as $\deltamin$ increases from $0.15 h_x$ to $0.4
h_x$, which can be explained as follows: even though the number of IB
points retained by the filter decreases with increasing $\deltamin$, the
quality of those points is high because they are separated further from
the smearing effects of the interface.

A non-dimensional flow rate can be computed by
integrating the computed velocity vertically along the channel inlet
\begin{gather}
  \label{eq-flowrate}
  Q = \frac{12 \mu L}{H^3 \Delta P} \int_{y=\half H}^{y= \thalf H}
  u(0,y) \; dy,
\end{gather}
yielding a value of $Q=0.554$ that agrees exactly with the result of
WFG~\cite{williams2009} for the same grid resolution.  We also compute
the dimensionless drag
\begin{gather}
  \label{eq-dragforce}
  f^\ast_D = \frac{L}{RH\Delta P} \int_{\Gamma}  T_1 \; d\surf,
\end{gather}
where $T_1$ is the $x$--component of the traction force and $\Gamma$
represents the portion of the bottom wall corresponding to the
semi-circular bump.  Our simulations yield $f^\ast_D=4.662$ and $5.195$
for filter parameters $\deltamin=0.15h_x$ and $0.40h_x$ respectively,
which should be compared with the WFG result of 5.617.  This discrepancy
of 7--17\%\ is acceptable in view of the increased flexibility we gain
from our modified filter in terms of being able to compute stress along
strongly-curved biofilm interfaces.

\subsection{Simulating biofilm deformation: Flow structure and forces}
\label{sec:flow-structure}

In this section, we investigate the response of deformable biofilm
colonies to a shear flow by studying the effect of changes in various
parameters on the biofilm shape, flow structure, hydrodynamic drag and
interfacial shear stress.
\begin{figure}[bthp]
  \footnotesize\sffamily
  \centering
    \includegraphics[width=0.45\textwidth,clip,bb=125 35 675 410]{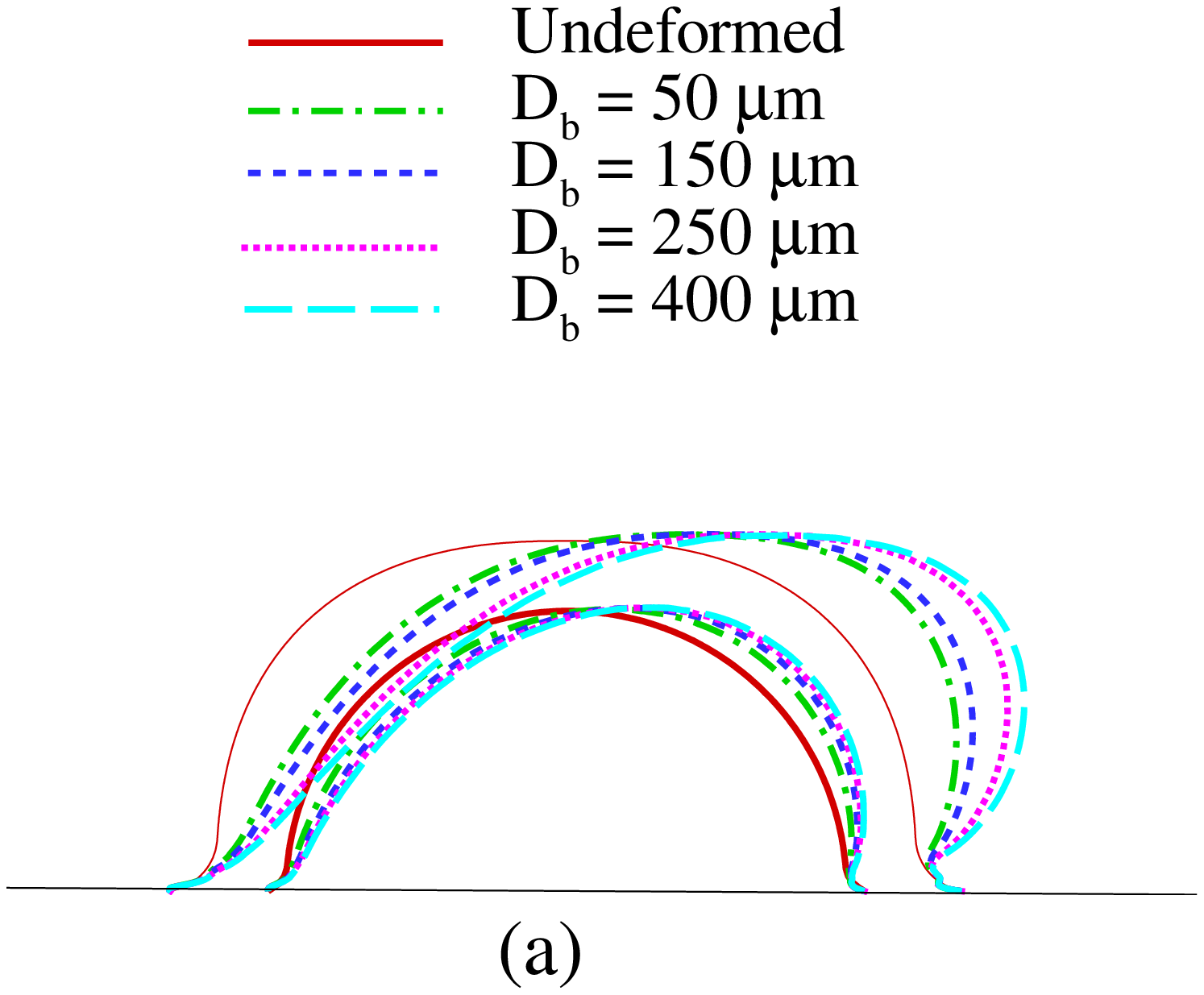}
    \qquad
    \includegraphics[width=0.45\textwidth,clip,bb=125 35 660 545]{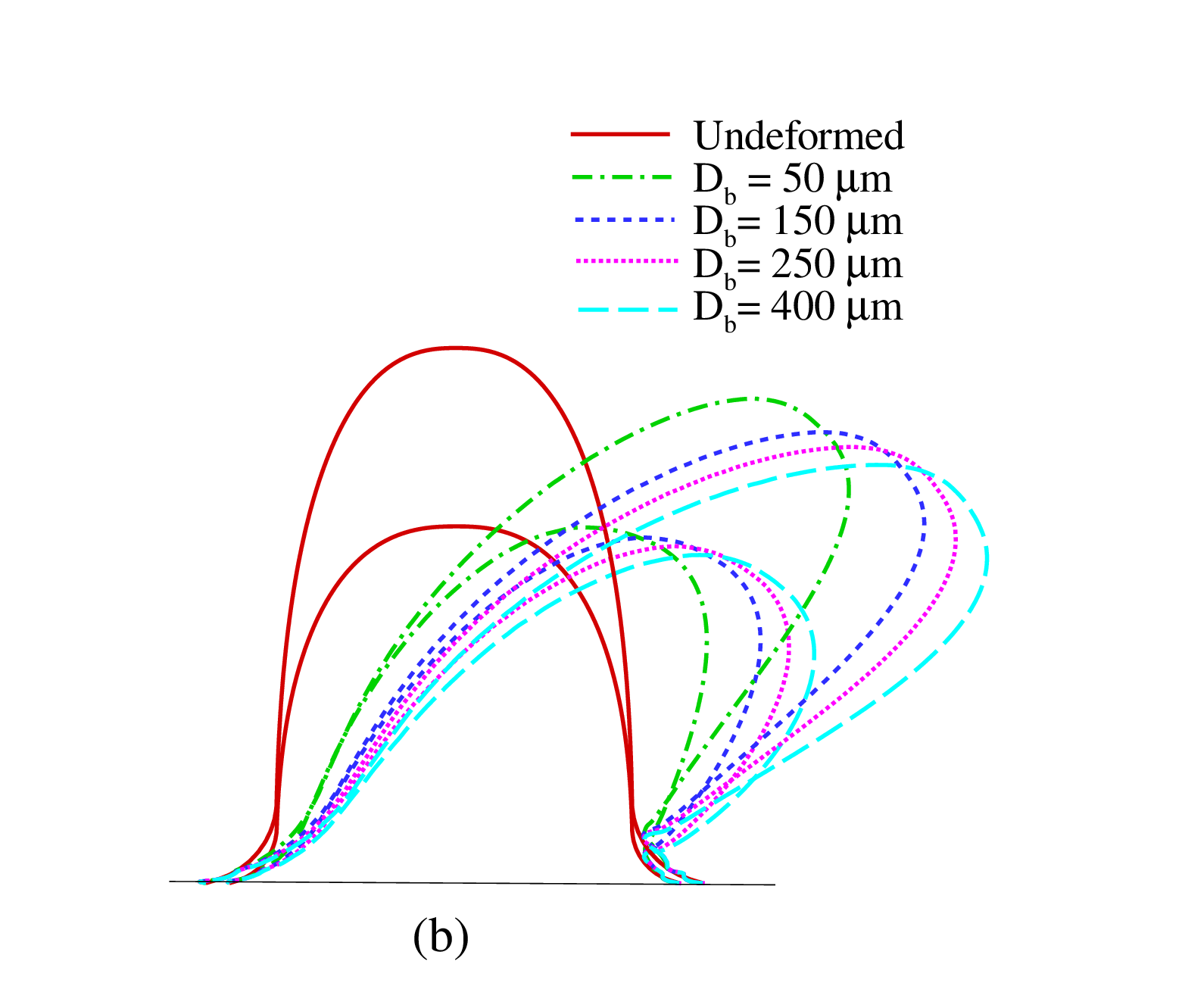}
    \caption{Initial (undeformed) and steady-state (deformed) biofilm
      shapes with $\kodzero=0.75$ as the spacing between colonies $\Db$
      is varied. (a) Test cases \testcase{Semi20} and \testcase{Sup25}.
      (b) Test cases \testcase{Sup50} and \testcase{Sup75}.}
  \label{Fig8-biofilmshape_vs_db}
\end{figure}

We begin by varying the inter-colony spacing $\Db$ for the four initial
colony shapes \testcase{Semi20}, \testcase{Sup25}, \testcase{Sup50},
\testcase{Sup75}.  Fig.~\ref{Fig8-biofilmshape_vs_db} depicts the
initial and final (steady-state) biofilm profiles for values of $\Db$
lying between 50 and $400\;\mymum$, holding the stiffness
parameter $\kodzero=0.75$.  The extent of the deformation
clearly increases as the biofilm height is increased, with the
largest deformations occurring for the \testcase{Sup75} colony.  This
behaviour is physically reasonable because longer structures are not
only more flexible but also extend further into the shear flow where
they experience a higher flow velocity.  The extent of deformation also
increases with the spacing parameter $\Db$, which is to be expected
since the shear flow is more able to impinge between
colonies having a greater separation.

Note that our simulated biofilm colonies appear to simply shear to the
right without exhibiting any of the elongation or vertical lifting that
is observed in the numerical simulations of Vo et~al.~\cite{vo2010,vo2011}.
This discrepancy can be attributed to several sources: (a) Vo et al.\
used different colony shapes having a sinusoidal profile that is much
taller (corresponding to $H=180\;\mymum$); (b) their spring stiffness is
roughly 100 times larger than ours; and (c) they also considered a much
faster flow corresponding to $Re=230$ (based on hydraulic diameter
of the square capillary reactor as length scale) whereas we have $Re={\rho G \Wb^2}/{\mu}\approx
10^{-3}$ (based on a much shorter length scale $\Wb$ corresponding to
the biofilm width).  It was also suggested in~\cite{vo2011}
that a significant contributor to elongation from lifting at
high enough Reynolds number is inertial effects arising from biofilm
deformation as well as the nature of the flow surrounding the colony.

\begin{figure}[bthp]
  \footnotesize\sffamily
  \centering
  \includegraphics[width=0.4\textwidth,clip,bb=115 175 585 625]{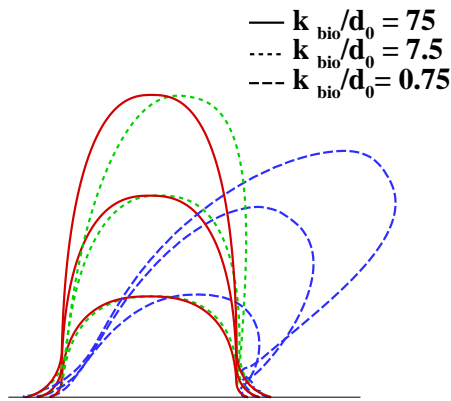}
  \caption{Final (steady state) biofilm shapes for different values of
    $\kodzero$ and constant colony spacing $\Db = 250\;\mymum$.  The
    super-ellipse test cases \testcase{Sup25}, \testcase{Sup50} and
    \testcase{Sup75} are shown.}
  \label{Fig8-biofilmshape_vs_sigmab}
\end{figure}
We next investigate the effect of changes in the spring
stiffness on the steady-state deformation, taking values of
$\kodzero=0.75$, 7.5 and 75 for three different initial colony shapes
while holding $\Db=250\;\mymum$ constant.
Fig.~\ref{Fig8-biofilmshape_vs_sigmab} depicts the final deformed
shapes from which we observe that at the lowest shear rate, even a
relatively modest value of biofilm stiffness $\kodzero\approx 7.5$ is
sufficient to resist deformation.  Indeed, it is only when stiffness is
reduced to $\kodzero=0.75$ that any significant bending of the biofilm
colony occurs.

\begin{figure}[bthp]
  \footnotesize\sffamily
  \centering
  \includegraphics[width=0.50\textwidth,clip,bb=30 90 535 690]{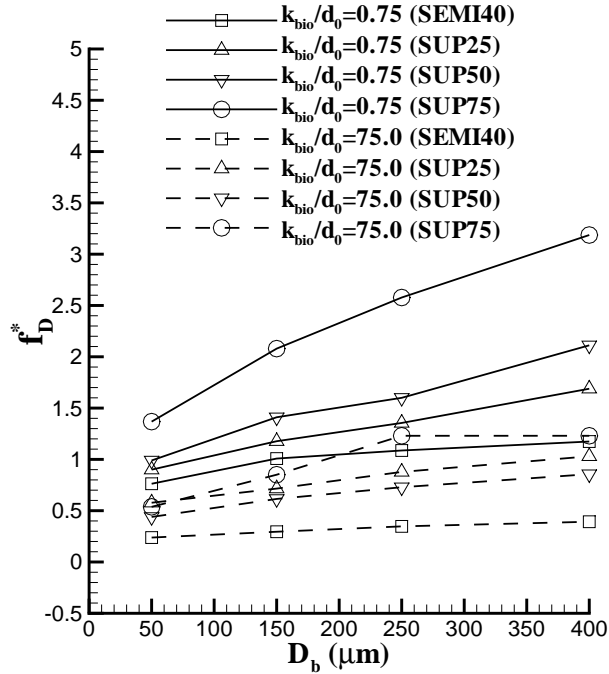}
  \caption{Variation of dimensionless drag force $f^\ast_D$ (at steady
    state) with colony spacing $\Db$.  Curves are shown for four
    different initial colony shapes and for $\kodzero = 0.75$, $75$.}
  \label{Fig10-drag_vs_db}
\end{figure}

The drag force acting on the biofilm at steady state is then computed
for all of simulations above.  Fig.~\ref{Fig10-drag_vs_db} plots the
drag force in each case as a function of colony spacing $\Db$.  Here,
the drag force has been non-dimensionalized using $f^\ast_D=f_D/(\mu G
\Wb)$, where the reference value $\mu G \Wb$ can be thought of as the
force exerted by a linear shear flow with shear rate $G$ acting on a
very thin biofilm colony with $\Hb\ll \Wb$.  The corresponding lift
force is not shown because lift is much smaller than drag (by at least a
factor of 10), not to mention that lift force is much less affected by
changes in colony spacing.  We observe from Fig.~\ref{Fig10-drag_vs_db}
that $f^\ast_D$ is an increasing function of $\Db$, with the rate of
increase being largest for the weakest biofilms having $\kodzero =
0.75$.  In particular, as $\Db$ increases from 50 to 400~$\mymum$, the
drag force increases by a factor of 50\%\ for cases \testcase{Sup25} and
\testcase{Sup50}, with the largest colony in \testcase{Sup75} exhibiting
a roughly 100\%\ increase in drag.  These results should be contrasted
with the simulations of rigid biofilms in~\cite{sudarsan2005} where the
drag force was non-monotonic, attaining a local minimum at some
intermediate value of $\Db$.  Because the drag fails to level out at the
upper end of the range $\Wb \leqslant \Db \leqslant 8\Wb$ considered
here, we would have to explore significantly larger values of colony
separation in order to obtain results consistent with an isolated
biofilm colony.  Because a much larger domain size would be required, we
have not investigated this high-$\Db$ regime for reasons of high
computational cost.

Another observation is that for stiffer biofilms ($\kodzero=7.5$ and 75)
organized into the closest-spaced colonies ($\Db=50\;\mymum$), the drag
force increases by at most 20\%\ as the colony height increases from
\testcase{Semi20} to \testcase{Sup75}, whereas the drag on the weaker
biofilms ($\kodzero=0.75$) increases by more than 50\%.  This should be
contrasted with the widest-spaced colonies ($\Db=400\;\mymum$) where the
drag increase for stiff biofilms is roughly 100\%\ and over 150\%\ for
weak biofilms.  In the latter case, the majority of the drag increase
occurs when the biofilm height increases from 50 to 75~$\mymum$, which
cannot be accounted for by the increase in colony surface area alone
(i.e., perimeter in 2D).  We therefore conclude that biofilm colonies may be
able to grow into tall structures even if they are weak mechanically
because of the protection they gain from having other colonies in close
spatial proximity.

\begin{figure}[bthp]
  \footnotesize\sffamily
  \centering
  \subfigure[$\Db=50\;\mymum$, $\kodzero=0.75$]{%
    \includegraphics[bb=125 200 415 605,clip,height=0.20\textheight]{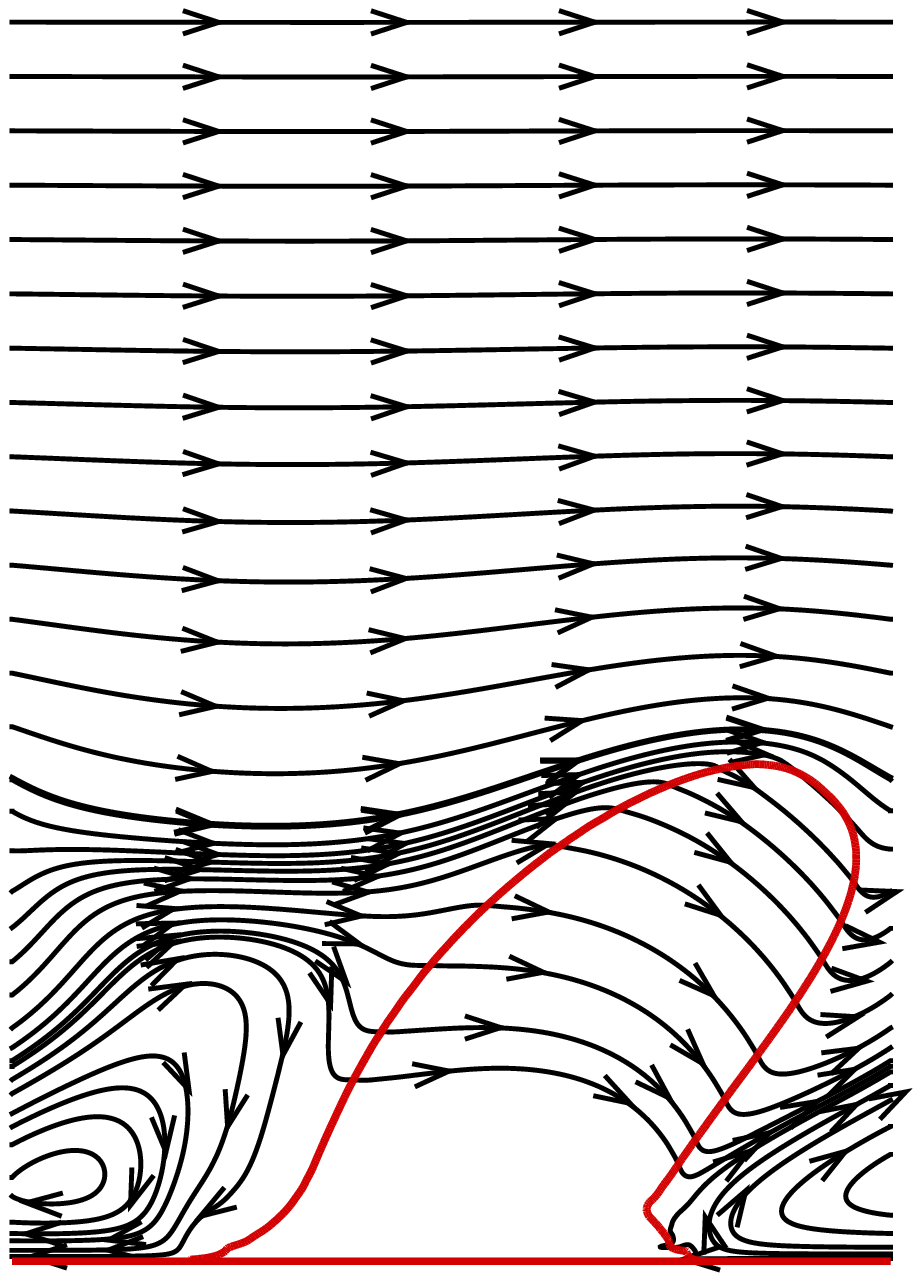}
    \label{fig:subfigure1}}
  \quad
  \subfigure[$\Db=400\;\mymum$, $\kodzero=0.75$]{%
    \includegraphics[bb=85 170 530 410,clip,height=0.23\textheight]{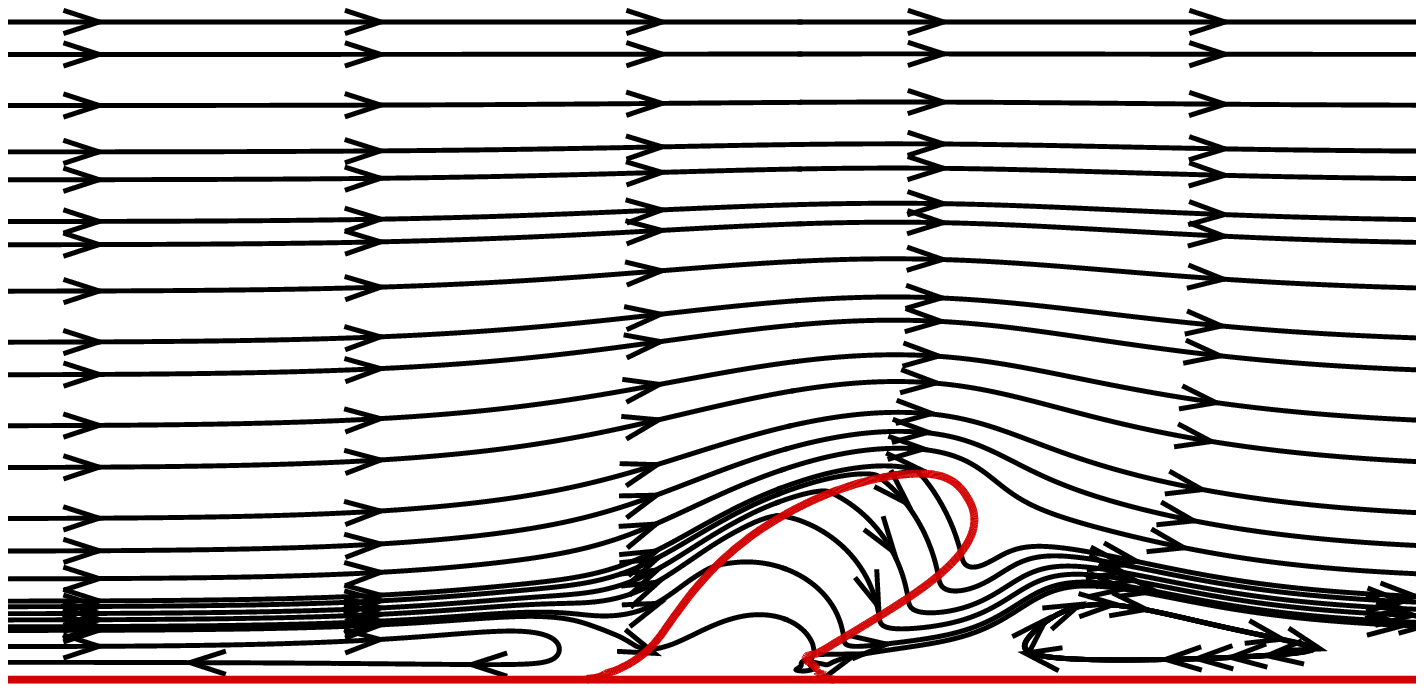}
    \label{fig:subfigure2}}
  \\
  \subfigure[$\Db=50\;\mymum$, $\kodzero=75$]{%
    \includegraphics[bb=125 220 415 615,clip,height=0.25\textheight]{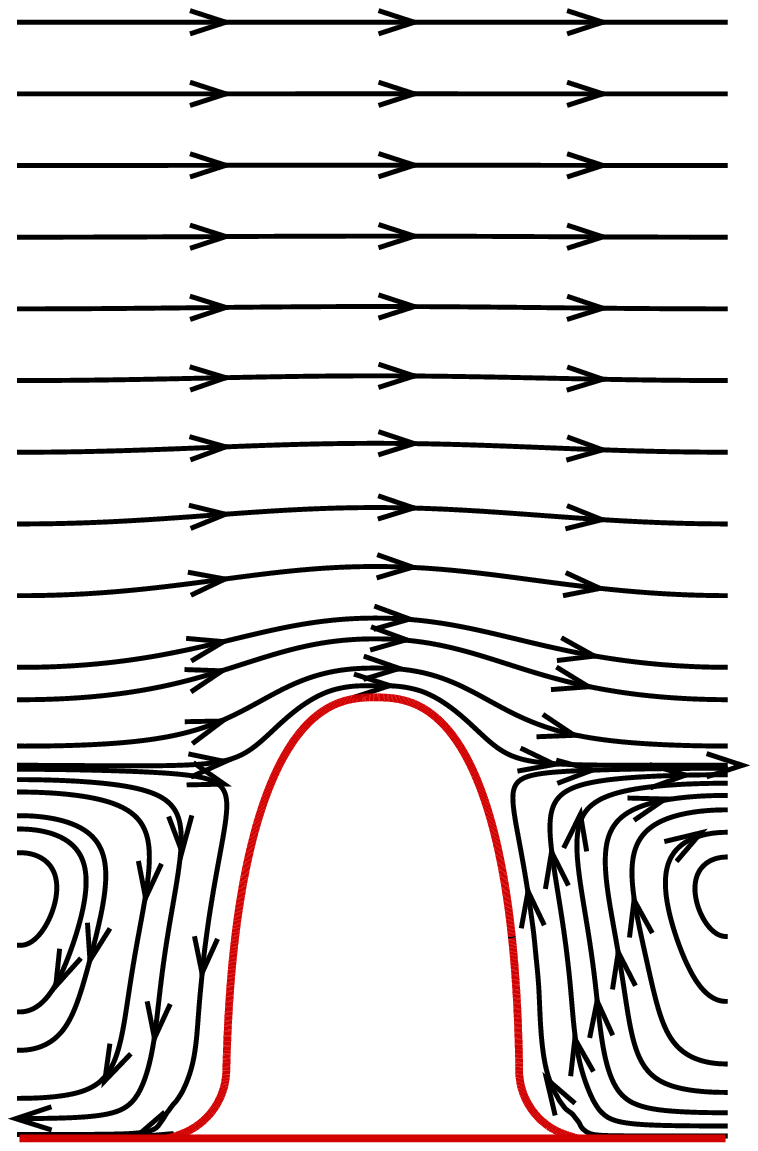}
    \label{fig:subfigure3}}
  \quad
  \subfigure[$\Db=400\;\mymum$, $\kodzero=75$]{%
    \includegraphics[bb=85 160 530 410,clip,height=0.23\textheight]{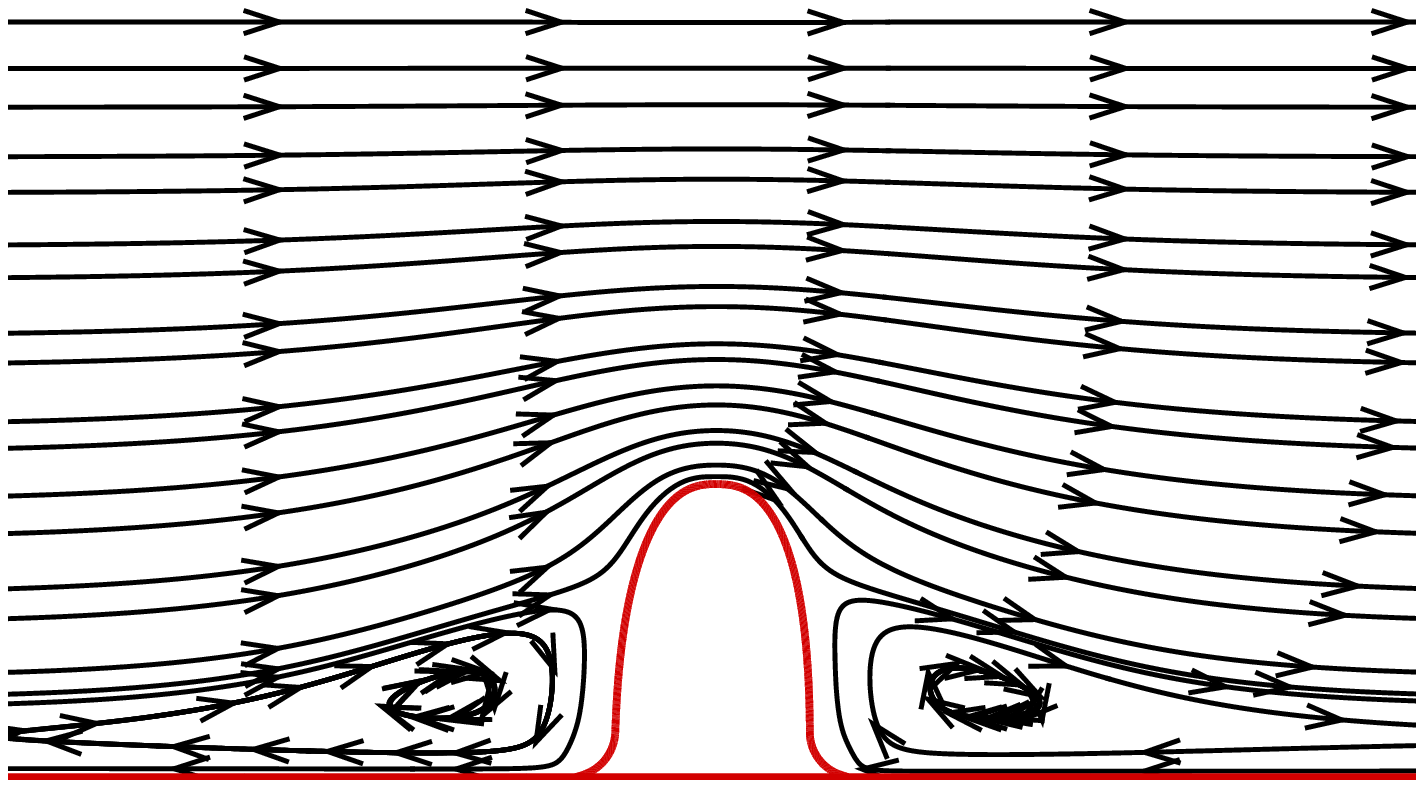}
    \label{fig:subfigure4}}
  \caption{Pathlines for \testcase{Sup75} showing the flow structure and
    biofilm deformation as a function of stiffness and colony
    separation.}
  \label{Fig12b-pathline_sup75}
\end{figure}

More insight into the causes and extent of fluid-induced deformation can
be derived by visualizing the flow structure using path-lines or fluid
particle trajectories.  Fig.~\ref{Fig12b-pathline_sup75} depicts four
path-line plots for the \testcase{Sup75} case, corresponding to biofilms
that are weak/stiff (with $\kodzero=0.75$, 75) and spaced closely/widely
($\Db=50$, 400).  For the stiffer biofilm, the flow over the
widest-spaced colony in Fig.~\ref{Fig12b-pathline_sup75}(d) clearly
exhibits flow separation both upstream and downstream of the colony.  As
the spacing between the stiff biofilm colonies is reduced, the
up/downstream eddies merge to form a single large eddy as pictured in
Fig.~\ref{Fig12b-pathline_sup75}(c), which is similar to what has been
observed in numerical simulations of rigid biofilms at high shear
rates~\cite{sudarsan2005,xu2008computational}.

In the case of a weak biofilm, the eddy structure and dynamics are more
interesting.  In narrow-spaced colonies ($\Db=50$) the deflection of the
weak biofilm causes the single eddy located between the neighbouring
colonies to become distorted and ``climb'' the upstream face as seen in
Fig.~\ref{Fig12b-pathline_sup75}(a).  For more widely-spaced colonies
($\Db=400$) with two distinct eddies present, the increased deformation
in the weak biofilm causes the upstream eddy to shrink while the
downstream eddy grows, as shown in Fig.~\ref{Fig12b-pathline_sup75}(b).

We remark that for the weak biofilms in
Figs.~\ref{Fig12b-pathline_sup75}(a,b), path-lines clearly traverse the
interior of the colony which corresponds to a very slow flow (slower by
several orders of magnitude than the flow outside the biofilm region).
The spring network making up our simulated biofilm therefore behaves
like a porous medium, which has a very small
permeability~\cite{stockie-2009} that can be attributed to small volume
conservation errors that are well-studied in the context of the IB
method~\cite{griffith-2012}.  The porous flow is so small that it has
minimal effect on the biofilm deformation or the flow, but it cannot be
ignored when computing shear stress along the interface using the FS
method described earlier in Section~\ref{comp-force}. The porosity is
particularly important along portions of the biofilm--fluid interface
that experience flow separation adjacent to recirculating eddies on the
upstream and downstream faces of the colony. The effect of the weak
porous flow inside the biofilm colony is accounted for in the FS method
by including its contribution to the jump term in
Eq.~\ref{eqn-shearstress}. This requires performing the filtering
operation twice: once to select IB points suitable for calculating the
fluid stress component outside the biofilm colony, and second time for
the stress inside. As mentioned earlier in Section~\ref{comp-force},
different IB points are selected for the stress calculations inside and
outside the biofilm colony, necessitating an interpolation between the
two sets of IB points. This is in contrast with the stress calculations
of WFG~\cite{williams2009}, where the interior flow was assumed to be
weak and its contribution to the interfacial stress was neglected. This
assumption did not affect their calculated interfacial shear stress,
since there was no flow separation at the low Reynolds numbers they
considered.

\begin{figure}[bthp]
  \footnotesize\sffamily
  \centering
  \subfigure[Case \testcase{Semi20}]{%
    \includegraphics[width=0.45\textwidth,clip,bb=13 133 596 664 410]{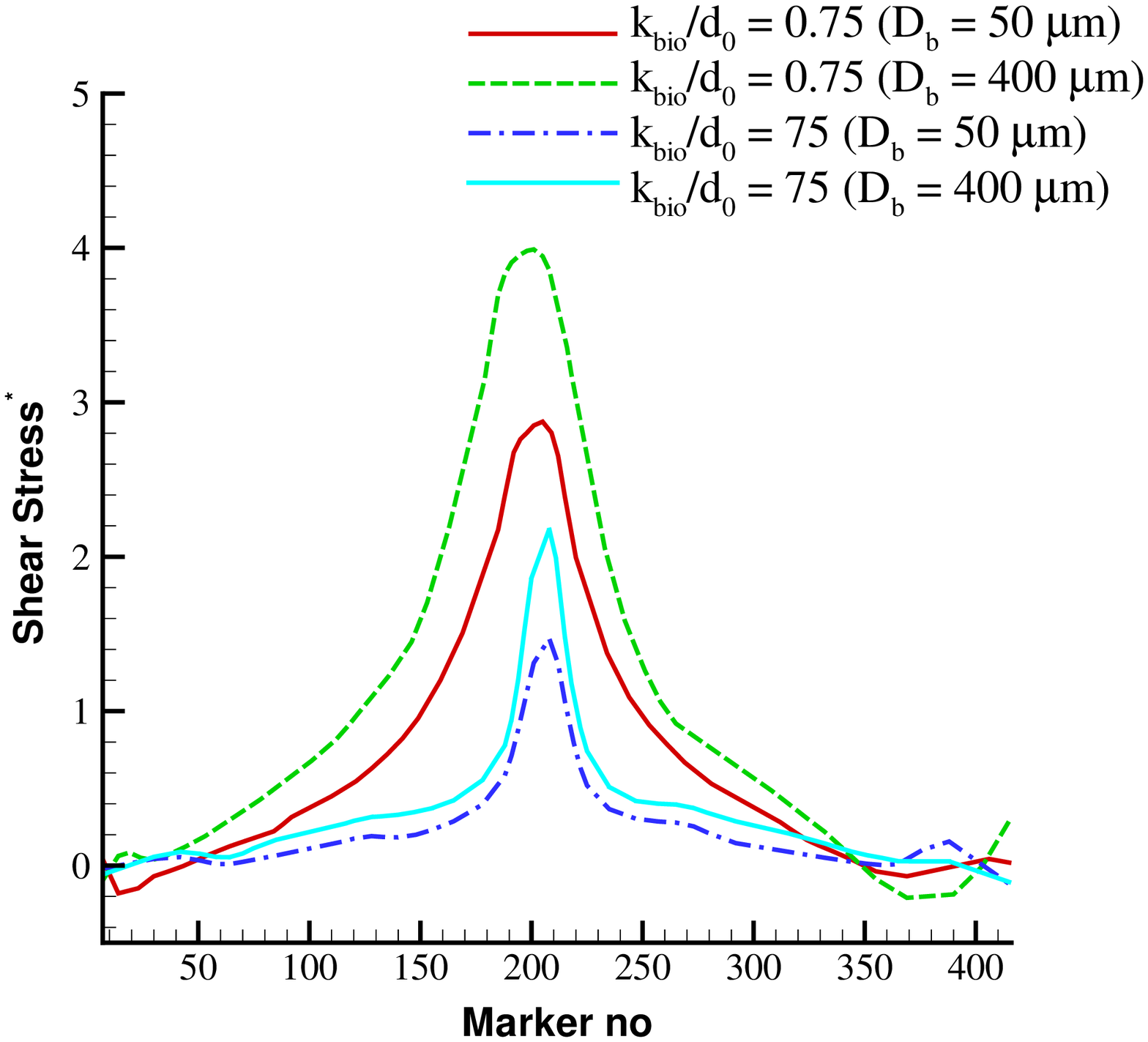}
    \label{fig13:subfigure1}}
  \quad
  \subfigure[Case \testcase{Sup25}]{%
    \includegraphics[width=0.45\textwidth,clip,bb=20 131 598  662]{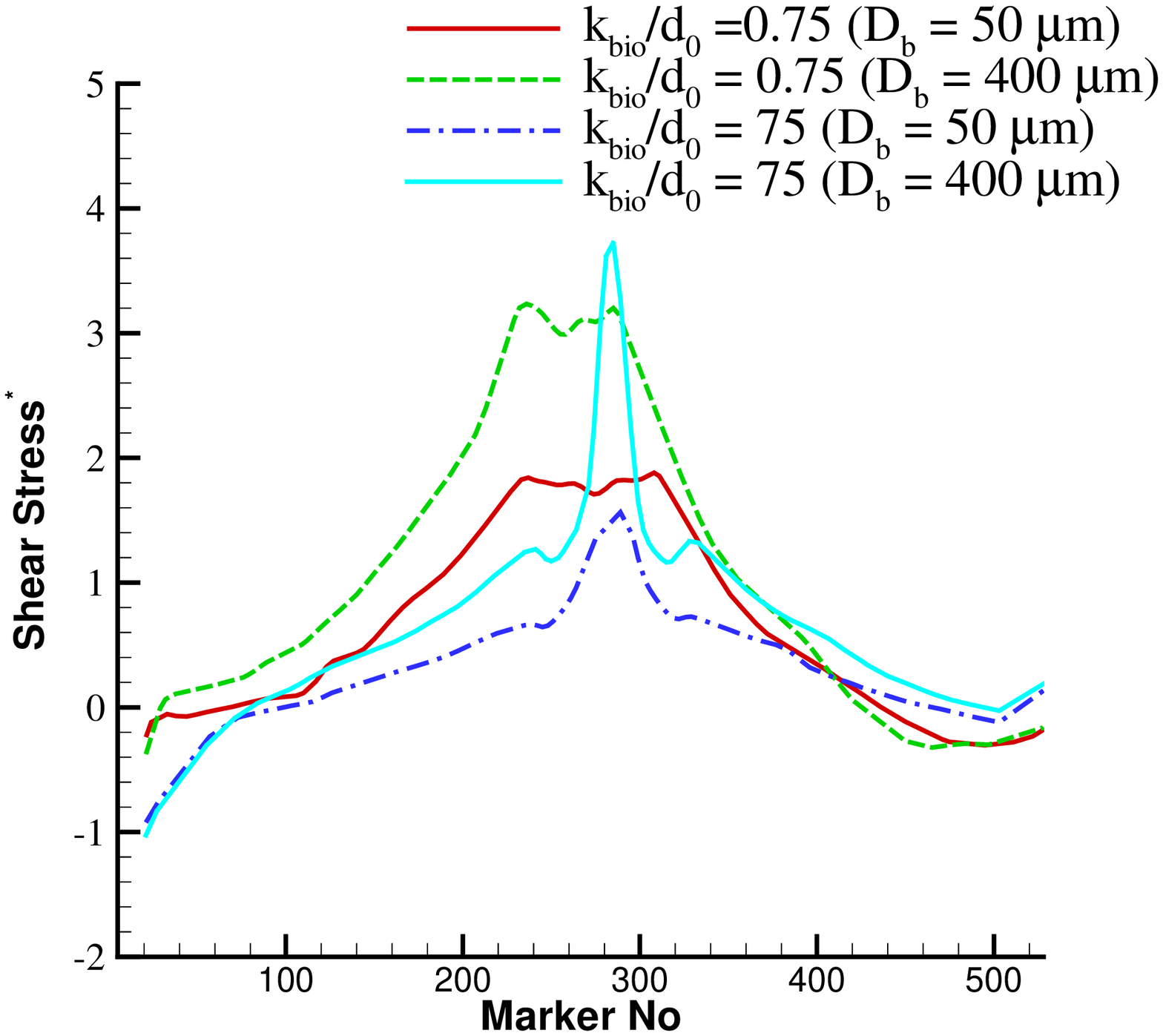}
    \label{fig13:subfigure2}}\\
  \subfigure[Case \testcase{Sup50}]{%
    \includegraphics[width=0.45\textwidth,clip,bb=24 129 597 662]{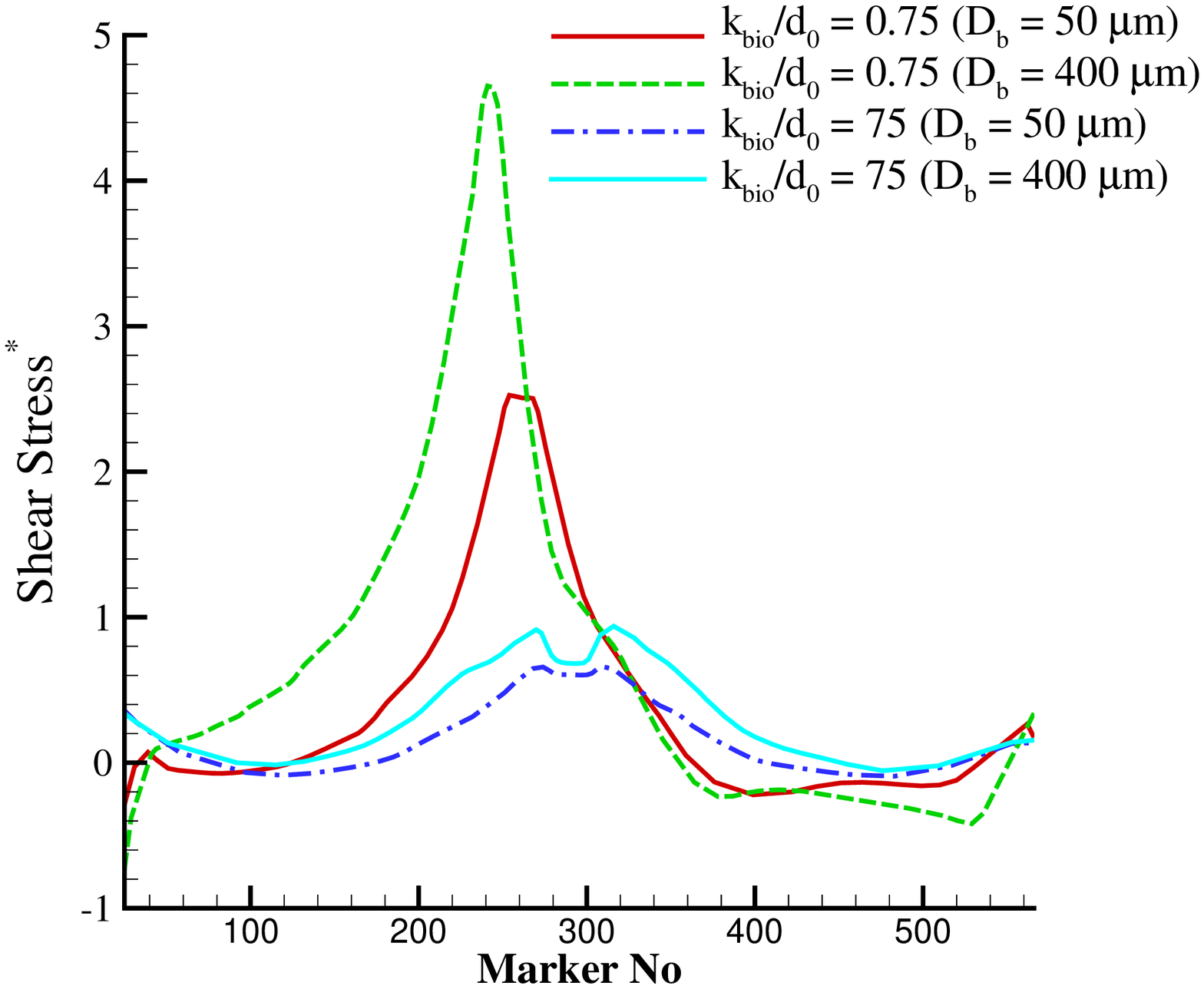}
    \label{fig13:subfigure3}}
  \quad
  \subfigure[Case \testcase{Sup75}]{%
    \includegraphics[width=0.45\textwidth,clip,bb=29 131 585 662]{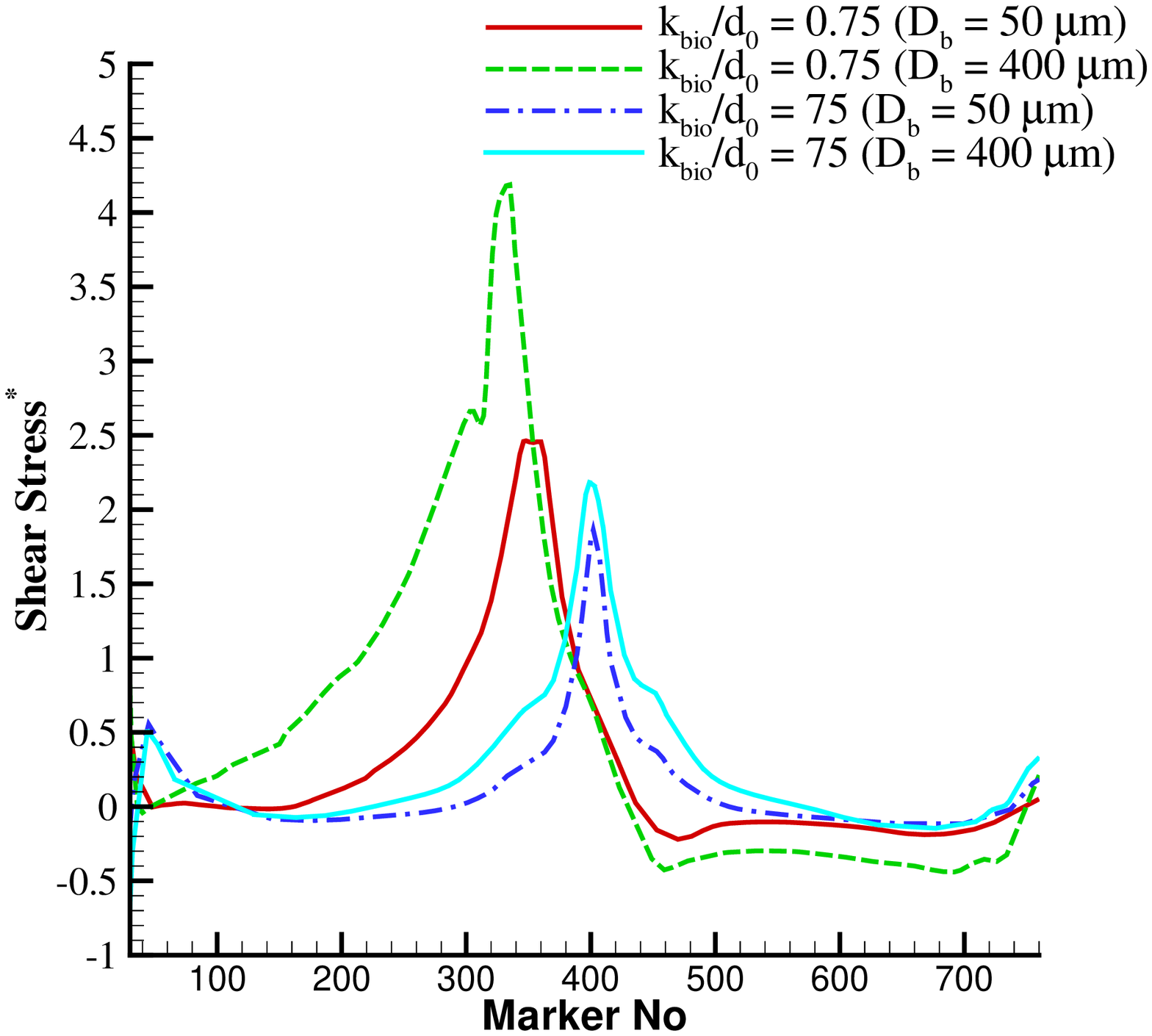}
    \label{fig13:subfigure4}}
  \caption{Dimensionless interfacial shear stress plotted as a function
    of IB point index, numbered from left to right along the
    biofilm-fluid interface.  Results are shown for cases
    \testcase{Semi20}, \testcase{Sup25}, \testcase{Sup50} and
    \testcase{Sup75}, with two values of colony spacing ($\Db = 50$,
    $400\;\mymum$) and two stiffness values ($\kodzero=0.75$, 75).}
  \label{Fig13-intstress_vs_db_bothksp}
\end{figure}

To conclude this section, we investigate the interfacial
shear stress distribution along the biofilm-fluid interface, which is
depicted in Fig.~\ref{Fig13-intstress_vs_db_bothksp} for weak/stiff
biofilms ($\kodzero = 0.75$, 75) that are closely/widely spaced ($\Db=
50$ and $400\;\mymum$).
The shear stress is plotted against the IB point index numbered from
left to right along the interface (note that the total number of IB
points increases with the size of the colony).  In all cases, the shear
stress is non-dimensionalized using the reference value $\mu G$, which
corresponds to the steady uniform shear flow that would occur in the
absence of any obstacle.  We adopt a sign convention that assumes stress
is positive when the flow adjacent to the biofilm-fluid interface is in
the same direction as the primary channel flow (i.e., from left to
right).  A zero shear stress indicates a point where flow separates and
a recirculating eddy attaches to the biofilm surface.

For each test case depicted in
Figs.~\ref{Fig13-intstress_vs_db_bothksp}(a)--(d) there are four curves,
which can be separated into two pairs having similar shape:
one pair corresponding to weak biofilms with high shear stress, and the
second to stiff biofilms with low shear stress.  Within each pair of curves,
increasing the colony spacing from 50 to 400\;$\mymum$ causes the
maximum shear stress to increase but leaves the general shape of the stress
curve unchanged; however, there is a slight downstream shift of the
location of the maximum stress for the three super-ellipse cases
\testcase{SupNN}.

In summary, these results indicate that both the magnitude of the shear
stress and its variation along the interface can change significantly
when the biofilm colony deforms.  We will see next that this has
important implications for biofilm detachment.

\subsection{Simulating biofilm detachment using equivalent continuum
  stress}

To demonstrate the effectiveness of our equivalent continuum
stress-based detachment strategy, we now consider a different colony
shape pictured in Fig.~\ref{Fig1-geometry}.  This mushroom-shaped colony
features wide head and base sections connected by a relatively narrow
stem.  The reason for using this shape is two-fold: first, these long
and thin structures are more realistic especially during the advanced
stages of deformation just before a detachment event; and second, that
it allows comparison with other IB studies of biofilm deformation and
detachment that use similar mushroom shapes~\cite{alpkvist2007,
  hammond2012}.  Based on solid mechanics principles and experimental
evidence, we know that when such an elongated colony is subjected to a
sufficiently strong shear flow, it will rupture at a location inside the
narrow stem region where the cross-sectional area is smallest.
Consequently, this scenario serves as a simple validation of our
detachment algorithm.

The precise mushroom shape used in our simulations is extracted
digitally from \cite[Fig.~3a]{alpkvist2007} which in turn was taken from
experimental images.  We use the following values of the physical and
geometric parameters: $H=300\;\mymum$, $\Db=240\;\mymum$,
$h_x=h_y=5\;\mymum$, $\dzero=1.25\;\mymum$, $G=0.625\;\myunit{s^{-1}}$
and $\kodzero=15$.  As before, \DistMesh\ is used to triangulate the
initial biofilm region, yielding a total of 6155 IB nodes connected by
6500 edges (springs).  In the remainder of this section, we present
results using two different detachment strategies: the first based on
spring strain; and the second based on equivalent averaged continuum
stress.  In particular, we aim to identify the drawbacks of the
strain-based approach that in turn highlight the advantages of our new
equivalent averaged continuum stress approach developed in
Section~\ref{model-stress-force}.

\myparagraph{Spring strain-based detachment}

We start by considering a detachment strategy based on spring strain,
$\strain_{\ell m} = d_{\ell m}/\dzero_{\ell m}-1$, where $d_{\ell m}$
represents the length of a spring and $\dzero_{\ell m}$ is the
corresponding unstressed (or resting) length.  The central parameter in
this detachment model is the critical strain, $\straincrit$, beyond
which the spring will break.  Other studies employing a similar
detachment criterion~\cite{alpkvist2007, hammond2012} have used
$\straincrit\equiv 1$, which coincides with detachment occuring when a
spring is stretched to twice its resting length.  The reason for this
choice of critical strain was not justified, even though it is evident
on physical grounds that $\straincrit$ should not be constant but rather
depend on the strength of the biofilm matrix (which in our IB model is
expressed by $\kodzero$).

\begin{figure}[bthp]
  \footnotesize\sffamily
  \centering
  \subfigure[]{%
    \includegraphics[width=0.45\textwidth,clip,bb=50 200 550 590]{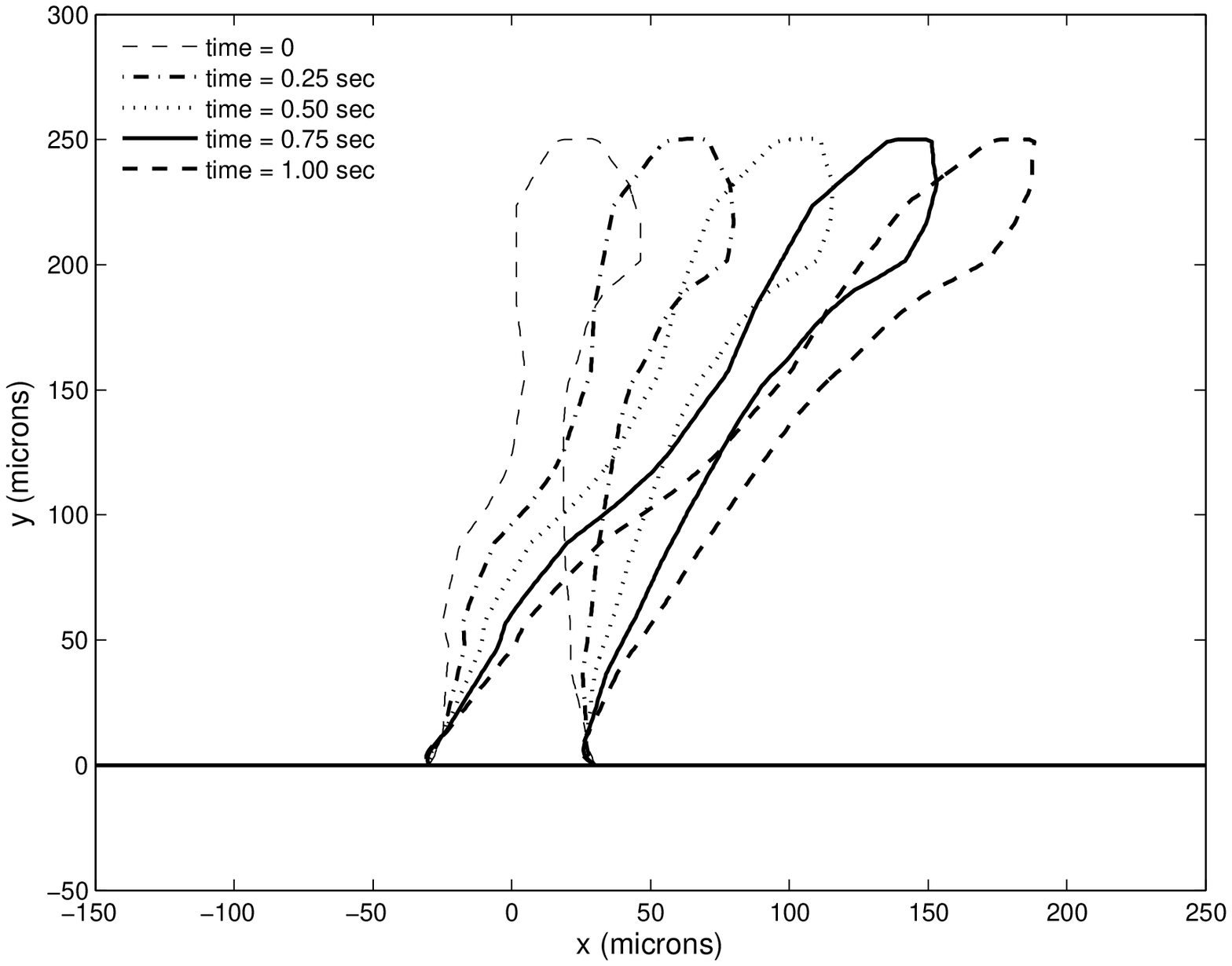}
    \label{fig15:subfigure1}}
  \quad
  \subfigure[]{%
    \includegraphics[width=0.45\textwidth,clip,bb=50 180 570 595]{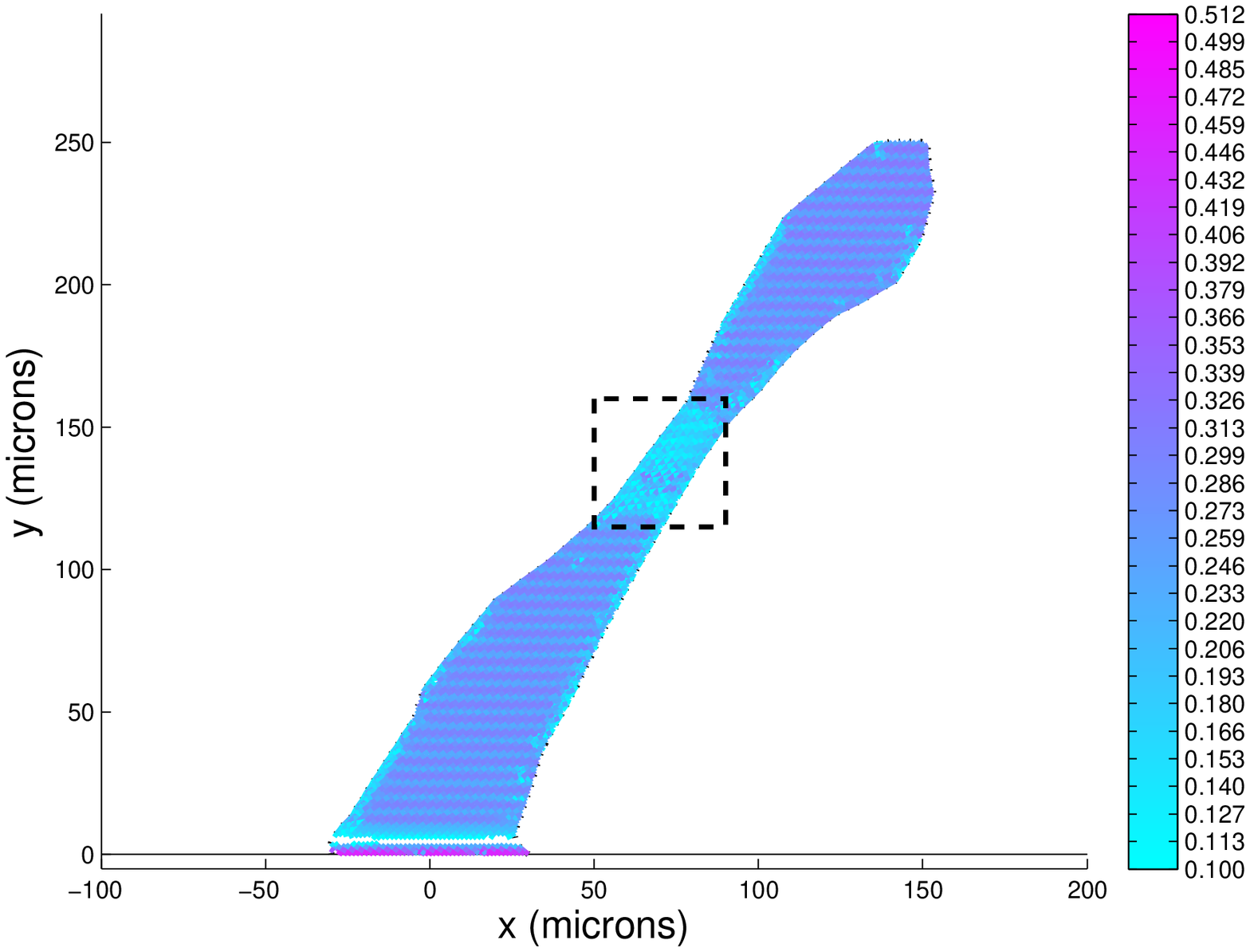}
    \label{fig15:subfigure2}}
  \subfigure[]{%
    \includegraphics[width=0.45\textwidth,clip,bb=50 185 557 601]{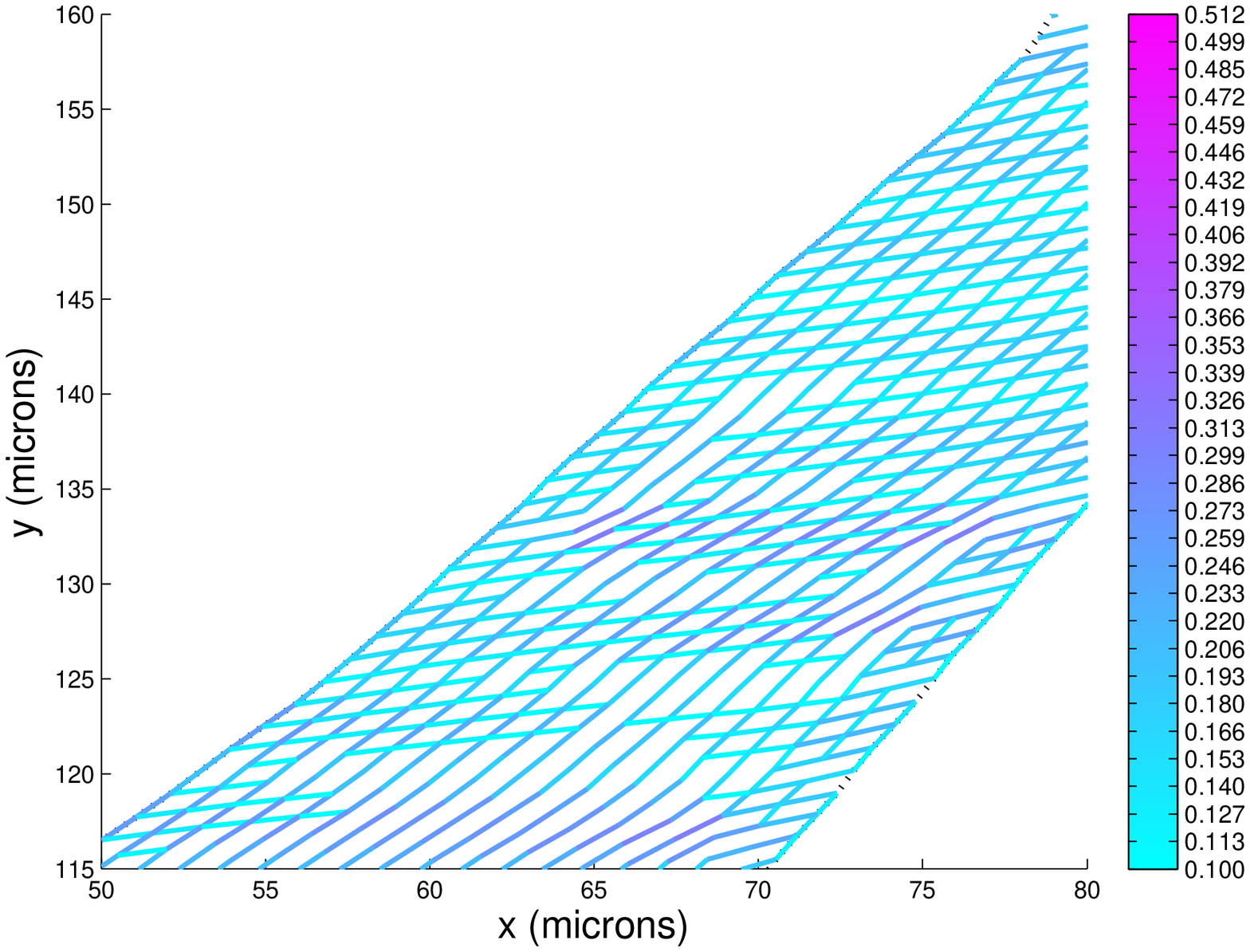}
    \label{fig15:subfigure3}}
  \caption{(a) Deformation of the mushroom-shaped biofilm colony at
    various times in the interval $t\in[0,1]\;\myunit{s}$.  (b) Edges
    colored according to strain at $t=0.75\;\myunit{s}$. (c) zoomed view
    showing edges inside dotted box in (b).}
  \label{Fig15-springstrain_mushroom}
\end{figure}

\begin{figure}[bthp]
  \footnotesize\sffamily
  \centering
  \subfigure[]{%
    \includegraphics[width=0.45\textwidth,clip,bb=50 185 557 601]{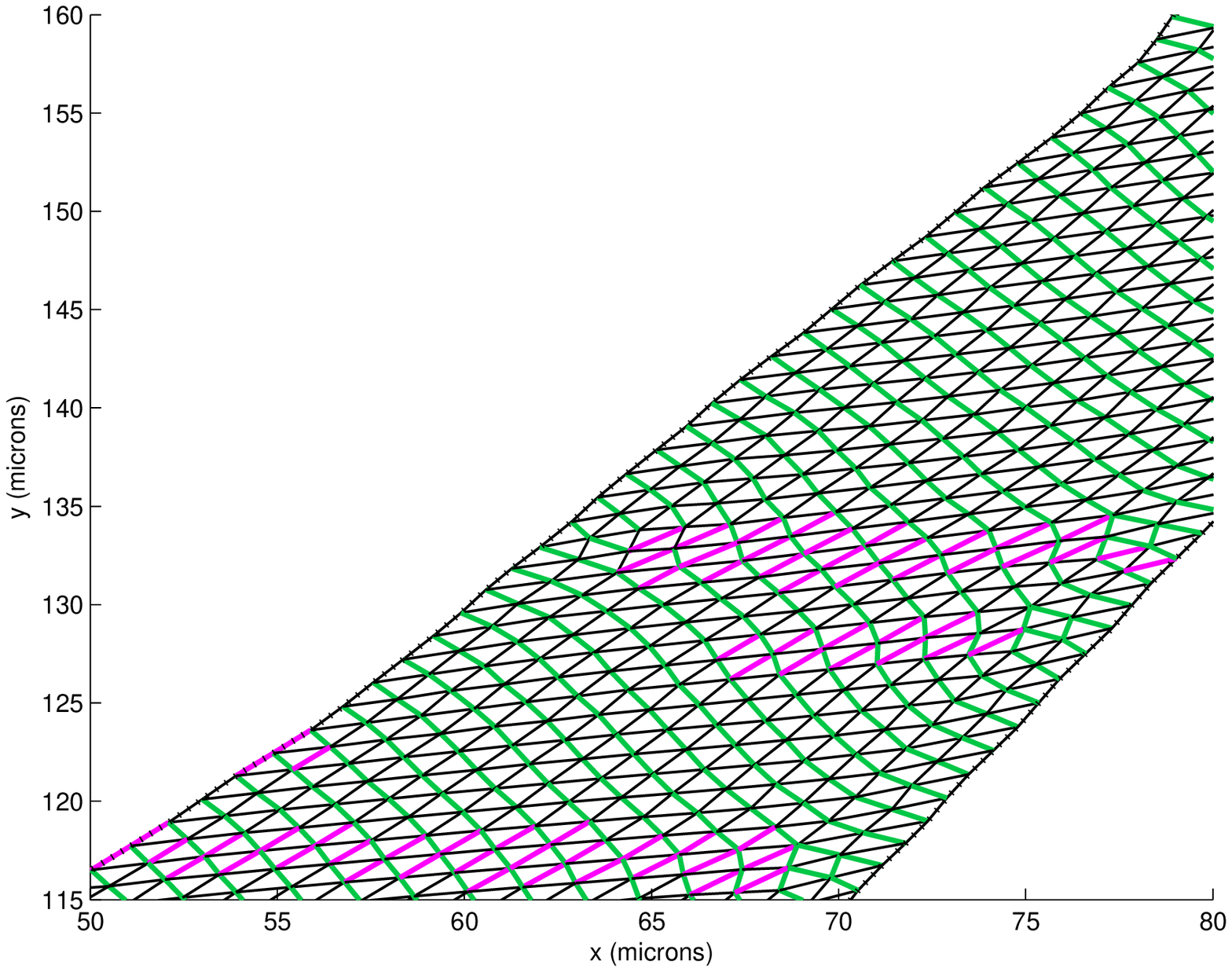}
    \label{fig15new:subfigure1}}
  \quad
  \subfigure[]{%
    \includegraphics[width=0.45\textwidth,clip,bb=50 185 557 601]{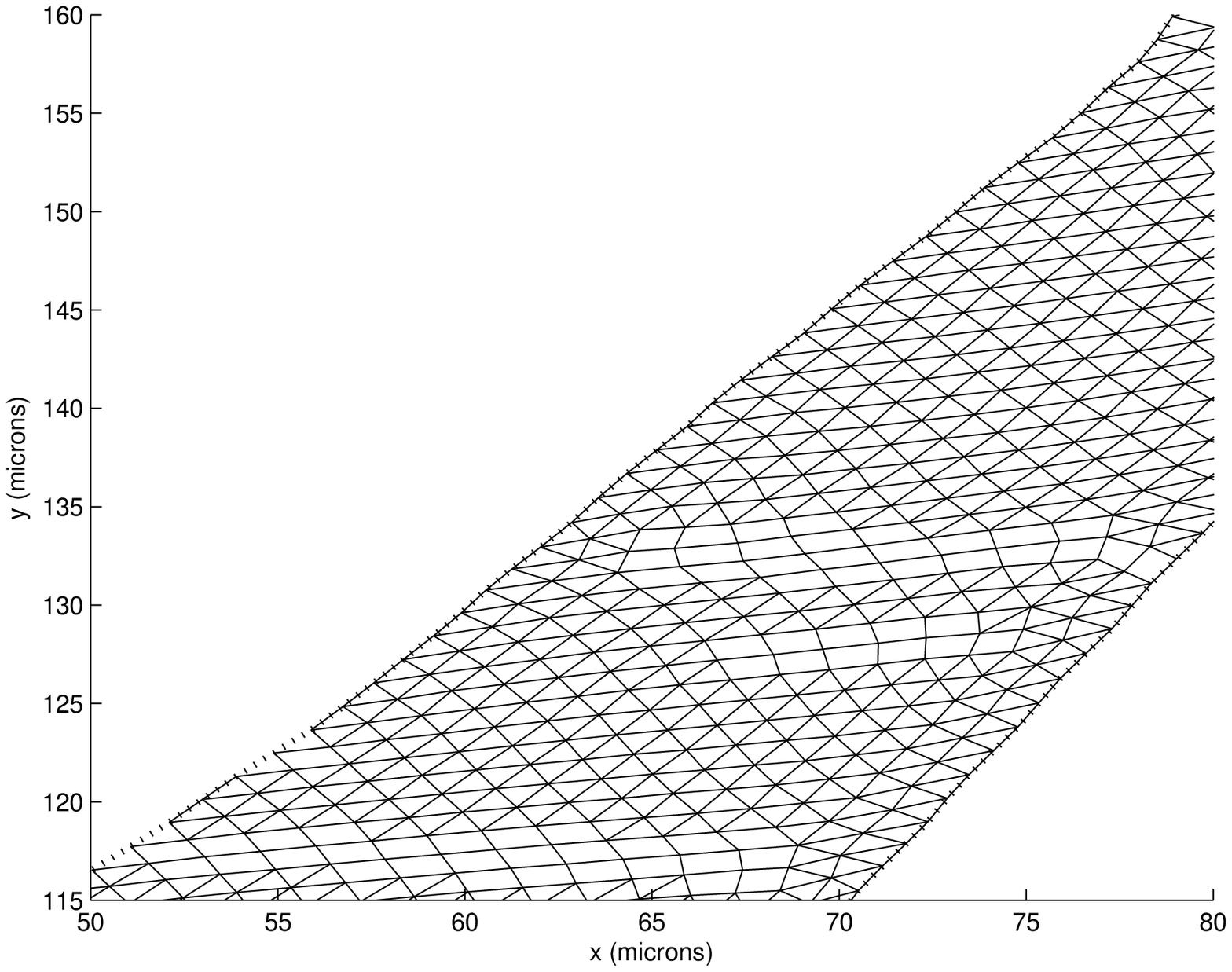}
    \label{fig15new:subfigure2}}
  \caption{(a) Edges in the stem region inside the dotted box in
    Fig.~\ref{fig15:subfigure2}.  Magenta identifies edges with strain
    above 0.25 and green indicates negative strain, while all other
    edges are colored black. (b) Spring connectivity after cutting
    all edges with strain greater than 0.25 at $t=0.75\;\myunit{s}$.}
  \label{Fig15new-springstrain_mushroom}
\end{figure}

As an illustration of strain-based detachment, we simulate flow over the
mushroom-shaped biofilm using the parameters indicated above.
Fig.~\ref{fig15:subfigure1} depicts the resulting biofilm configuration
at four equally-spaced times between $t=0$ and $1\;s$.  Unlike the
\testcase{SupNN} (super-ellipse shaped) test cases, we have not
simulated the mushroom shaped colony for long enough time to reach a
quasi-steady state. Up to time $t=1\;\myunit{s}$, the mushroom shaped
colony exhibits continuous deformation characterized by a near-constant
IB point displacement velocity of approximately
60$\;\myunit{{\mymum}/{s}}$ (not shown in the
figure). Fig.~\ref{fig15:subfigure2} shows the edges in the deformed
triangulation after 0.75 seconds, colored according to the local strain
value. For the sake of clarity, we have only shown the edges with strain
greater than 0.1 (where positive strain corresponds to a spring that is
stretched relative to the resting configuration).  Clearly, the base of
the colony near the substratum experiences the highest strain, with
values near 0.5.  The next highest edge strains (between 0.25 and 0.5)
occur in sizable portions of the head and base as well as a few points
near the midsection of the stem region, which can be seen in the zoomed
view in Fig.~\ref{fig15:subfigure3}.  An alternate view of the stem is
shown in Fig.~\ref{fig15new:subfigure1}, with the edges having negative
strain colored green, while edges with strain greater than 0.25 are
colored magenta (and all other edges shown in black).  The high strain
(magenta) edges are clearly aligned along the long axis of the colony,
whereas those experiencing compression (negative strain, green) are
aligned at an angle of 45--60 degrees to the main colony axis.

We now illustrate a critical drawback of the spring strain-based
detachment methodology that has not received attention in earlier
biofilm IB studies \cite{alpkvist2007, hammond2012}.  We assume that
detachment is initiated after 0.75~s of deformation and apply a critical
strain threshold of $\straincrit=0.25$ that is significantly lower than
the value 1.0 used in these other studies.  Performing a single
detachment step yields the modified spring network shown in
Fig.~\ref{fig15new:subfigure2} (noting that in an actual detachment
scenario, these springs would be severed gradually over time instead of
all at once).  On comparing Figs.~\ref{fig15new:subfigure1} and
Fig.~\ref{fig15new:subfigure2}, we see that the edge connectivity around
many IB nodes in the stem has been altered so that there are now a
significant number of rectangular elements in the place of triangles.
Based on the work of Lloyd et al.~\cite{lloyd2007identification} we know
that two spring networks, one built of triangles and the other with
rectangles, (but otherwise having the same edge length and spring
stiffness) will approximate equivalent elastic continua that have
different Young's modulus.  Therefore, the spring cutting operation we
just performed has effectively introduced an instantaneous change
in the local mechanical stiffness of the biofilm, which is
clearly undesirable and hence is a major disadvantage of the spring
strain-based detachment strategy.

\myparagraph{Equivalent continuum stress-based detachment}

\begin{figure}[bthp]
  \footnotesize\sffamily
  \centering
   \subfigure[time $= 0.375\;\myunit{s}$]{%
    \includegraphics[width=0.40\textwidth,clip,bb=88 170 475 591]{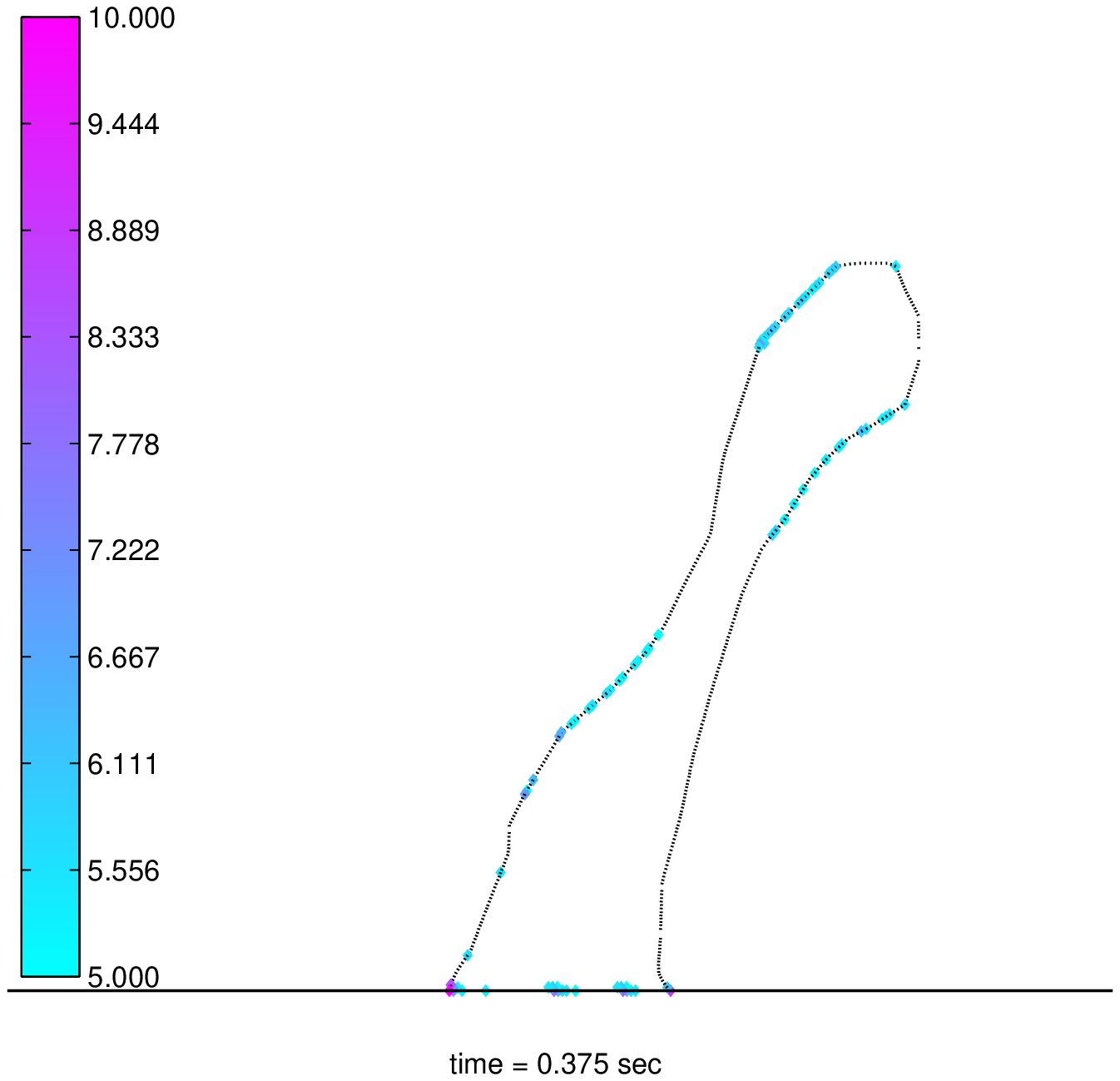}
    \label{fig16:subfigure1}}
  \quad
  \subfigure[time $=0.45\;\myunit{s}$]{%
    \includegraphics[width=0.40\textwidth,clip,bb=88 170 475 591]{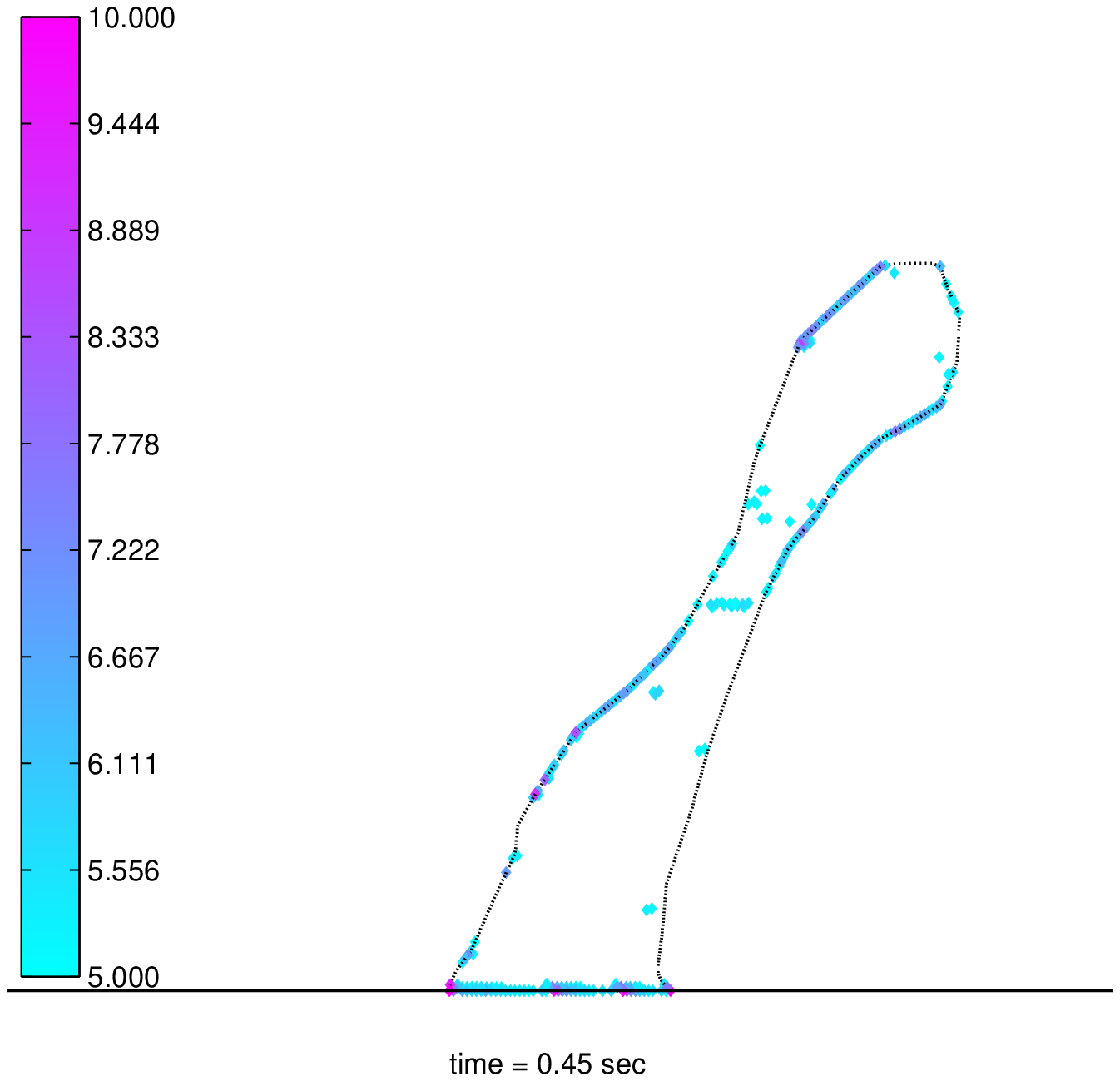}
    \label{fig16:subfigure2}}
  \quad
  \subfigure[time $= 0.50\;\myunit{s}$]{%
    \includegraphics[width=0.40\textwidth,clip,bb=88 170 475 591]{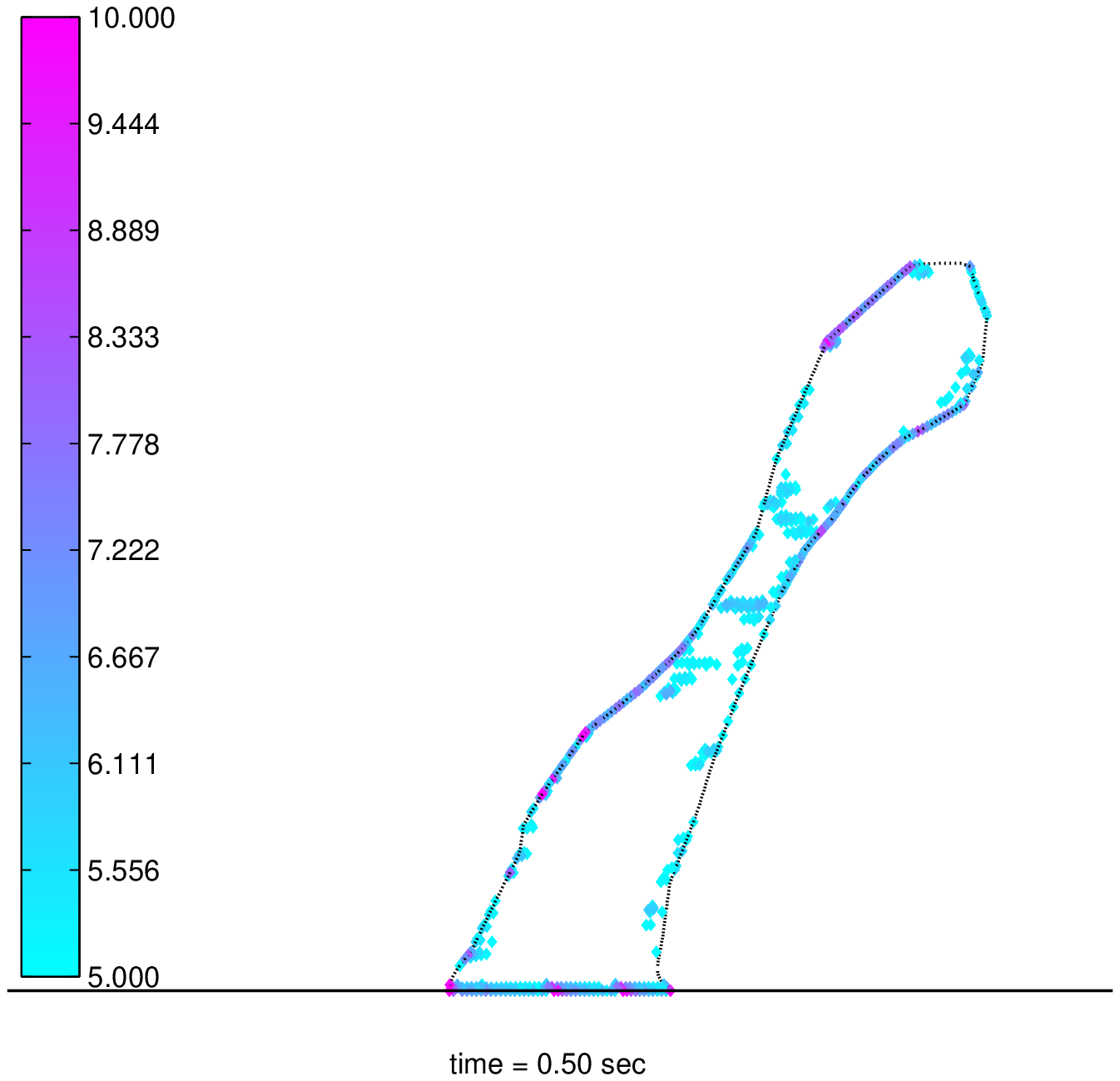}
    \label{fig16:subfigure3}}
   \quad
  \subfigure[time $= 0.60\;\myunit{s}$]{%
    \includegraphics[width=0.40\textwidth,clip,bb=88 170 480 591]{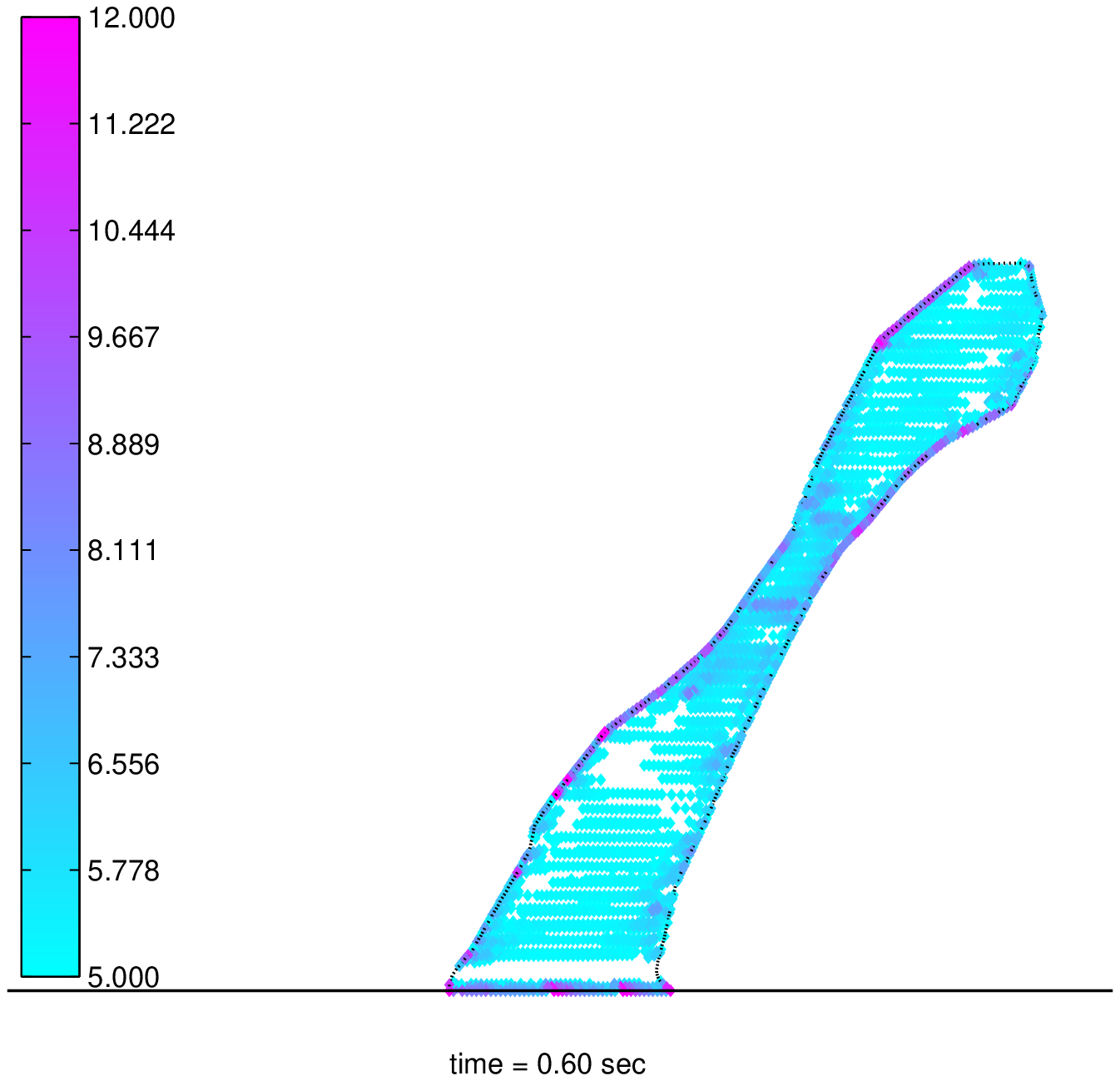}
    \label{fig16:subfigure4}}
  \caption{Von~Mises stress inside the deformed biofilm at a sequence of
    times. For purpose of clarity, only points where the stress exceeds
    $5\myunit{dyne/cm^2}$ are colored.}
  \label{Fig16-vonmises_mushroom}
\end{figure}
We next apply the equivalent continuum stress-based detachment strategy
to the same problem. The stress tensor components are computed using
Eq.~\eqref{eq-avgeqstress_final} after which the von~Mises yield stress
is computed at each IB node using Eq.~\eqref{eq-vonmises}.
Fig.~\ref{Fig16-vonmises_mushroom} displays the von~Mises stress value
inside the biofilm region at four time instants during its deformation,
where we only color those points with stress above the threshold
$5~\myunit{dyne/cm^2}$.  The von~Mises stress has relatively low values
throughout most of the colony except in three areas: the stem region,
near the base where the colony attaches to the substratum, and in
portions of the biofilm-fluid interface that are subject to large fluid
shear.  The von~Mises stress is discontinuous as well as noisy, which is
consistent with other numerical simulations of microstructural stress
inside granular materials~\cite{balevivcius2011investigation,
  fortin2003construction}. However, we emphasize that this behaviour is
in stark contrast with the apparently smooth von~Mises stress field
obtained by Towler et al.~\cite{towler2007} for a simpler biofilm shape
using a finite element simulation.

\begin{figure}[bthp]
  \footnotesize\sffamily
  \centering
   \subfigure[]{%
    \includegraphics[width=0.47\textwidth,clip,bb=49 195 550 583]{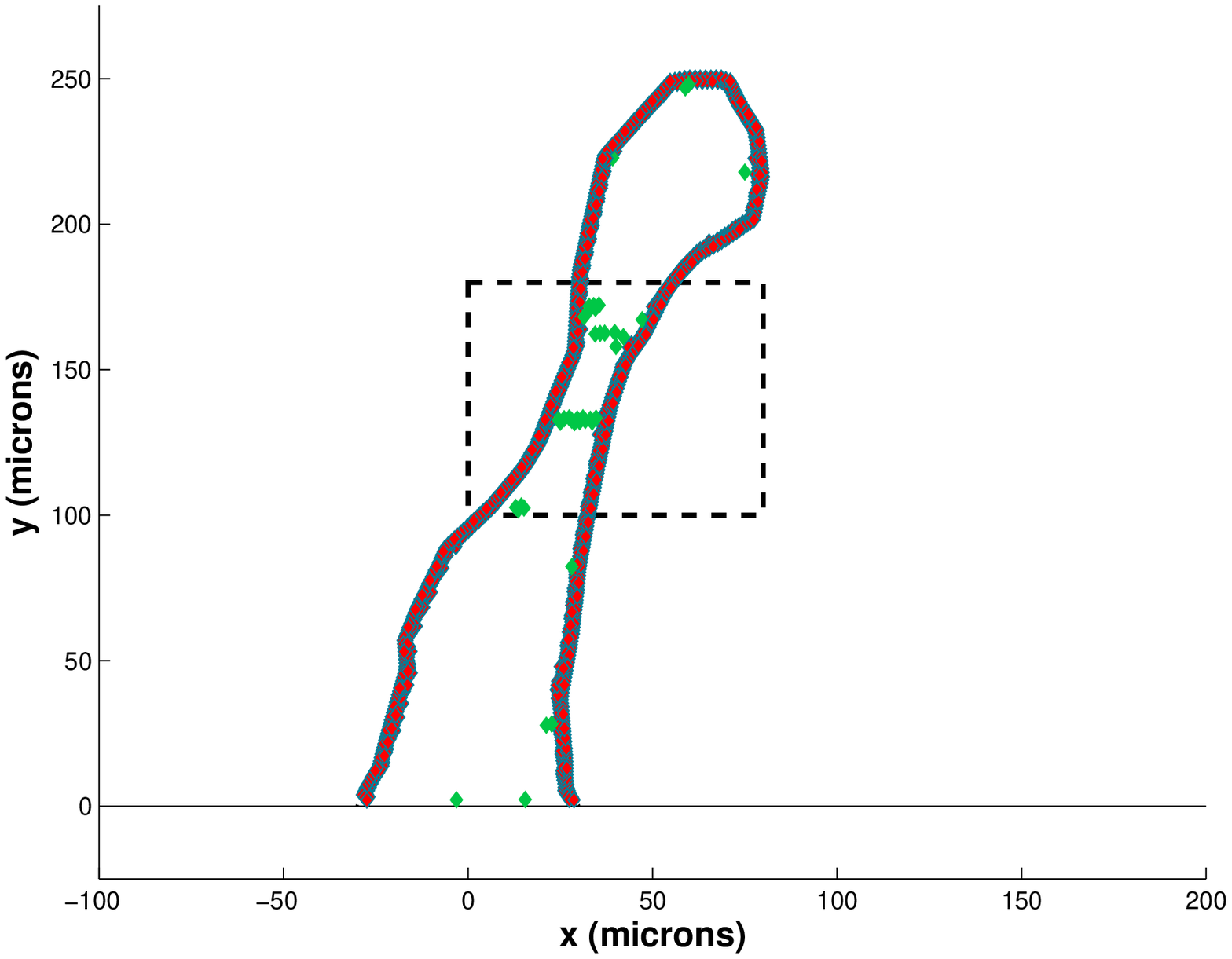}
    \label{fig17:subfigure1}}
  \quad
  \subfigure[]{%
    \includegraphics[width=0.47\textwidth,clip,bb=49 195 547 589]{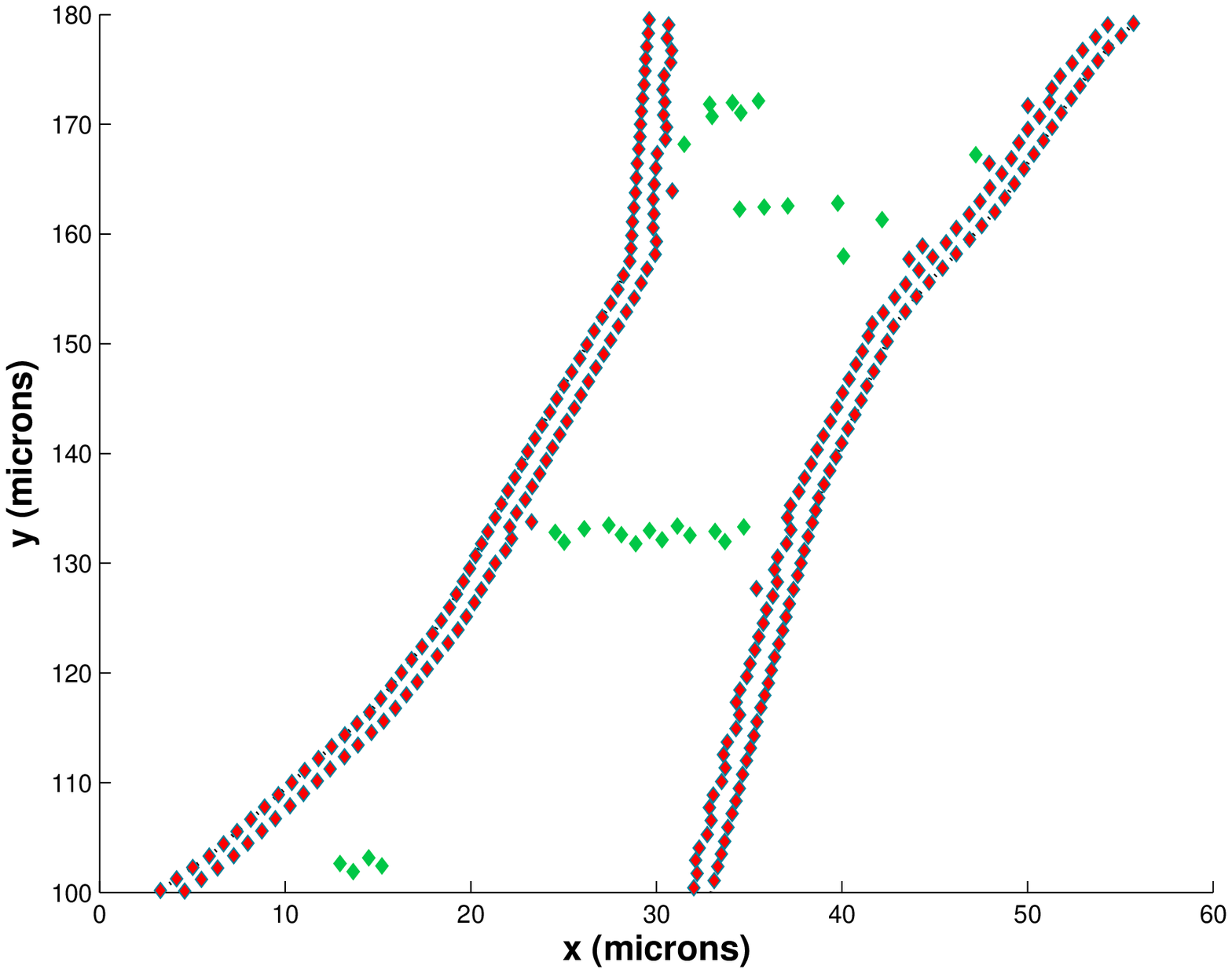}
    \label{fig17:subfigure2}}
  \caption{Points where detachment occurs in the biofilm colony at time
    $t=0.25\;\myunit{s}$ (zoomed-in view on the right).  Detachment
    parameters are $\deltasub=\deltaext=2$, $\stressAdh=10$,
    $\stressCohInt=2.5$ and $\stressCohExt=0.1$.}
  \label{Fig17-detachment-vonmises}
\end{figure}

We make no attempt here to draw any explicit correspondence between
results from the strain- and stress-based detachment strategies because
the spring strain-based threshold parameter $\straincrit$ cannot be
translated into a von~Mises yield stress value. However, we can still
compare the two by simulating detachment for the equivalent continuum
stress methodology using the same parameters.  This time, we initiate
detachment after $0.25\;\myunit{s}$ of deformation and choose the two
distance threshold parameters $\deltasub=\deltaext=2\;\mymum$ (refer to
Section~\ref{algo-detachment}).  This divides the mushroom shaped colony
into three zones: in zone~1 near the substratum, the adhesive strength
is $\stressAdh=10~\myunit{dyne/cm^2}$; for zone~2 near the biofilm-fluid
interface, the interfacial cohesive strength is
$\stressCohExt=0.1~\myunit{dyne/cm^2}$; and for zone~3 in the interior,
the cohesive strength is $\stressCohInt=2.5~\myunit{dyne/cm^2}$.  Based
on these parameter values, Fig.~\ref{Fig17-detachment-vonmises} depicts
those IB points that will be severed according to the local value of
von~Mises stress, with the red points indicating detachment in zone~2,
while the green points corresponding to detachment in zones~1
and~3. From this figure it is clear that two regions within the colony
neck experience complete rupture, which is in agreement with what one
would expect from solid mechanics.  Furthermore, the detachment
criterion is active along the entire biofilm-fluid interface, which
corresponds to an erosion process.

In summary, our new detachment strategy based on equivalent continuum
stress provides an unambiguous method for performing biofilm detachment
that is also consistent with methods employed by other continuum biofilm
models.  The detachment of any IB point proceeds by cutting of all
springs attached to it so that the topology of the spring network
remains triangular, thereby avoiding a major disadvantage of the
spring strain-based detachment strategy. However, this approach does
introduce some additional computational work in each time step for
evaluating the equivalent continuum stress and distance functions at
each IB node, not to mention maintaining all of the relevant data
structures.

\subsection{Biofilm deformation and internal stress}
\label{biofilm_internalstress}

This section presents a final series of simulations that
study the biofilm stress distribution in more detail and draw
specific conclusions regarding the various modes of detachment
(sloughing or erosion).  We also provide evidence that
questions the validity of another class of detachment models based on a
detachment speed function.

\begin{figure}[bthp]
  \footnotesize\sffamily
  \centering
  \begin{tabular}{cc}
    \includegraphics[width=0.45\textwidth,clip,bb=50 0 430 350] {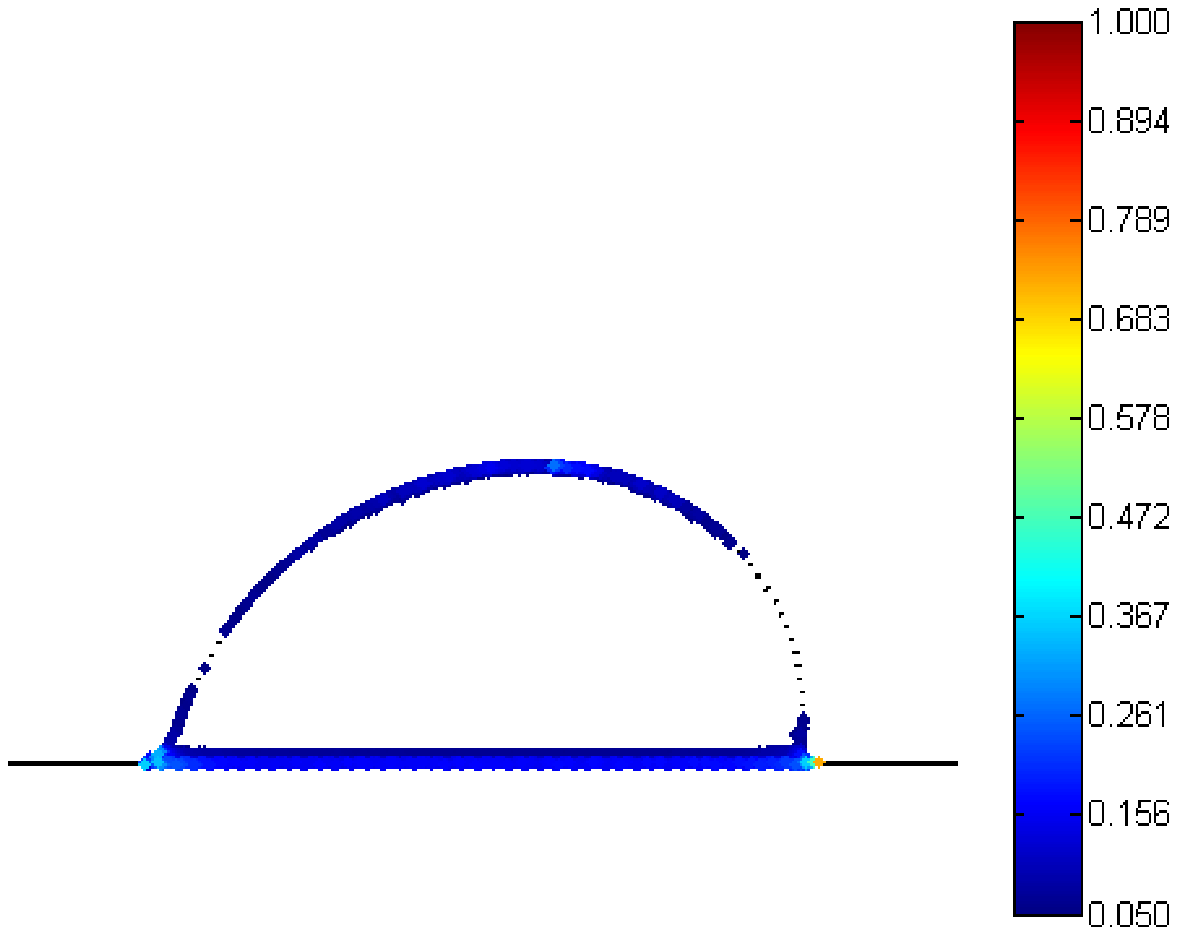} &
    \includegraphics[width=0.45\textwidth,clip,bb=50 0 430 350 ] {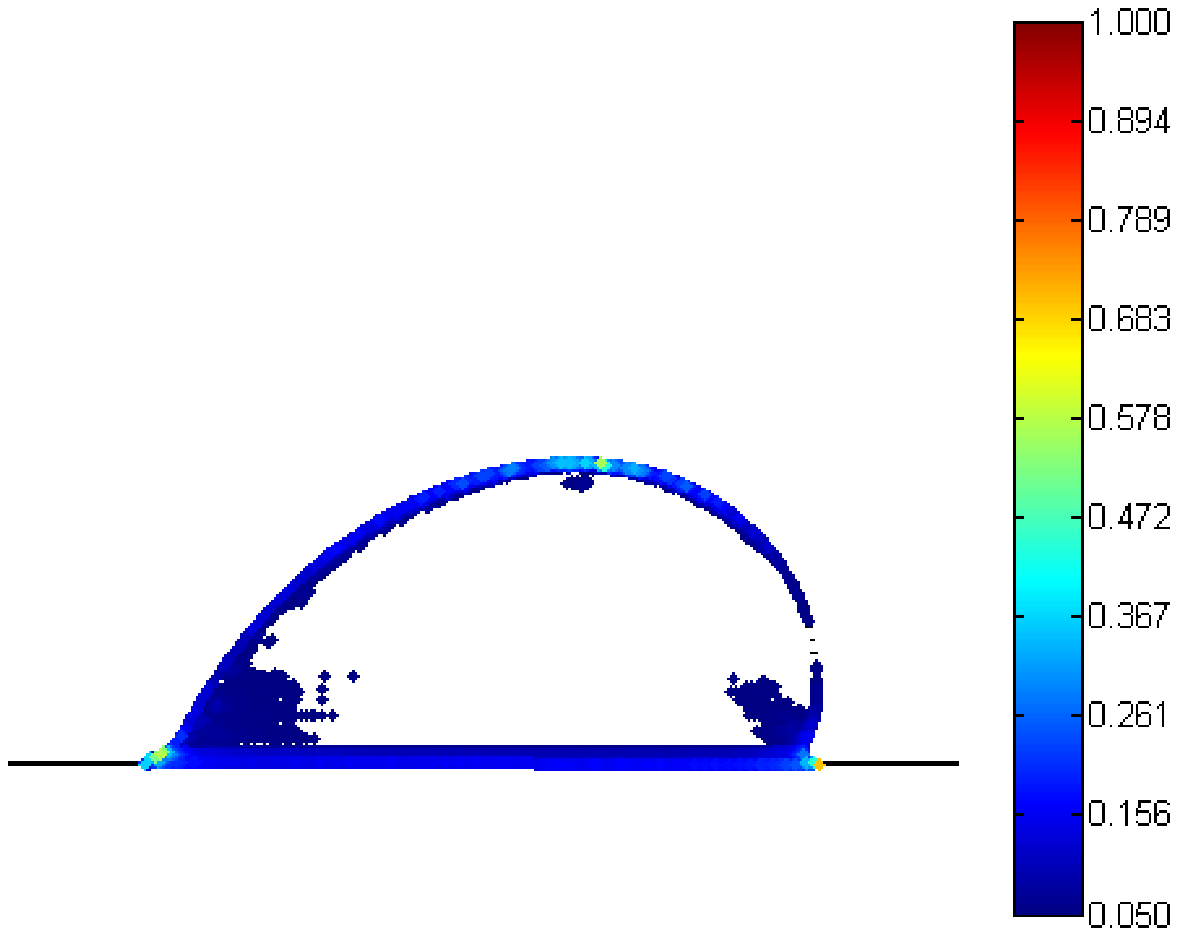}
    \\
    (a) \testcase{Semi20}, $\Db=50$ & (b) \testcase{Semi20}, $\Db=400$ \\
    \includegraphics[width=0.45\textwidth,clip,bb=50 0 430 350] {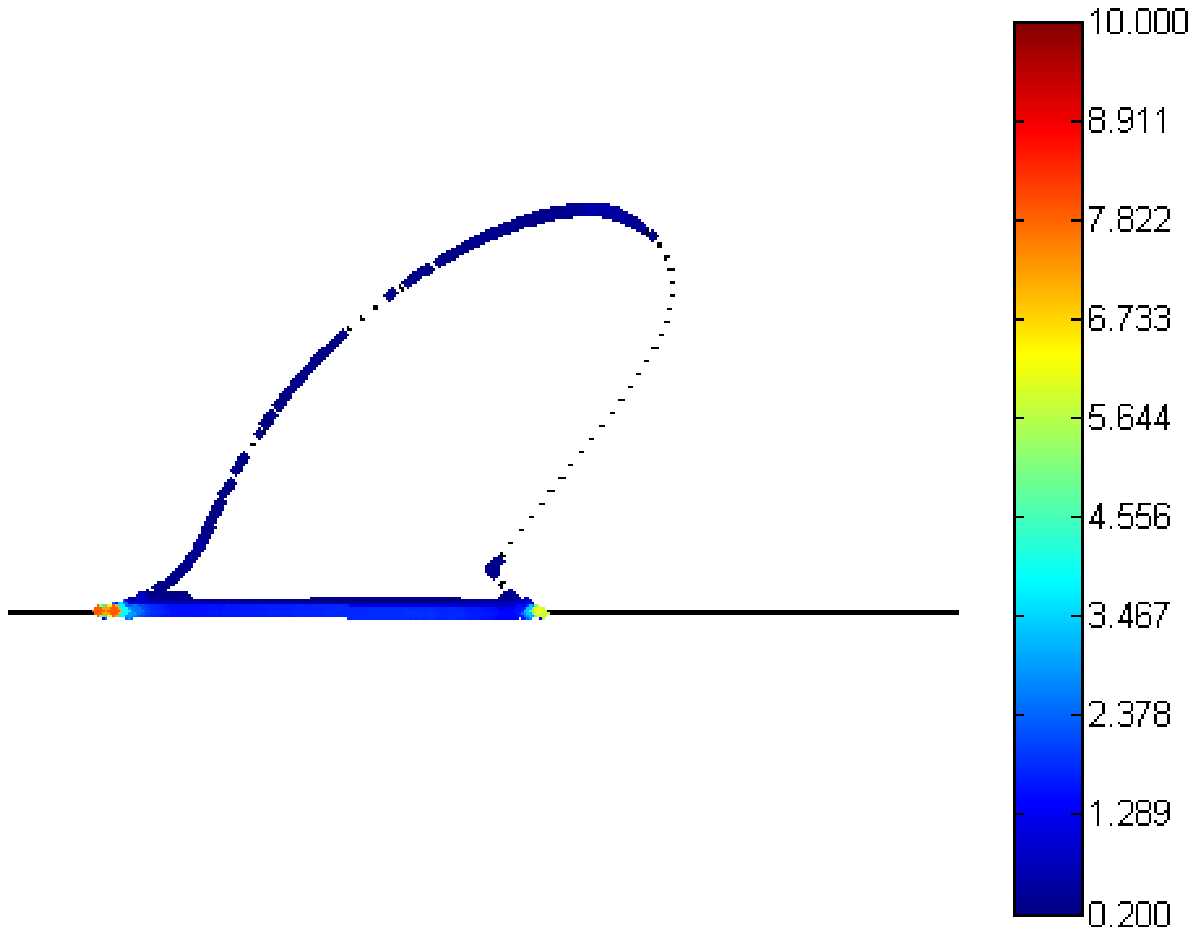} &
    \includegraphics[width=0.45\textwidth,clip,bb=50 0 430 350] {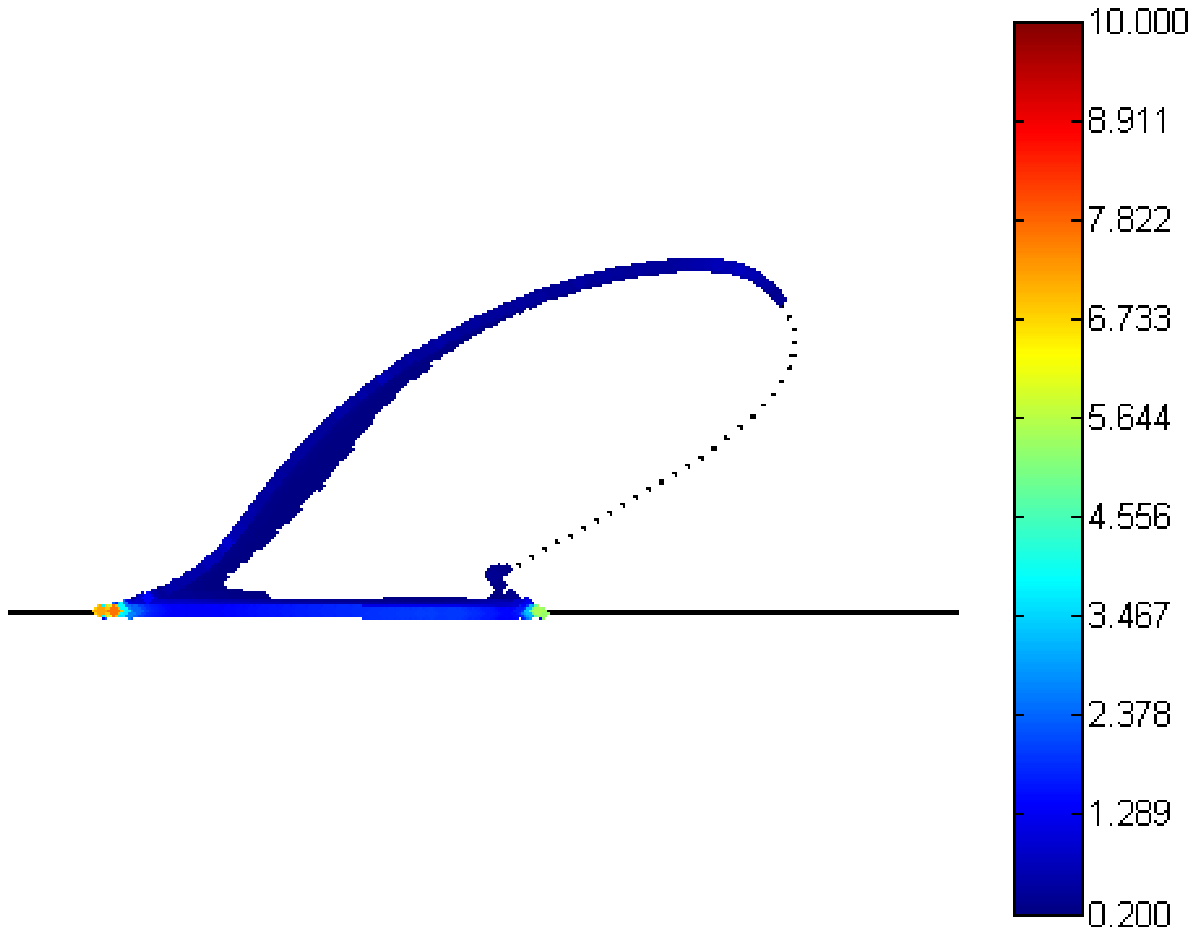}
    \\
    (c) \testcase{Sup75}, $\Db=50$ & (d) \testcase{Sup75}, $\Db=400$
  \end{tabular}
  \caption{Von~Mises yield stress at steady state for cases
    \testcase{Semi20} and \testcase{Sup75}, with $\kodzero = 0.75$ and
    $\Db = 50$ and $400\;\mymum$.  For ease of visualization, only
    IB points with stress above a threshold of $0.050 ~\myunit{dyne/cm^2}$ in
    (a) and (b) and $0.20 ~\myunit{dyne/cm^2}$ in (c) and (d) are shown.}
  \label{Fig18-vonstress_vs_db}
\end{figure}

The steady-state von~Mises stress for the \testcase{Semi20} and
\testcase{Sup75} simulations from the previous section are depicted in
Fig.~\ref{Fig18-vonstress_vs_db} for two values of biofilm spacing, $\Db
= 50$ and $400\;\mymum$. In all cases, the stress is lowest in the
interior of the biofilm and largest along the biofilm-fluid and
biofilm-wall interfaces, with the absolute maxima occurring on the wall
near the leading and trailing corners.  We also observe that the
proportion of the biofilm experiencing high von~Mises stresses increases
as the spacing parameter $\Db$ increases, which is consistent with the
results from Section~\ref{sec:flow-structure}. Finally, the stresses
experienced in a long, thin colony such as \testcase{Sup75} are
significantly larger (note the increase in the colormap scale by a
factor of 10 for plots in the bottom row in
Fig.~\ref{Fig18-vonstress_vs_db}).

\begin{figure}[bthp]
  \footnotesize\sffamily
  \centering
  \subfigure[\testcase{Semi20}, biofilm-wall interface]{
    \includegraphics[width=0.45\textwidth,clip,bb=53 146 559 653] {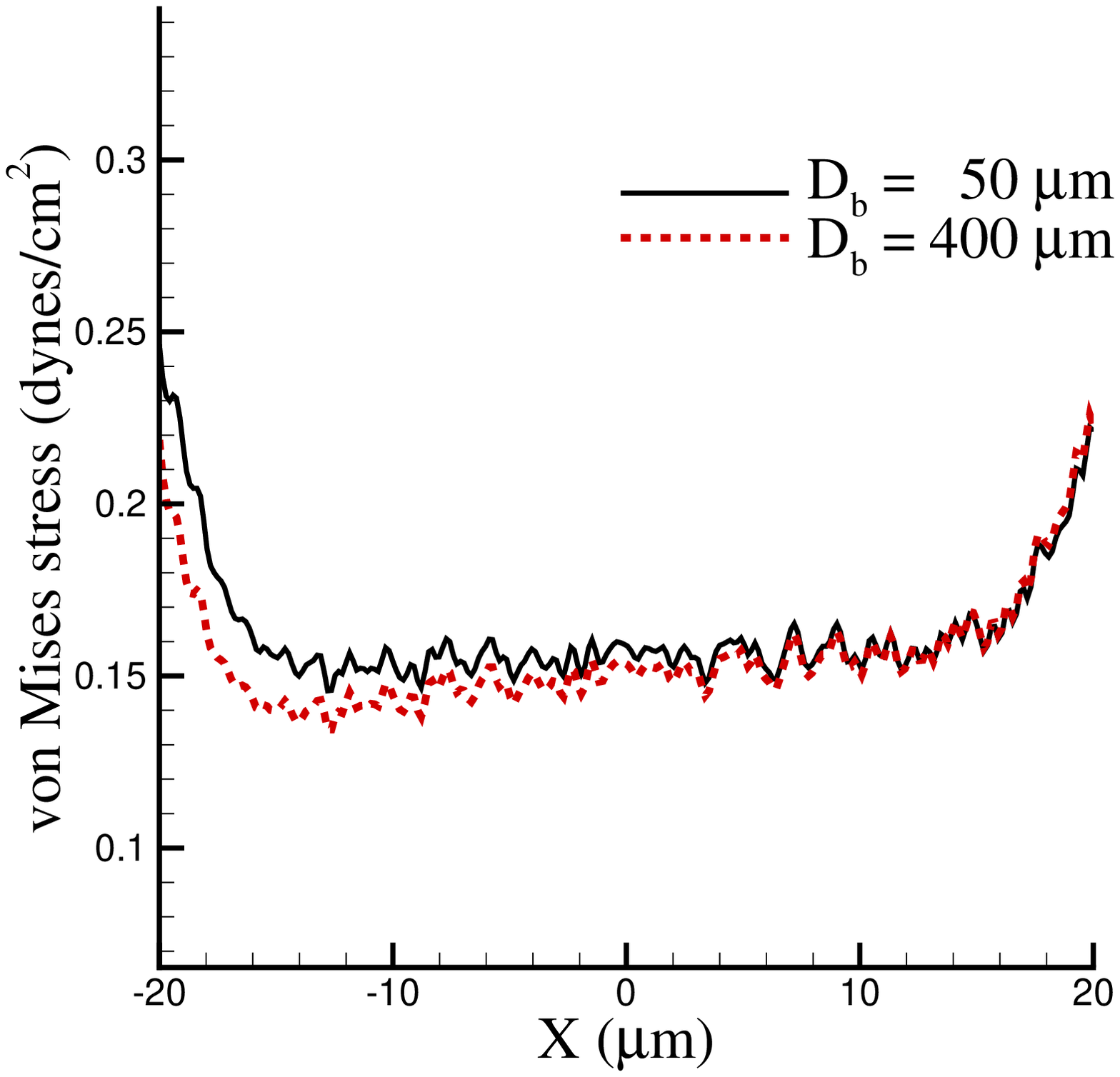}
    \label{fig19:subfigure1}
  }
  \qquad
  \subfigure[\testcase{Sup75}, biofilm-wall interface]{
    \includegraphics[width=0.45\textwidth,clip,bb=53 146 559 653] {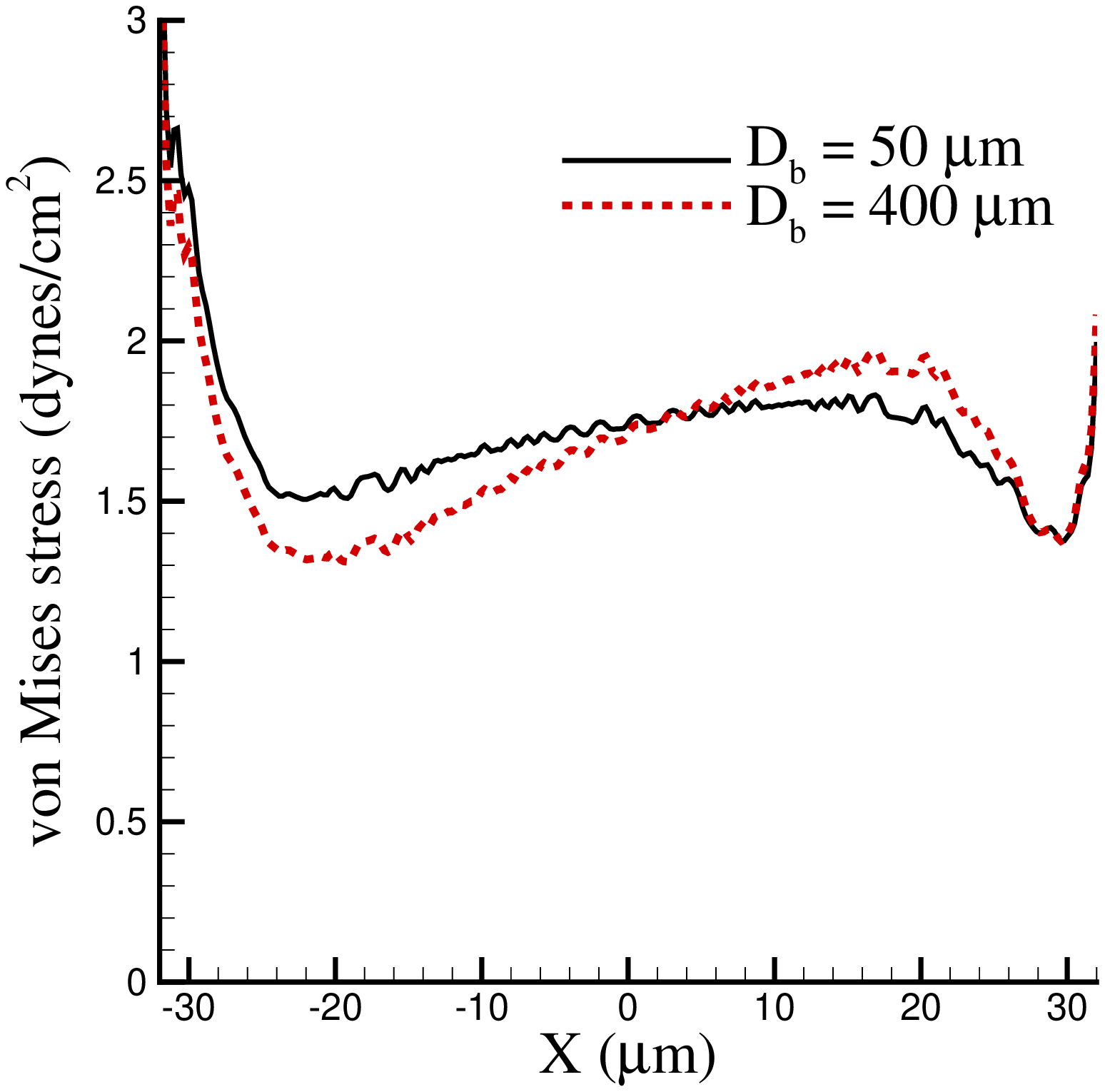}
    \label{fig19:subfigure2}
  }
  \subfigure[\testcase{Semi20}, biofilm-fluid interface]{
    \includegraphics[width=0.45\textwidth,clip,bb=53 146 559 653] {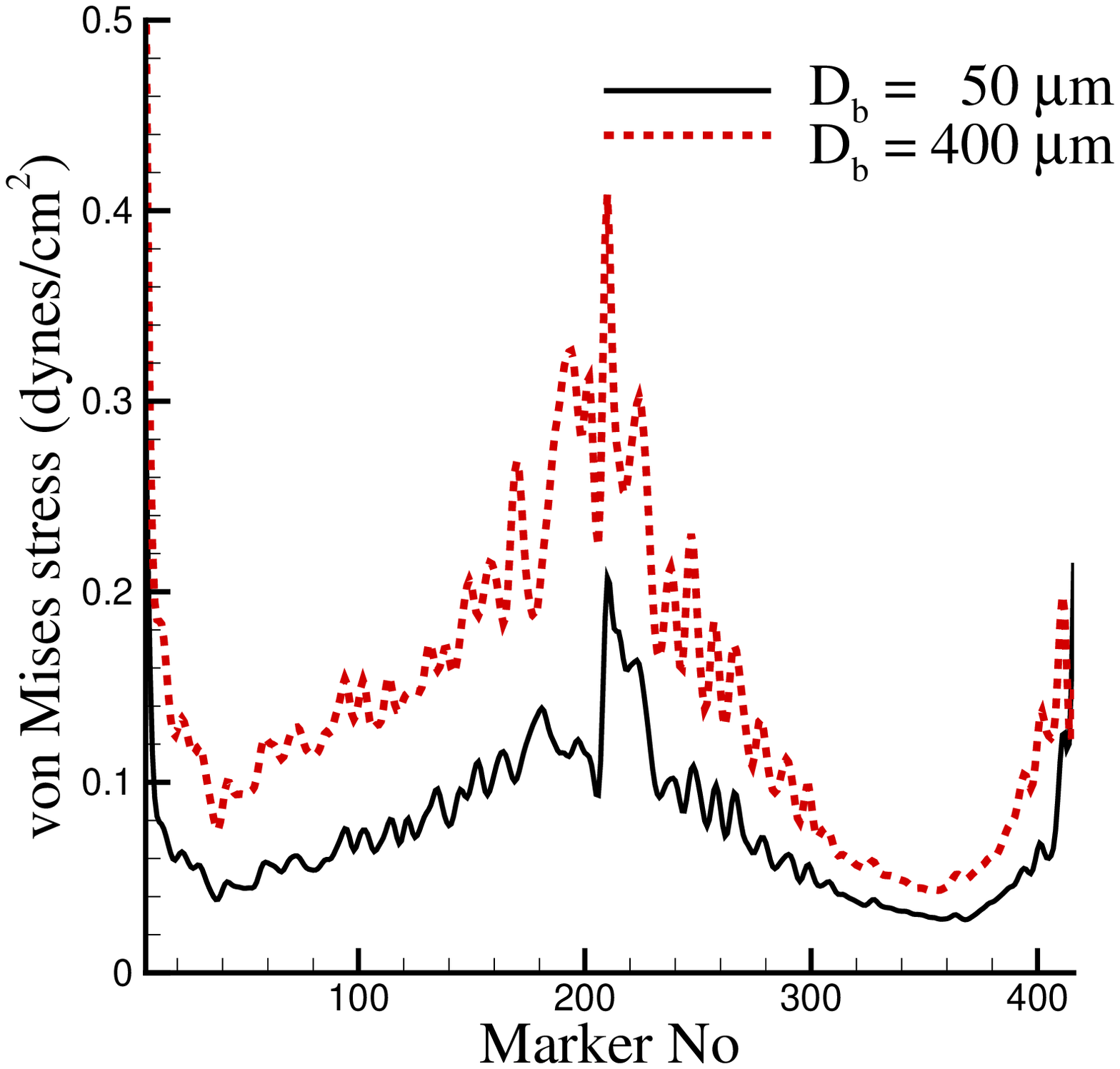}
    \label{fig19:subfigure3}
  }
  \qquad
  \subfigure[\testcase{Sup75}, biofilm-fluid interface]{
    \includegraphics[width=0.45\textwidth,clip,bb=53 146 559 653] {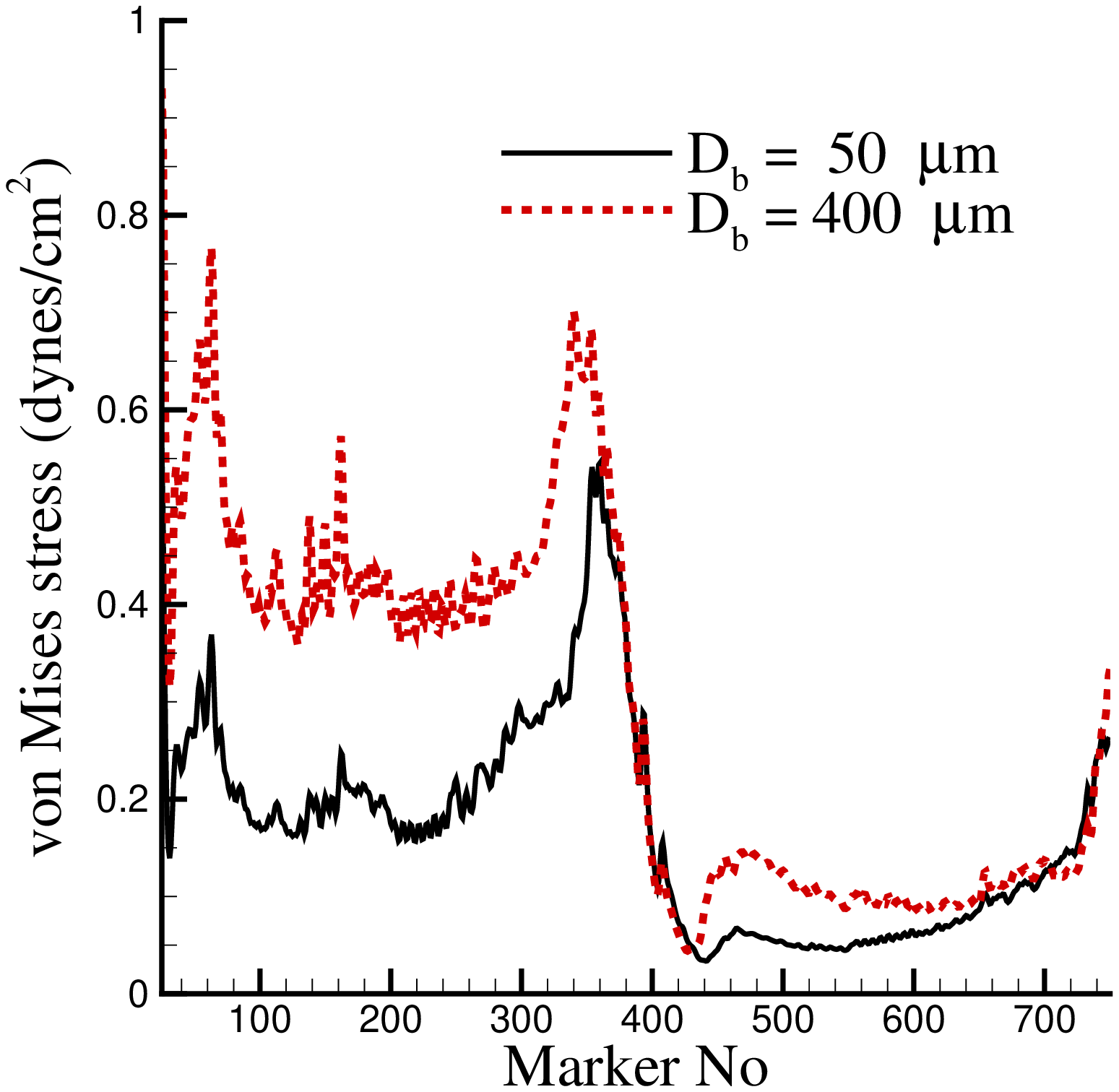}
    \label{fig19:subfigure4}
  }
  \caption{Von~Mises yield stress adjacent to the wall (a,b) and on the
    biofilm-fluid interface (c,d) for the same simulations as in
    Fig.~\ref{Fig18-vonstress_vs_db}.}
  \label{Fig19-wallvonstress_vs_db}
\end{figure}
The corresponding plots of von~Mises stress along the bottom edge of the
colony are shown in Figs.~\ref{Fig19-wallvonstress_vs_db}(a,b) from
which it is clear that the stress peaks at the up/downstream corners.
These plots actually terminate a short distance away (5--8 IB points)
from the corners because the stress maxima at the corners is 8 to 10
times the value in the interior.  We do the same for the stress along
the biofilm-fluid interface in
Figs.~\ref{Fig19-wallvonstress_vs_db}(c,d), from which we see that the
interfacial shear stress attains a local maximum near the biofilm tip
where the fluid shearing force is largest.  On comparing the stress
plots for the wall and biofilm-fluid interfaces, it is clear that
increasing the colony spacing $\Db$ has a much more pronounced effect on
the biofilm-fluid interface than on the wall stresses.  We conclude from
these results that for widely-spaced colonies, detachment is more likely
to occur by surface erosion than by sloughing from the substratum.  This
behavior is to be expected since we have used a relatively large value
of the spring stiffness connecting the wall and biofilm (compared to the
value inside the biofilm).  However, if we were to use a lower value to
mimic weak wall attachment then we could capture the competition between
surface erosion and sloughing modes of detachment, with the dominant
mode being determined by the relative sizes of the adhesive
($\stressAdh$) and exterior cohesive ($\stressCohExt$) stress
thresholds.

Upon closer investigation of the plots in
Fig.~\ref{Fig18-vonstress_vs_db}(c,d), we note that for case
\testcase{Semi20} the von~Mises stress increases along the entire
exposed biofilm surface as $\Db$ increases, whereas in case
\testcase{Sup75} only the left face experiences an increase.  This
suggests that surface erosion is more uniform for circular colonies,
whereas long thin colonies will tend to erode only along the upstream
face.  Based on this result, and the fact that von~Mises stresses are
much larger for \testcase{Sup75}, it seems reasonable to suppose that
the enhanced surface erosion observed in elongated colonies could be a
precursor to formation of \emph{streamers} that are observed in
experiments at low Reynolds number~\cite{rusconi2011secondary,
  valiei2012web}.  An in-depth investigation of streamer formation is
beyond the scope of this work but could form the basis for an
interesting future study based on our IB model.

Additional insight can be gained by comparing plots of
interfacial shear stress computed earlier in
Figs.~\ref{Fig13-intstress_vs_db_bothksp}(a,d) (with $\kodzero=0.75$)
with the corresponding von~Mises stress curves for the same biofilm
colonies in Figs.~\ref{Fig19-wallvonstress_vs_db}(c,d).  For the
\testcase{Semi20} case, the interfacial fluid stress and von~Mises
stress curves have the same general shape along the central portion,
with the main difference being that the von~Mises stress increases
towards the corners whereas the interfacial shear stress does not.  For
the elongated colony in \testcase{Sup75}, both stresses are asymmetric
about the colony apex (near IB point 360) and although both experience
a rapid decrease to the right (downstream) of the apex, the upstream
behaviour is very different.  In particular, the interfacial fluid
stress increases gradually on the left toward the apex, while the
von~Mises stress sustains a relatively large value on the left with a
more rapid rise to the maximum.

These differences just mentioned point to an important error in another
commonly-used detachment approximation based on a \emph{detachment speed
  function}.  In this approach, rather than explicitly solving the
continuum equations for mechanically-induced detachment, they account
for these effects instead by specifying a local speed at which the
biofilm-fluid interface recedes into the
biofilm~\cite{merkey2009modeling, xavier2005}. For example, this
detachment speed function may depend on the local interfacial shear
stress~\cite{duddu2009} or interfacial curvature~\cite{xavier2005}.
Based on the IB results above in which significant differences occur
between fluid shear stress and von~Mises stress along the biofilm-fluid
interface, it is clear that even for a reasonably stiff biofilm a
detachment speed function depending on interfacial shear stress is
incapable of correctly capturing detachment dynamics at all points along
the interface.  It is possible that an alternate speed function could be
found that accounts for the variation in von~Mises stress along the
interface that has been identified in this study and so  this would be an
worthwhile subject for further study.

\begin{figure}[bthp]
  \footnotesize\sffamily
  \centering
  \subfigure[\testcase{Sup25}, $\Db=50\;\mymum$ ] {
    \includegraphics[width=0.45\textwidth,clip,bb=91 234 542 359] {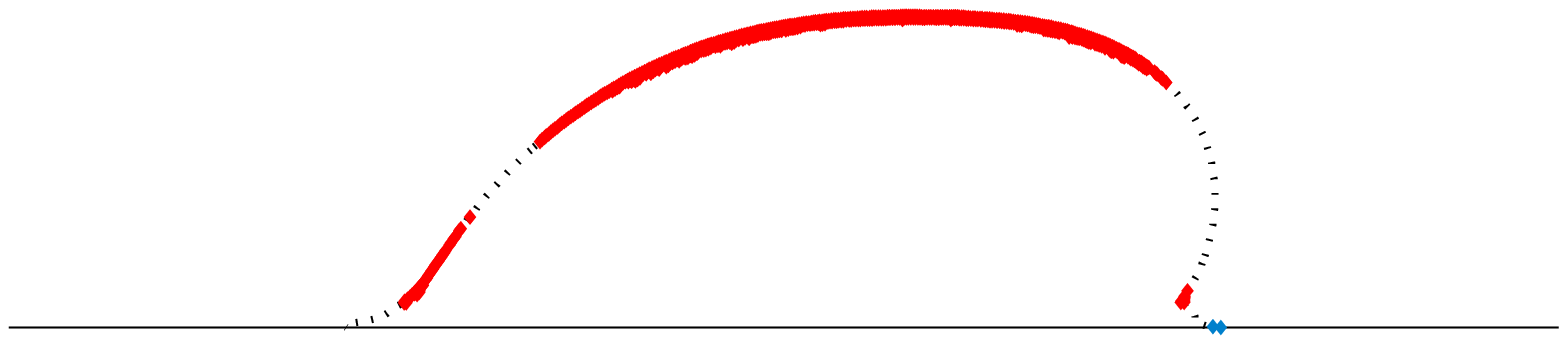}
    \label{fig20:subfigure1}
  }
  \qquad
   \subfigure[\testcase{Sup25}, $\Db=400\;\mymum$] {
    \includegraphics[width=0.45\textwidth,clip,bb=91 234 542 359] {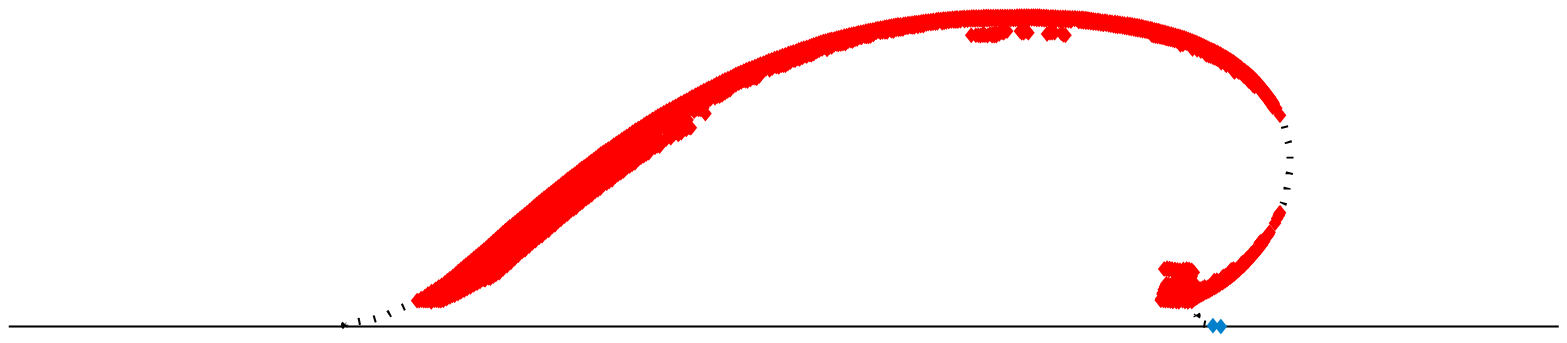}
    \label{fig20:subfigure2}
  }
  \qquad
    \subfigure[\testcase{Sup50}, $\Db=50\;\mymum$] {
    \includegraphics[width=0.45\textwidth,clip,bb=91 228 542 375] {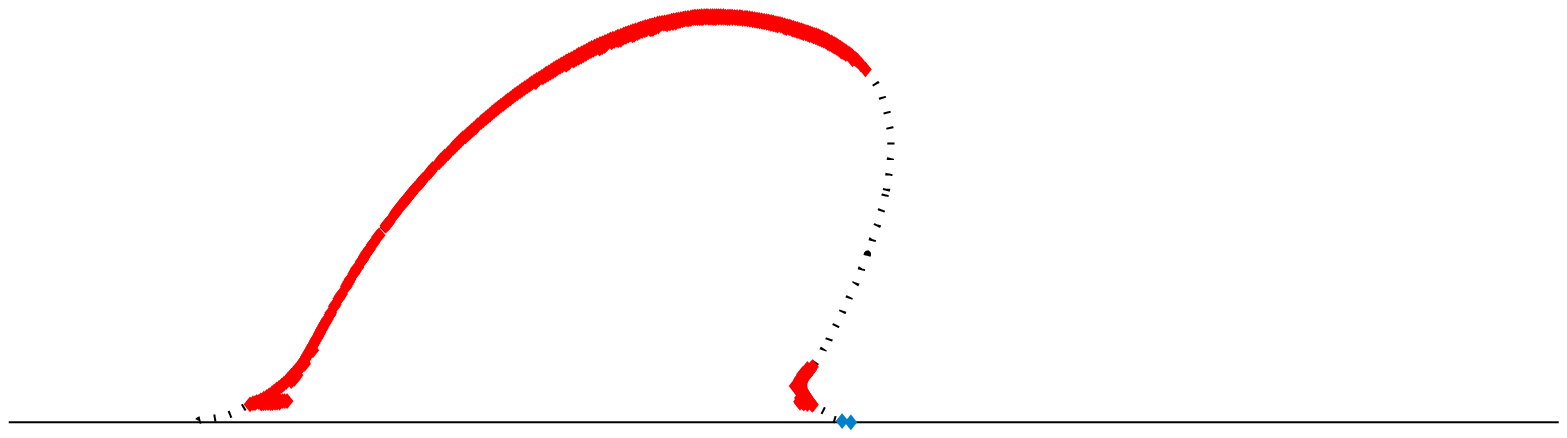}
    \label{fig20:subfigure3}
  }
  \qquad
    \subfigure[\testcase{Sup50}, $\Db=400\;\mymum$ ] {
    \includegraphics[width=0.45\textwidth,clip,bb=91 228 542 366] {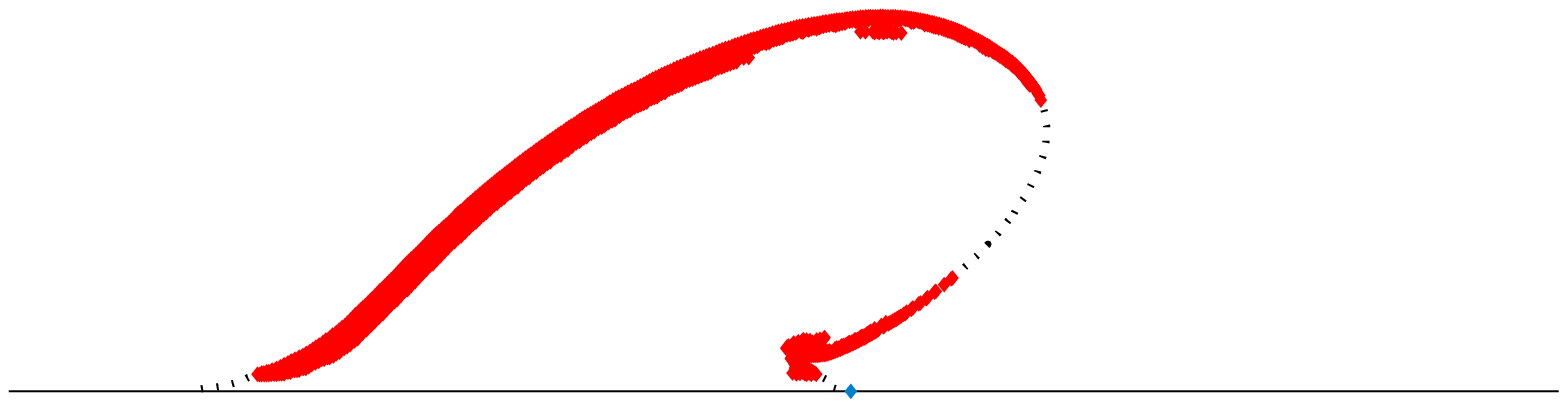}
    \label{fig20:subfigure4}
  }
  \qquad
    \subfigure[\testcase{Sup75}, $\Db=50\;\mymum$ ] {
    \includegraphics[width=0.45\textwidth,clip,bb=91 228 542 417] {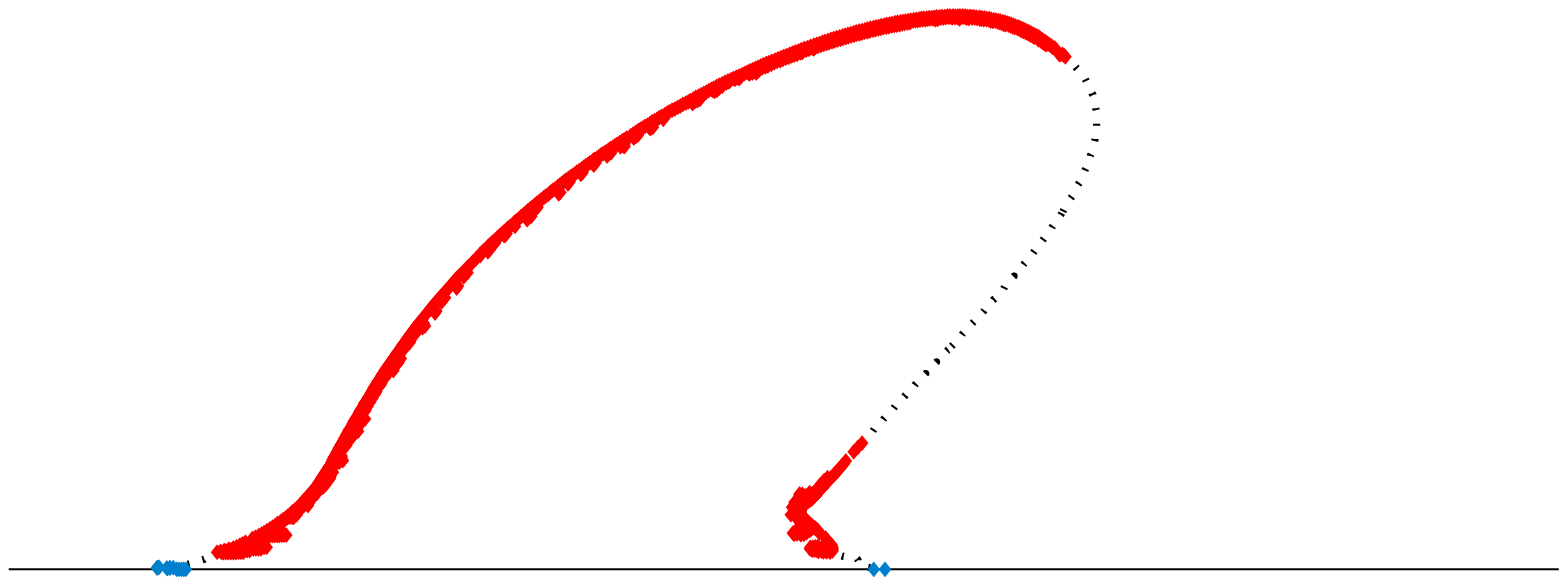}
    \label{fig20:subfigure5}
  }
  \qquad
   \subfigure[\testcase{Sup75},  $\Db=400\;\mymum$] {
    \includegraphics[width=0.45\textwidth,clip,bb= 91 228 542 395] {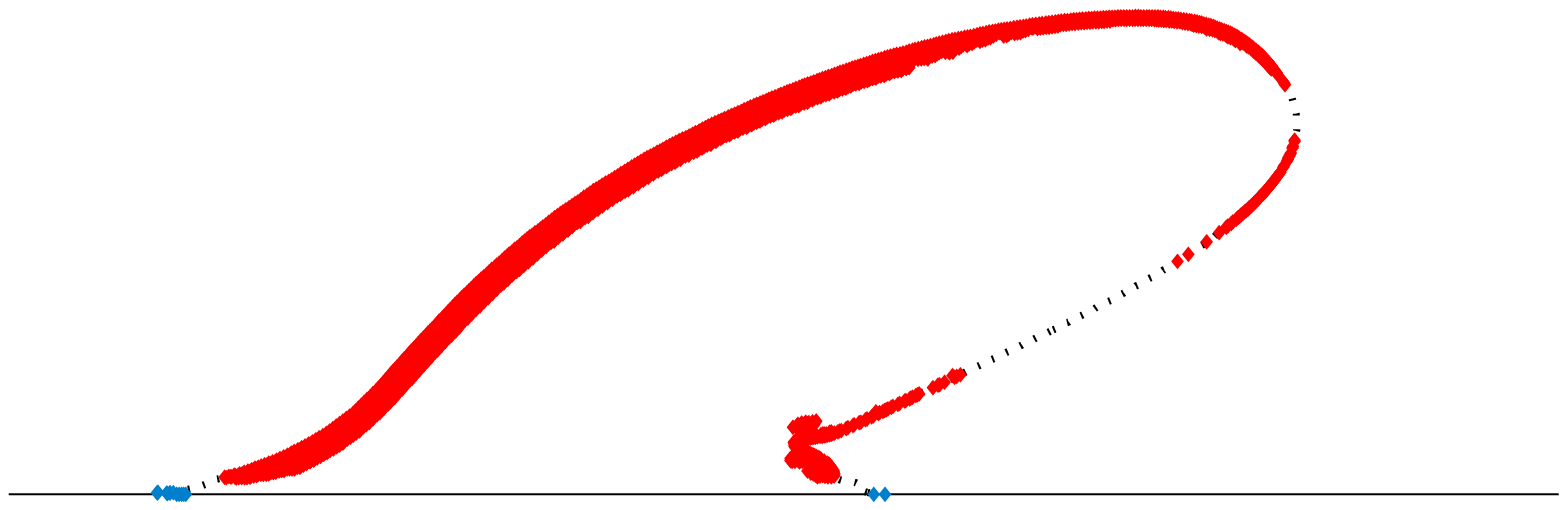}
    \label{fig20:subfigure6}
  }
  \caption{Plots indicating the portion of the biofilm colony (at steady
    state) that will detach for cases \testcase{Sup25}, \testcase{Sup50}
    and \testcase{Sup75} with $\kodzero = 0.75$.  Sub-figures (a,c,e)
    have colony spacing $\Db = 50\;\mymum$, while (b,d,f) have $\Db
    = 400\;\mymum$.  Other parameters: $\deltasub=\deltaext=2\;\mymum$,
    $\stressCohExt=0.1\;\myunit{dyne/cm^2}$,
    $\stressCohInt=1\;\myunit{dyne/cm^2}$ and
    $\stressAdh=5\;\myunit{dyne/cm^2}$.}
  \label{Fig20-detach_vonmises}
\end{figure}

As a further illustration of the detachment process,
Fig.~\ref{Fig20-detach_vonmises} shows the portion of the biofilm that
will detach at steady state for cases \testcase{Sup25}, \testcase{Sup50}
and \testcase{Sup75} with $\kodzero=0.75$ and $\Db=50$ or $400\;\mymum$.
The remaining parameters in the detachment algorithm from
Section~\ref{algo-detachment} are $\deltasub=\deltaext=2\;\mymum$,
$\stressCohExt=0.1\;\myunit{dyne/cm^2}$,
$\stressCohInt=1\;\myunit{dyne/cm^2}$ and
$\stressAdh=5\;\myunit{dyne/cm^2}$.  In practice, the removal of IB
points by detachment will alter the forces acting on the colony, which
in turn induces further deformation; this process repeats until no
further detachment is possible.  Simulations implementing this
alternating detachment/deformation process were conducted for a few
selected cases; however, the time step limitation in our IB algorithm
($\Delta t \approx 10^{-5}\;\myunit{s}$) precluded integrating the
solution over the long time intervals required.  Consequently, the scope
of this work is restricted to introducing the equivalent continuum
stress based detachment strategy and validating the results on a range
of colony shapes.  Efforts to develop a more efficient implementation
for a complete deformation and detachment strategy are currently
underway with a more efficient IB algorithm and will form the basis for
a future publication.

\section{Conclusions}
\label{conclusions}

We employed a 2D immersed boundary method to simulate the
deformation of a periodic array of uniformly-spaced, wall-bounded, weak
biofilm colonies in response to a linear shear flow.  In order to
capture different stages of biofilm growth under mass transfer-limited
conditions, we chose a family of biofilms having the same generic shape
(sections of a super-ellipse) but with increasing aspect ratios (fixed
width, increasing height).  Actual biofilm colonies behave mechanically
like viscoelastic solids and we mimic this behaviour by replacing the
biofilm with a network of Hookean springs corresponding to the edges in a
quasi-uniform triangulation of the colony.

We began by performing a parametric study that investigated the effect
of colony spacing and spring stiffness on the drag/lift forces and
interfacial shear stress acting on the biofilms.  The main results of
this parametric study can be summarized as follows:
\begin{itemize}
\item \emph{Varying spring constant:} At low shear rates, colonies with
  even a moderate spring stiffness of $\kodzero \geqslant 10$ are able
  of resisting large deformation forces.  Weak biofilm colonies with
  $\kodzero \leqslant 1$ experience increased drag and larger
  deformations with a maximum displacement in the 10's of microns.
  These larger deformations are accompanied by a change in the
  interfacial shear stress profile wherein stress increases along the
  upstream face and decreases downstream.

\item \emph{Varying colony spacing:} For low to moderate
  biofilm stiffness ($\kodzero < 10$)
  reducing the colony spacing from 400 to 50\;$\mymum$ reduces the drag
  by as much as 50-100\%, accompanied by a change in the shear stress
  profile along the biofilm-fluid interface.  The interfacial shear
  stress in colonies with large aspect ratio (\testcase{Sup75}) differs
  significantly from ones with small aspect ratio (\testcase{Semi20}).

\item \emph{Varying colony shape:} It is possible for
  biofilm colonies to grow into tall structures with large aspect ratio
  even if they are weak mechanically ($\kodzero < 10$) owing to the
  protection from surface erosion afforded by being in close spatial
  proximity to other colonies. We believe that this result
  will carry through to 3D (although to a lesser degree) and this is an
  issue that we plan to investigate further in a future study.
\end{itemize}

We also developed a new method for
initiating biofilm detachment using averaged equivalent continuum
stress, which we implemented within our IB framework.  We overcame
problems encountered in other spring strain based detachment
strategies such as in~\cite{hammond2012} and~\cite{alpkvist2007}. The
following conclusions can be drawn regarding detachment:
\begin{itemize}
\item \emph{Variation in von~Mises stress:} Increasing the spacing
  between colonies leads to an increased tendency
  for surface erosion instead of sloughing from the wall. Based on the
  von~Mises stress along the biofilm-fluid interface, we concluded that
  semi-circular biofilm colonies will undergo roughly uniform surface
  erosion, while colonies with larger aspect ratios erode predominantly
  along the upstream face.

\item \emph{Correlation between fluid stress and von~Mises stress:} The
  fluid stress and von~Mises yield stress along the biofilm-fluid
  interface differ substantially.  We conclude that biofilm dynamics
  based on a detachment speed function approach~\cite{duddu2009,
    merkey2009modeling} (where detachment speed is a function only of
  local interfacial fluid stress) cannot capture the actual detachment
  behaviour resulting from excessive straining. This highlights the
  importance of using detachment strategies that accurately capture the
  biofilm mechanics.
\end{itemize}

The main advantage of our detachment algorithm is that it provides a
uniform framework for handling biofilm deformation, surface erosion and
sloughing through the use of a continuum mechanics-based detachment
model that employs measured biofilm mechanical properties.  The primary
disadvantage of our approach as implemented herein is the high
computational cost; in particular, the small time step required for
stability reasons combines with the extra work of simulating detachment
to make long-time computations of simultaneous deformation and
detachment impractical.  We emphasize that this is not a limitation of
the immersed boundary approach, but rather our specific implementation
that uses a simple explicit time-stepping strategy.  We conclude
therefore that the full potential of our detachment algorithm can only
be realized in combination with either a (semi-)implicit time-stepping
approach or an efficient parallel implementation (such
as~\cite{wiensStockie2015}), which will be the subject of future work.

\section*{Acknowledgments}

We acknowledge funding for this research from the Natural Sciences and
Engineering Research Council of Canada, as well as the Mitacs and AFMNet
Networks of Centres of Excellence.

\bibliographystyle{plain}
\bibliography{cicp_sudarsan_biofilm_imb}

\end{document}